COVE: A PRACTICAL QUANTUM COMPUTER PROGRAMMING FRAMEWORK

By

Matthew Daniel Purkeypile

M.S. Computer Science, American Sentinel University, 2005

B.S. Computer Science, American Sentinel University, 2004

A Dissertation Submitted to the Faculty of Colorado
Technical University in Partial Fulfillment of the
Requirements for the Degree of Doctor of Computer
Science

Colorado Springs, Colorado

September 2009

COVE: A PRACTICAL QUANTUM COMPUTER PROGRAMMING FRAMEWORK

By
Matthew Daniel Purkeypile

This Dissertation Proposal is Approved

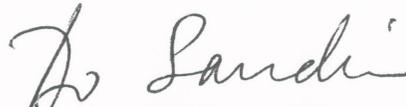

---
Bo I. Sandén, Ph.D., Professor of Computer Science
Director of Dissertation

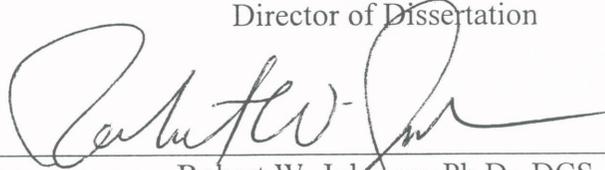

---
Robert W. Johnson, Ph.D., DCS

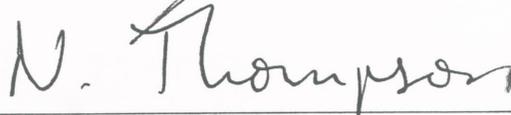

---
Nigel Thompson, DCS

9/15/09
---
Date Approved

# ABSTRACT

While not yet in commercial existence, quantum computers have the ability to solve certain classes of problems that are not efficiently solvable on existing Turing Machine based (classical) computers. For quantum computers to be of use, methods of programming them must exist. Proposals exist for programming quantum computers, but all of the existing ones suffer from flaws that make them impractical in commercial software development environments. Cove is a framework for programming quantum computers that extends existing classical languages to allow for quantum computation, thus providing a quantum computing toolkit for commercial software developers. Since the target users of Cove are commercial developers, it is an object oriented framework that can be used by multiple languages and also places emphasis on complete documentation. The focus of Cove is not so much on the software product, but on the fundamental concepts that make quantum computing practical for common developers.

DEDICATION

For Angela and Scarlett.

## ACKNOWLEDGMENTS

I'd like to thank Colorado Technical University for providing an excellent Doctoral program. It should go without saying that this work would not be possible without the extensive guidance and input from my committee: Dr. Sandén, Dr. Johnson, and Dr. Thompson. They've spent countless hours reading and rereading this dissertation, and their feedback and insights have resulted in a much more complete and solid dissertation. Thank you.

I would also like to thank Leo Chan and Stephane Dubois at Xignite for their support through this lengthy endeavor. They have fostered an excellent environment at Xignite that has allowed me to be involved in great projects while still being able to carry out this research.

Thanks also goes to Nine Inch Nails and Atmosphere, whose music I've listened to more times while working on this dissertation than my wife cares for.

My family and friends have also lent endless support and encouragement. My parents Bob and Jan have always supported my education and I'd like to thank them. Finally I'd like to thank my Wife Angela and Daughter Scarlett for their support. They have sacrificed much more than I have to make this possible.

TABLE OF CONTENTS

# LIST OF TABLES

# LIST OF FIGURES

CHAPTER I

INTRODUCTION

While still years away from being commercially viable, quantum computers hold the power to carry out computations that are not feasible on current "classical" computers. Quantum computers are different from the classical computers we are familiar with in very fundamental ways– in some respects they represent a new computing paradigm. Like any computer though, quantum computers are of little value if there is no useful software to run on them. In order to immediately take advantage of quantum computers when they are commercially available, practical techniques to program them must be developed.

While the idea of quantum computing first appeared in the 1980's, programming them did not receive much consideration until the 1990's. Nonetheless, many of the existing proposals for programming quantum computers suffer from one or more of the following flaws, making them impractical or expensive for use in a typical commercial software development environment:

- Foreign techniques – the proposal utilizes techniques that are foreign to a majority of commercial developers. The use of these foreign techniques represents a significant learning curve, and thus expense, for any software development organization wishing to utilize them.

- Not scalable– the proposal only works well for small "snippets" of code; beyond that it becomes difficult to manage and understand. Graphical languages and languages requiring formal proofs fall under this category. Commercial software is typically large and complex, and any approach that cannot be scaled up to these large software systems is impractical[1]. For life critical software the use of formal proofs is acceptable, but the cost to do so is prohibitive for typical commercial applications.

---

[1] The quantum part of the entire computation is often a subset, but still significant. As will be shown in Shor's factoring algorithm, the quantum part of the computation is far from trivial.

- Proprietary language– In addition to the quantum programming languages covered in this dissertation, over 8,500 classical languages have been developed to date [1], yet only a very select few see use in the commercial domain. Furthermore, languages developed for the purpose of quantum computing are unlikely to be adopted because they lack the features and power of popular classical languages that are already in use. It is also doubtful that the classical parts of a quantum language would keep pace with advances made in classical languages [2]. The focus of quantum programming techniques should be on quantum computation; also trying to include classical programming methods distracts from the goal of quantum computation and forces the designer to tackle classical issues that are already well addressed in existing languages.

- Difficult to integrate with existing software– it is unlikely that entire code bases will be rewritten solely so certain parts, such as finding records in a database, can take advantage of quantum computers. Quantum computers will likely be a resource that is used by a classical computer [3]. Consequently the programming techniques must easily integrate with classical computers.

- General usability/unconventional framework design– languages and libraries that are easy to use utilize many common conventions. Application Programming Interfaces (APIs) and frameworks that do not follow established conventions are thus difficult to work with and prone to being used incorrectly while also having a significant learning curve. While the programming community adapts to poor frameworks and APIs if needed, this adaption comes at a cost. Included under this point is thorough documentation. Without it there is even more of a learning curve.

- Remoteness – quantum computers may be remote resources [3], much like web services. Programming techniques for quantum computers should take this into account by providing the ability to specify the location of the quantum resource. As an example, one may change the location of a web service call or a file without altering any other code that utilizes that resource.

To carry out quantum computing, three things must be done: initialization to a classical state, manipulation through reversible operations, and measurement to obtain a classical result. Although quantum computer programming may appear simple given what it must accomplish, these various flaws in existing programming proposals show that developing quantum programming techniques is far beyond a trivial task. David

Deutsch, one of the fathers of quantum computing, said in his seminal 1985 paper: "Quantum computers raise interesting problems for the design of programming languages…" [4]. More than twenty years later we are still encountering problems with creating methods to program quantum computers.

Most software is written by commercial developers, and a majority of commercial developers do not even possess a Bachelors degree in a software related discipline [5]. This means that they likely have not been exposed to more exotic programming techniques such as functional languages or formal proofs of software. It isn't unreasonable to expect commercial developers to also utilize quantum computers in the future. If quantum computers are to be used in real world applications instead of being idling black boxes, any technique to program them must make these practitioners the primary target group.

This dissertation outlines a proposal for the construction of a quantum computing framework called Cove[2], which focuses primarily on usability[3]. Frameworks give programmers the tools to solve certain problems [6]; in this case the problem is how to program quantum computers to carry out computations that cannot be efficiently done classically. Specifically this project entails:

- An object oriented design of a framework. By utilizing an object oriented framework users may extend parts as needed while also making use of the supplied components as provided.

- Examples that show the framework can carry out common quantum computing tasks and algorithms. Three examples are utilized throughout this

---

[2] Cove is named after the author's hometown of Shady Cove, Oregon, United States of America.
[3] While usability measurements are largely subjective, the existing literature on quantum computer programming neglects the topic. It is hoped by making this the primary focus of Cove that there will be more emphasis on the usability of quantum computer programming methods.

dissertation: tossing quantum coins, entanglement, and factoring (Shor's algorithm).

- An implementation of the framework that simulates a quantum computer[4] on existing classical computers as a proof of concept.

Furthermore this dissertation makes the following contributions to the field:

- A summary of framework design literature. This has not yet been encountered in the literature.

- A list of functional properties a quantum programming approach must satisfy.

- A design of a quantum programming framework, called Cove.

- A prototype of a classical simulation for the implementation. This allows Cove to be used on a limited scale on existing classical computers.

One may ask why the proposed framework itself does not contain all the known algorithms that a quantum computer could carry out. It is certainly possible that it could contain most or all of the currently known algorithms, such as Shor's factoring algorithm. However, the purpose of the framework is to provide the components necessary to implement these known algorithms and future ones. Simply exposing existing algorithms does not allow for new ones to be implemented using the framework, although those algorithms should still be exposed to enhance the usefulness of the framework. Even if the algorithms are exposed, this framework will also allow those exploring quantum computing to actually implement the algorithms themselves– much the same way a student in a data structures course will build data structures that may already be provided.

Figure 1 shows where Cove and other proposals fit into the various levels of abstraction between the hardware and application. In this figure the rectangles are meant to illustrate various levels of abstraction ranging from the actual hardware (bottom), which is at a very low level of abstraction, to an application (top), which is at the highest level of abstraction since a user does not even need to be aware they are utilizing quantum hardware.

---

[4] As will be detailed later, this simulation faces an exponential slow down.

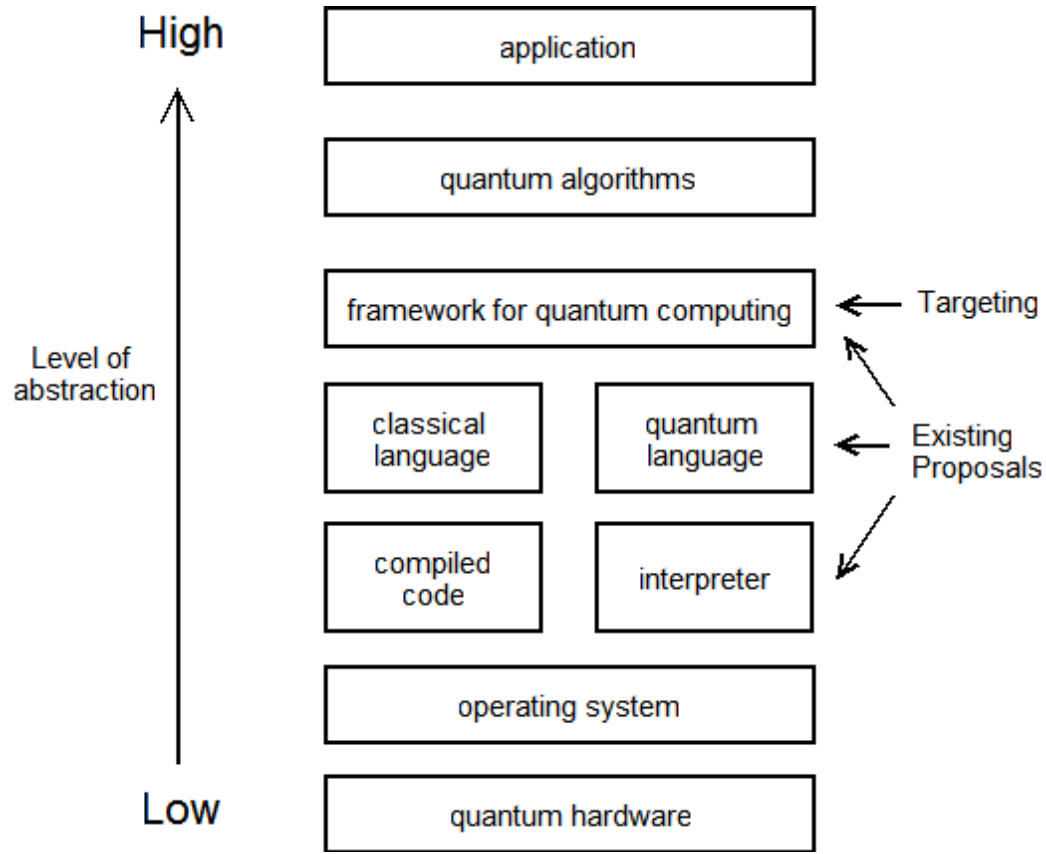

Figure 1. Level of abstraction targeted by Cove.

The core challenge lies not only in developing a language or framework capable of programming a quantum computer, but in developing a method that is usable for the common commercial programmer. While quantum computing is not needed for every application, making it more usable reduces development costs when it is needed.

The remainder of this dissertation is organized as follows:

- Chapter 2– Covers the literature related to quantum programming framework design. This generally falls into three categories: quantum computing and background (section 2.1), quantum computer programming (section 2.2), and framework design (section 2.3).

- Chapter 3– Lays out the goals, as well as criteria for judging they have been successfully met.

- Chapter 4– Outlines the methodology, which is a prototype of the framework that includes a locally simulated quantum computer to execute the code on.

- Chapter 5– Discusses the design and implementation of Cove, in addition to the rationale behind the decisions. This chapter also includes how the proof criteria are satisfied.

- Chapter 6– An analysis of Cove, including stumbling blocks and some alternates considered.

- Chapter 7– A summary of this dissertation.

- APPENDIX A– Describes how various electronic resources which are outside the scope of this paper can be obtained. Resources include the complete source code (written in C#), help, development web log (blog), and presentations.

CHAPTER II

BACKGROUND AND RELATED WORK

Several areas of computer science and physics form the foundation for creating a practical quantum computer programming framework. General quantum computing and quantum mechanics are covered in section 2.1, which provides the background needed for the rest of the dissertation– quantum computers must be covered before programming them can. Section 2.2 covers work related to quantum computer programming, including a survey of different methods that have been developed for programming quantum computers. Since the focus of this dissertation is the design of a framework for programming quantum computers, section 2.3 covers work related to framework design and use.

## 2.1 Quantum Computing

As computer technology marches forward in accordance with Moore's Law, it is estimated that useful quantum computers will appear in the year 2021, plus or minus five years [7]. Quantum computers operate differently than Turing machines which were introduced in the 1930s [8, 9], and on which all modern computers are based. This section provides the background necessary to understand quantum computation so that the programming challenges can be intelligently discussed.

*2.1.1   Basics of Quantum Mechanics*

A basic understanding of quantum mechanics is necessary to understand how quantum computers work, as well as to help comprehend their limitations and the difficulties constructing them. This section is not intended to be a thorough primer on quantum mechanics, but only the minimal introduction necessary to understand some key examples and how they pertain to quantum computation. As this dissertation is on quantum computer programming and not on quantum mechanics or quantum information, more complex topics including Hilbert spaces, quantum information protocols (such as superdense coding), evolution of quantum systems, and quantum states have little coverage in this dissertation, if at all. The reader is referred to [10-13] for a more thorough introduction to quantum mechanics as it relates to quantum computing.

Quantum generally refers to things at very small sizes: electrons, photons, and so on. At this scale nature is non-deterministic: given an arbitrary quantum state there is no way to determine with certainty exactly how it will evolve. This is fundamentally different than the Newtonian (frequently called "classical" in the literature) world we are familiar with, where the evolution of a system is deterministic. A Newtonian example would be the orbit of a planet around a star: given a specific location and velocity we can determine where the planet will be at a future point in time. At the quantum level this determinism doesn't hold; as an example, a photon can be in two places at once. The fact that things at the quantum scale are not absolute and cannot be determined with certainty is deeply disturbing and confusing to many people. Einstein spent the later part of his life trying to disprove this counterintuitive nature of quantum mechanics, referring to entanglement as "spooky action at a distance" [10, 14]. As odd and unnatural as quantum mechanics may seem it has been experimentally proven, even with experiments that can be performed at home [15].

Classical physics are familiar to us because they govern what we see in our daily lives. Extending the classical view point further, if we somehow had a computer that could track every bit of matter and all the forces acting upon them, then the Universe would be completely predictable and thus deterministic. Under this worldview the only thing that keeps the Universe from being predictable is the sheer complexity of the system.

This classical view of the world remained for hundreds of years until the late nineteenth and early twentieth centuries [14]. It was at this time that the quantum view of the world was uncovered. If the classical view can be described as deterministic, then the corresponding description for the quantum view is nondeterministic. At the quantum level nature itself is unpredictable– identical situations can lead to different outcomes. Furthermore, we cannot even gain complete knowledge of many systems. As an example Heisenberg's famous uncertainty principle states: "we cannot know the precise position and momentum of a quantum particle at the same time" [14], illustrating that we cannot obtain complete knowledge of an arbitrary system.

Perhaps even stranger than not being able to measure both position and momentum of a particle is the fact that particles can literally be in two places at once. This is commonly illustrated by the two slit experiment that will be detailed and shows that a photon can pass through two slits at once. The only catch is that as soon as we try to observe the particle it is said to "collapse" into a single position [16].

It should be pointed out that this term of "collapsing to" a particular state is taken from what is known as the Copenhagen interpretation of quantum mechanics, which is a popular and well known view. Although not as popular, another interpretation of quantum mechanics is the many worlds interpretation, first proposed by Everett in 1957, with objections by famous physicists such as Niels Bohr [17, 18]. In the many worlds interpretation when "collapsing", the universe splits, and what we observe is one of those branches of the split. This means that the universe is constantly splitting, and as a result every possible outcome exists in some branch. While not as popular as the Copenhagen interpretation, the many-worlds interpretation is advocated by one of the fathers of quantum computing, David Deutsch [4, 19]. In fact he sees it as the obvious interpretation of quantum mechanics [19]. Regardless of the interpretation of quantum mechanics one subscribes to, in order to be more consistent with the popular terms in the literature the term "collapse" will be used throughout this dissertation when referring to the outcome of a quantum measurement.

The Bohr model of the atom that many students learn in high school is easy to understand yet contains a serious deficiency. The deficiency is that it does not take into account the uncertainty of the locations of the particles that make up an atom, such as electrons. In the Bohr model electrons orbit the nucleus of the atom much like a planet orbits a star [14, 20]. Figure 2 is the Bohr model of a heavy hydrogen (deuterium) atom, where all particles have an absolute location. In this figure the solid green circle (in orbit) is an electron, solid blue (the right one in the nucleus) a neutron, and solid red a proton (the left on in the nucleus). The outer light green circle represents the orbit of an electron around the nucleus.

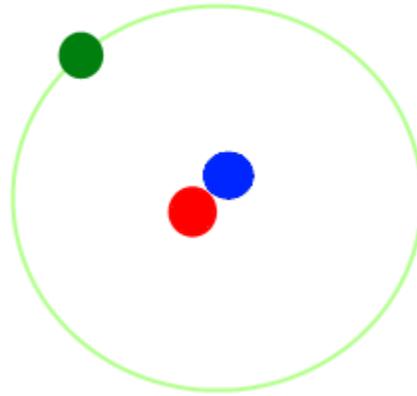

Figure 2. Bohr model of a heavy hydrogen (deuterium) atom.

The reality is that there is not a particular location of where the electron is, but merely a cloud of probability indicating where it is likely to be [14]. Figure 3 is a more accurate representation than Figure 2, where the location of the particles is defined by a probability distribution. In Figure 3, the darker the shade the higher the probability is that the particle will be in that location when observed. The density of the shading in the outer green circle represents the likelihood of observing the electron at that location[5]. The shaded blue and shaded red areas represent the locations of the neutron and proton, respectively. Figure 3 is based in part on an example in [20].

---

[5] There is a very small, but nonzero, probability that the electron could be well outside the shown area.

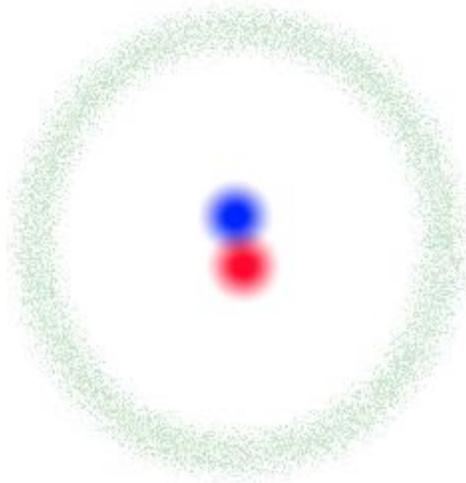

Figure 3. The more accurate probability distribution of heavy hydrogen (deuterium).

An often used example in illustrating quantum weirdness is a photon of light. When a photon is observed it acts like a particle; in the absence of observation it acts more like a wave. When people talk about the dual nature of light, it is the wave and particle nature of light they are referring to. This dual nature has been experimentally verified by what is known as the two slit experiment, called so because light passes through two slits in a plane. The two slit experiment is also the example often used in illustrating these two properties of light.

The following example of the two-slit experiment is based on the example given in [14] by Al-Khalili. In the two-slit experiment we take a light source that emits a photons one at a time. A stream of photons is then shot at a plane that has two slits in it. Behind the plane is a photon detector. Figure 4 shows the setup of the experiment, with the photon source on the left, the plane with two slits as the broken line in the center, and the detector on the far right. In this example the photons are fired from the left and travel to the right.

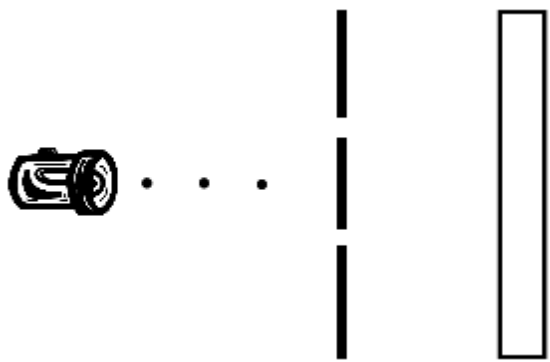

Figure 4. Setup of the two-slit experiment.

When the photons are *not* observed going through the splits, they act like waves. In this sense the photon can be considered to be going through both splits at the same time. This would be much like dropping a rock in a pond– the waves would radiate outward and each wave would pass through the two slits. Outward from each of the slits a new set of waves radiates. These waves interfere with each other and create an interference pattern on the photon detector. On the photon detector the waves reinforce each other where photons are detected; where none are detected they cancel each other out. It is this constructive and destructive interference of waves that can be utilized for computation, as will be later shown. It should be noted that this isn't due to multiple photons going through the slits. If the emitter is slowed down to one photon at a time then the result is the same– an interference pattern emerges. Figure 5 illustrates this case when the photons are not observed going through the slits, as you can see two sets of waves emerging from the plane with the slits and creating an interference pattern on the detector. They key point in Figure 5 is that we see this interference pattern even though the photons are being emitted one at a time.

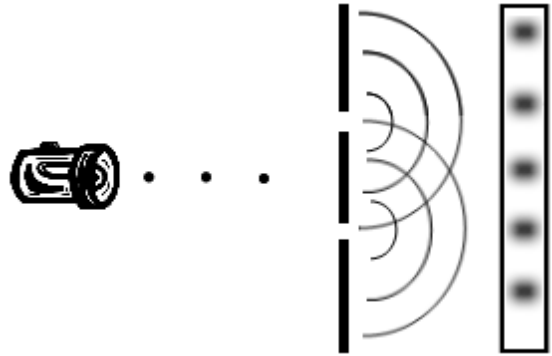

Figure 5. Two-slit experiment when the photons are not observed going through the slits.

An interesting thing happens if we try to observe the photons going through the slits: in this case the photons will only go through one slit or the other. When observed the photons act like particles instead of waves. Since each photon now goes through one of the slits there is no longer an interference of wave patterns. Instead we will only see two areas on the photon detector where each photon hits. In this case the image of the photon is more like sand is being poured through each slit instead of waves– each photon passes through one slit and not the other as Figure 6 shows. As soon as we stop trying to record which slit the photon goes through the interference pattern returns, even though the photons are being emitted one at a time[6].

---

[6] The pattern on the wall could be considered an observation, but the interference pattern emerges when we cannot determine which slit the photon passed through.

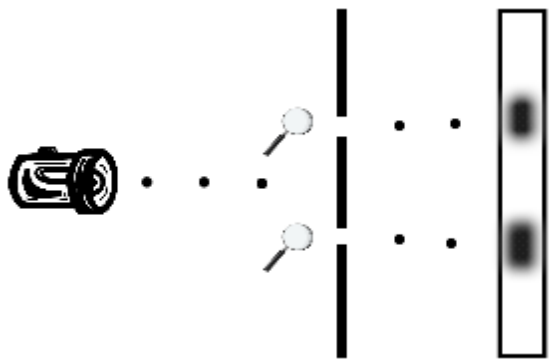

Figure 6. Two-slit experiment when observing the photons going through the slits.

The idea of interference of waves when not observed can be expanded to illustrate behavior that cannot be carried out classically. A beam splitter can be placed in front of the photon source. This splitter will randomly send the photon in one of two possible directions where a photon detector is placed. The photon detectors observe the photons, so the photons behave like particles instead of waves. (As long as the path of the photon can be determined, it is considered an observation and thus cannot be in two places at once.) Figure 7 details this scenario where a beam of photons is split and each path observed by a detector.

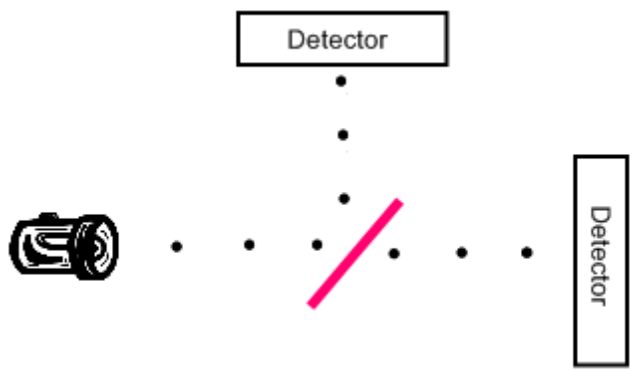

Figure 7. Splitting a stream of photons.

If the detectors are removed and the photons are not observed, then the behavior reverts to that of a wave. In this case the photon can be considered to travel both paths. In the real world the photon will eventually be observed[7], so we are only interested in the case of it not being observed in the theoretical sense at this point.

If the stream of photons is split twice without observing them along the way they end up only being detected on one of the detectors. This is because the stream will act like waves– for one of the paths they will cancel each other out and reinforce the other path. This is only possible because of the wave like behavior when not being observed. This concept plays an important role in quantum computation, as will later be shown in the quantum coin toss example. If an observation is placed before the detectors then the stream again acts like particles and the photons will be evenly distributed between the two detectors.

Figure 8 shows this case when splitting twice results in detection at only one of the two detectors. In Figure 8 the photons are not detected at the top detector because the probabilities have canceled out, while on the right detector they have reinforced. This figure can be easily traced through. The photon emerges from the emitter with a rotation of "up", and it should be noted that these up and down labels are arbitrary. When encountering a beam splitter the photon will retain its orientation if it passes through; if it is reflected then it is rotated by 90 degrees counter clockwise, as in "up" to left. The rotation is flipped when a mirror is encountered, from "left" to "right" for example. As an example we can trace one of the four possible paths.

---

[7] That observation does not necessarily have to be by a detector.

1. The photon leaves the emitter with an orientation of "up".

2. The first beam splitter is encountered. The photon is redirected instead of passing through, so its orientation is rotated 90 degrees and becomes "left". At this point it is headed to the upper mirror.

3. The upper mirror is encountered, flipping the photon from "left" to "right".

4. The second beam splitter (the upper right one) is encountered. The photon passes through so it retains its orientation of "right".

5. The photon is detected at the right detector.

The behavior of the other three paths can be traced out following the same logic: a mirror flips the orientation, a splitter rotates it 90 degrees counter clockwise if redirected, else a splitter leaves the orientation unchanged. The entire example in Figure 8 is based off a similar example given in [21].

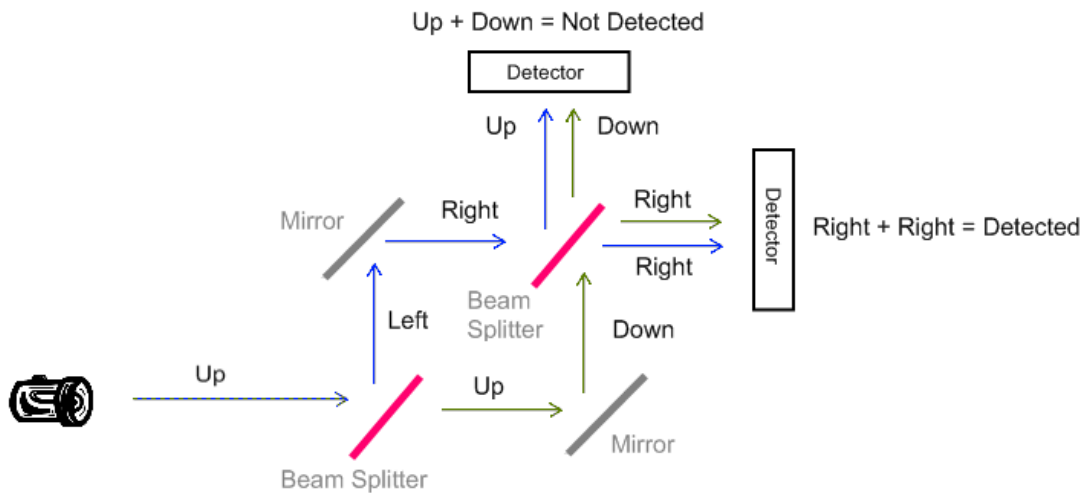

Figure 8. Splitting a beams of photons twice, resulting in photons only being detected at one of the detectors.

Quantum mechanical states can be represented by matrices of complex numbers [22]. Dirac notation is a more condense form that is often used to express these states in quantum mechanics [12], and by extension also quantum computing. Thus Dirac notation and matrix notation are different representations of the same states. In Dirac notation column vectors are expressed using "kets". So the column ($1 \times n$ matrices) vector "a" would be expressed as $|a\rangle$. Row vectors ($n \times 1$ matrices) are expressed using "bras" [23], so the row vector "b" would be expressed as $\langle b|$. States are always represented by these vectors, but operations are not and will be detailed later. Figure 9 shows the general form of these vectors of complex numbers and is based in part on [23].

$$|x\rangle = \begin{bmatrix} x_1 \\ x_2 \\ . \\ . \\ . \\ x_n \end{bmatrix} \qquad \langle y| = \begin{bmatrix} y_1 & y_2 & . & . & . & y_n \end{bmatrix}$$

Figure 9. General representation of kets (x) and bras (y) and their equivalent matrix form.

The matrix representation represents the probabilities of the system being in the particular state when observed. For a single (qu)bit the topmost entry in the matrix represents the probability of being in state 0 when observed, the second entry is the probability of collapsing to state 1. For classical bits the probability is 1 that it is in a particular state, as a bit is either one or zero and thus can be expressed in matrix form as Figure 10 demonstrates. In this dissertation matrices will be enclosed in brackets in accordance with the notation in [13, 24, 25], although parenthesis are also used in the literature [23].

$$0 = |0\rangle = \begin{bmatrix} 1 \\ 0 \end{bmatrix} \qquad\qquad 1 = |1\rangle = \begin{bmatrix} 0 \\ 1 \end{bmatrix}$$

Figure 10. Representation of the classical bits 0 and 1.

Since the entries in the matrix are probabilities, they must add up to 1, as in the case of Figure 10, $P_0 + P_1 = 1$. Bits are absolutely 0 or 1, hence the probability of 1 in either the top position for the bit value of 0, or in the bottom position for the bit value of 1. Since a bit is either 0 or 1, and not some other state, Figure 10 shows the only possible states for bits.

Throughout this dissertation, column vectors, and thus kets, will be more frequently used. Dirac notation is used because it is more compact than the corresponding matrix representation, and thus much more readable for larger systems than the corresponding matrix notation.

Qubits[8][9] are the smallest unit of quantum information, as their name is an abbreviation of "quantum bit". The smallest unit of information in a classical system is a bit. Qubits are different than bits and probabilistic bits (bits that are 0 or 1 with certain probabilities), in that the entries in the matrices describing the state are complex numbers. Recall that a complex number takes the form of $a + bi$, where $a$ and $b$ are real numbers and $i$ is the imaginary part where $i^2 = -1$. Complex numbers can also be drawn on the complex plane, sometimes also called the Argand plane [26]. In this plane the $x$ axis represents the real part (*a*) and the *y* axis represents the imaginary part (*b*) [26], as shown in Figure 11. Not only are these complex numbers, but they can be negative in the matrices that represent a qubit. This potential of negative values is what causes possibilities to then cancel out.

---

[8] Occasionally in the literature the term qbit is used instead of qubit, as in [26]. When the term qbit is used cbit is also used in conjunction, meaning "classical bit".

[9] Occasionally qutrits are encountered as well. A qutrit is the quantum equivalent of a trit, a unit of information that can be in one of three states instead of just two as in a bit.

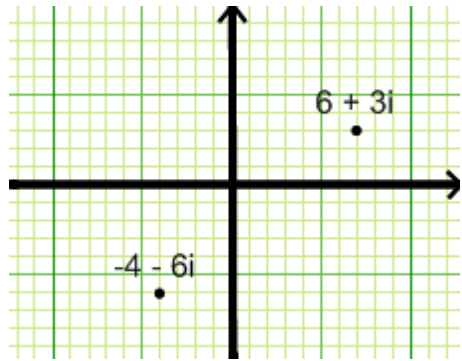

Figure 11. The complex plane (Argand plane) showing two example complex numbers.

For qubits, the squares of the absolute values of the complex numbers in the matrix must add up to 1, thus the values of the entries in the matrix representing the state are not restricted to 1 and 0 like they are for classical bits. The absolute value (also called magnitude) of a complex number $a + bi$ is defined as $\sqrt{a^2 + b^2}$ [26]. For any complex number $z$ the absolute value is expressed as $|z|$. The real numbers are also a subset of complex numbers and fall on the x axis in Figure 11. Consequently the general state of a qubit can be expressed in Dirac notation and matrix forms, where the complex numbers $\alpha_0$ and $\alpha_1$ are often referred to as the probability amplitudes, as shown in Figure 12.

$$|\psi\rangle = \alpha_0 |0\rangle + \alpha_1 |1\rangle = \begin{bmatrix} \alpha_0 \\ \alpha_1 \end{bmatrix}$$

$$|\alpha_0|^2 + |\alpha_1|^2 = 1$$

Figure 12. General state of a qubit, from [27].

Since real numbers are a subset of the complex numbers, it is also possible for a qubit to take the same state as a classical bit, as shown in Figure 13. In this case the qubit is in an absolute state, there is no chance of it being the opposite value[10]. Thus classical information, bits, can be considered a subset of quantum information, qubits.

$$|0\rangle = \begin{bmatrix} 1 + 0i \\ 0 \end{bmatrix} = \begin{bmatrix} 1 \\ 0 \end{bmatrix} \qquad\qquad |1\rangle = \begin{bmatrix} 0 \\ 1 + 0i \end{bmatrix} = \begin{bmatrix} 0 \\ 1 \end{bmatrix}$$

Figure 13. A qubit in Dirac and matrix notation [11] that has the same state as classical bits.

While the complex number $a + bi$ can easily be plotted on the complex plane given in Figure 11 when expressed in Cartesian form, complex numbers can also be expressed in polar form. When expressed as polar coordinates $a = r\cos\theta$ and $b = r\sin\theta$, where $\theta$ is referred to as the phase of a complex number $c$ [28]. More formally, $c$ can be expressed given the following figure also from [28]:

$$c = a + bi = r\cos\theta + ir\sin\theta = r(\cos\theta + i\sin\theta) = re^{i\theta}$$

Figure 14. Expressing complex numbers in Cartesian and polar form.

---

[10] Of course, when absolutely in a state it can also be expressed classically.

Given this polar expression of a complex number, it can be visualized in the complex plane given in Figure 15, which is based on a similar figure in [28].

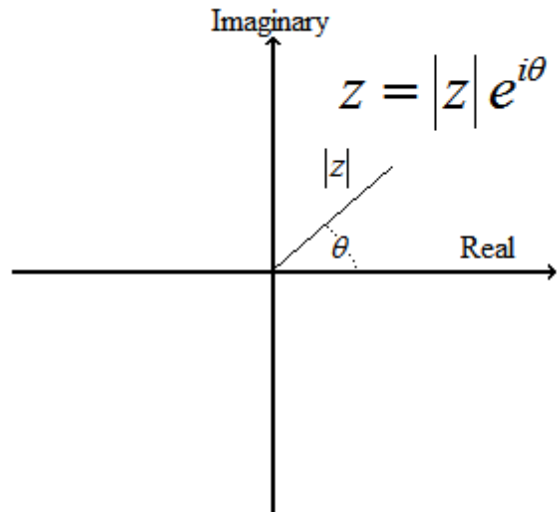

Figure 15. Visualization of the polar form of a complex number.

When a quantum system (one or more qubits) is in multiple states at a time it is said to be in superposition. Consequently the term superposition generally means that the qubits do not have a single probability amplitude of 1 in the matrix and the rest zero, but rather there are two or more elements with probability amplitudes values less than 1. This superposition will exist until the system is observed, at which time it will randomly "collapse" to a particular state. This collapse sets the probability amplitude of the state observed to 1 and all others to 0. Since all knowledge of the state of the system before observation is lost, the observation cannot be undone.

To express the entire system in superposition, the probability amplitudes of particular states (which are complex numbers) are placed in front of each state when in Dirac notation (Figure 17). Since these are probability amplitudes, to get the actual probability of collapsing to a particular state one must square the absolute value of them.

Thus Figure 17 is an example of a single qubit with a 50% chance of collapsing to $|0\rangle$ and a 50% chance of collapsing to $|1\rangle$. Note that the odds are 50-50 because the absolute value for $|0\rangle$ and $|1\rangle$ is $\left(\dfrac{1}{\sqrt{2}}\right)^2 = \dfrac{1}{2}$. Due to the fact that the amplitudes are absolute values that are squared, these amplitudes values can be negative and the result is still 50-50.

$$\begin{bmatrix} \dfrac{1}{\sqrt{2}} \\ \dfrac{1}{\sqrt{2}} \end{bmatrix}$$

Figure 16. A qubit in a 50-50 superpostition of 0 and 1 (matrix notation).

$$\dfrac{1}{\sqrt{2}}|0\rangle + \dfrac{1}{\sqrt{2}}|1\rangle$$

Figure 17. A qubit in a 50-50 superposition of 0 and 1 (Dirac notation).

A qubit can be put into superposition by what is known as the Hadamard operation. Figure 18 shows a qubit put in superposition via a Hadamard operation starting from the state of $|0\rangle$ (absolutely 0) on the left and $|1\rangle$ (absolutely 1) on the right. As will be done elsewhere, in Figure 18 the imaginary parts have been omitted since they are $0i$. Figure 18 shows the qubit in superposition in Dirac form on the left and matrix form on the right. Due to the fact that the entries in the matrix are squared, the qubits Figure 18 will be observed as 0 or 1 with equal chances.

$$\frac{1}{\sqrt{2}}\ket{0} + \frac{1}{\sqrt{2}}\ket{1} = \begin{bmatrix} \dfrac{1}{\sqrt{2}} \\ \dfrac{1}{\sqrt{2}} \end{bmatrix} \qquad\qquad \frac{1}{\sqrt{2}}\ket{0} - \frac{1}{\sqrt{2}}\ket{1} = \begin{bmatrix} \dfrac{1}{\sqrt{2}} \\ -\dfrac{1}{\sqrt{2}} \end{bmatrix}$$

Figure 18. A qubit put in superposition as a result of a Hadamard operation.

While it does not scale beyond a single qubit, the Bloch sphere can be used to help visualize a qubit. In this sphere straight up (along the Z axis) represents $\ket{0}$ while straight down represents $\ket{1}$. The edge of the sphere is always 1 away from the origin, and the state of a qubit is a point along the edge of this sphere[11]. Thus the Bloch sphere is given in Figure 19, where the dark red line represents the vector of length one from the origin to the edge of the sphere.

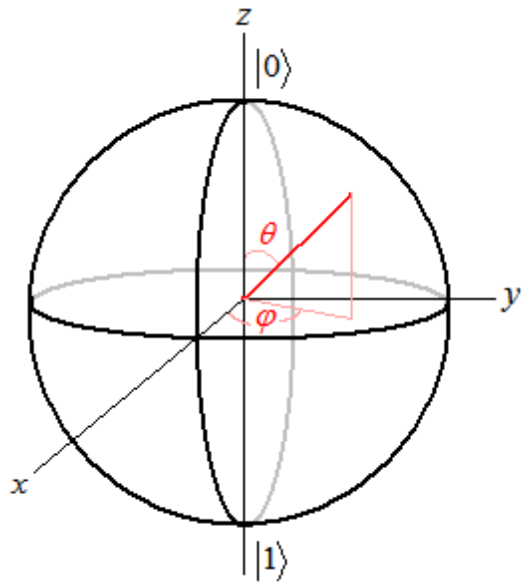

Figure 19. State of a qubit in a Bloch sphere.

---

[11] The magnitude will always equal 1, as this is a probability and must total 1.

Within Figure 19 the vector representing the state of the qubit can be expressed by two angles, $\theta$ and $\varphi$. Going back to the polar form of complex numbers earlier, we can then see how the state of a single qubit can be expressed by two complex numbers. In polar form the state of a qubit can then be represented by Figure 20 and translated to Cartesian coordinates in Figure 21. Both these are given in [12].

$$\cos\left(\frac{\theta}{2}\right)|0\rangle + e^{i\varphi}\sin\left(\frac{\theta}{2}\right)|1\rangle$$

Figure 20. Expressing the state of a qubit in polar form.

$$(x, y, z) = (\sin\theta\cos\varphi, \sin\theta\sin\varphi, \cos\theta)$$

Figure 21. Translation from polar to Cartesian coordinates for a qubit.

Given the Bloch sphere it can be seen how the polar form of complex numbers is useful in visualizing the state of a qubit. One can imagine the state of the qubit represented by two complex numbers. One of the Argand planes corresponds with the X-Y plane in Figure 19, while the other plane contains the angle $\theta$. As will be elaborated on, a Not operation is a flip about the X axis and it can be visualized from the figure how this would flip $|0\rangle$ to $|1\rangle$. This example of the Bloch sphere also helps to demonstrate why complex numbers are needed to express the states of qubits in the first place.

While the Bloch sphere does not scale beyond a qubit, the mathematical state of a qubit can be expanded to $n$ qubits, where each possible state will be preceded by a complex number– always with the restriction that all squared absolute values of the complex numbers add up to 1. It is important to point out that such a complex number cannot be extracted– doing so collapses it to one of the possible values based on their probabilities [27]. The key points are that the probability amplitudes of all possible states are interrelated and when a qubit is observed it will probabilistically "collapse" to one of the possible states. For a more in depth explanation of qubits in the context of computer science the reader is referred to Mermin's text <u>Quantum Computer Science: An Introduction</u> [27]. Multiple qubits logically grouped together are also referred to as quantum registers [10, 11, 29, 30].

Sometimes quantum registers also contain extra qubits that are required to make the computation reversible; these extra qubits are called ancilla qubits [13]. This means that inputs to quantum operations may contain extra qubits, sometimes in prepared states. Likewise, the output may contain extra qubits not relevant to the result the user is interested in. Thus in a certain sense these ancilla qubits can be thought of scratch space for the quantum computation. As an example, a user can construct a register of *x* qubits, where the true input (non-ancilla) is a subset of *y* qubits, and the output the user is concerned with is the subset of *z* qubits. More explicitly a user may want to And something. The register would be 3 qubits (*x*), where the true input would be 2 qubits (*y*), and the output they are concerned about is the result of the And, which is one qubit (*z*).

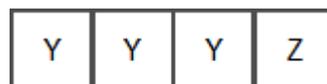

Figure 22. A register of 4 qubits with a single (*z*) ancilla qubit.

This matrix notation becomes cumbersome when there are multiple (qu)bits, as there will be $n^2$ entries in the matrix, where *n* is the number of (qu)bits [12]. A matrix is not complete without all of the entries, hence the large number of zero entries. In Dirac notation the zeros can be left off, making it more concise. Figure 23 shows the general state of two qubits, while Figure 24 shows the representation for the states of two qubits that correspond to pairs of classical bits since they have a 100% chance of being observed as 1 or 0. Since the chance of each state of Figure 24 is absolute the other members in Dirac notation can be left off, but this cannot be done in the matrix form because it would be incomplete. Figure 25 expands this example to three qubits that correspond to three classical bits.

$$|\psi\rangle = \alpha_0|00\rangle + \alpha_1|01\rangle + \alpha_2|10\rangle + \alpha_3|11\rangle = \begin{bmatrix} \alpha_0 \\ \alpha_1 \\ \alpha_2 \\ \alpha_3 \end{bmatrix}$$

Figure 23. General representation of two qubits.

$$|00\rangle = \begin{bmatrix} 1 \\ 0 \\ 0 \\ 0 \end{bmatrix} \qquad |01\rangle = \begin{bmatrix} 0 \\ 1 \\ 0 \\ 0 \end{bmatrix} \qquad |10\rangle = \begin{bmatrix} 0 \\ 0 \\ 1 \\ 0 \end{bmatrix} \qquad |11\rangle = \begin{bmatrix} 0 \\ 0 \\ 0 \\ 1 \end{bmatrix}$$

Figure 24. Two qubits represented in Dirac notation and matrix form.

$$|000\rangle = \begin{bmatrix} 1 \\ 0 \\ 0 \\ 0 \\ 0 \\ 0 \\ 0 \\ 0 \end{bmatrix} \quad |001\rangle = \begin{bmatrix} 0 \\ 1 \\ 0 \\ 0 \\ 0 \\ 0 \\ 0 \\ 0 \end{bmatrix} \quad |010\rangle = \begin{bmatrix} 0 \\ 0 \\ 1 \\ 0 \\ 0 \\ 0 \\ 0 \\ 0 \end{bmatrix} \quad |011\rangle = \begin{bmatrix} 0 \\ 0 \\ 0 \\ 1 \\ 0 \\ 0 \\ 0 \\ 0 \end{bmatrix}$$

$$|100\rangle = \begin{bmatrix} 0 \\ 0 \\ 0 \\ 0 \\ 1 \\ 0 \\ 0 \\ 0 \end{bmatrix} \quad |101\rangle = \begin{bmatrix} 0 \\ 0 \\ 0 \\ 0 \\ 0 \\ 1 \\ 0 \\ 0 \end{bmatrix} \quad |110\rangle = \begin{bmatrix} 0 \\ 0 \\ 0 \\ 0 \\ 0 \\ 0 \\ 1 \\ 0 \end{bmatrix} \quad |111\rangle = \begin{bmatrix} 0 \\ 0 \\ 0 \\ 0 \\ 0 \\ 0 \\ 0 \\ 1 \end{bmatrix}$$

Figure 25. Three qubits represented in Dirac notation and matrix form.

While listing out the representations of a few qubits in Dirac notation and matrix form, it is also worth pointing out some common qubit states that are often encountered: the Bell states [13]. These are also known as EPR pairs, after the authors of the paper that first described them, Einstein, Podolsky and Rosen in [31]. Figure 26 shows the Bell states in abbreviated form as defined by [13]. For clarity the topmost Bell state is also represented in expanded form at the bottom of the figure. The Bell state $|\beta_{00}\rangle$ can be created with a Hadamard gate followed by a controlled not (CNot). The Hadamard gate essentially puts one of the qubits in superposition, while the CNot entangles the second qubit with the first. (In a CNot the target qubit is flipped or Not'd if the control qubit is $|1\rangle$.) The Bell states are often used to illustrate the principle of entanglement. Entanglement and quantum operations are covered below.

$$|\beta_{00}\rangle = \frac{|00\rangle + |11\rangle}{\sqrt{2}}$$

$$|\beta_{01}\rangle = \frac{|01\rangle + |10\rangle}{\sqrt{2}}$$

$$|\beta_{10}\rangle = \frac{|00\rangle - |11\rangle}{\sqrt{2}}$$

$$|\beta_{11}\rangle = \frac{|01\rangle - |10\rangle}{\sqrt{2}}$$

$$|\beta_{00}\rangle = \frac{1}{\sqrt{2}}|00\rangle + 0|01\rangle + 0|10\rangle + \frac{1}{\sqrt{2}}|11\rangle$$

Figure 26. The Bell states (EPR pairs)

Not only are strings of (qu)bits represented by matrices or Dirac notation, but operations can be expressed as matrices also and applied to bits. Figure 27 shows a Not operation being performed on a (qu)bit, both in Dirac notation and matrix form, as shown in [13].

<u>Not of 0</u>

$$not(|0\rangle) = |1\rangle$$

$$not(\begin{bmatrix} 1 \\ 0 \end{bmatrix}) = \begin{bmatrix} 0 \\ 1 \end{bmatrix}$$

$$\begin{bmatrix} 0 & 1 \\ 1 & 0 \end{bmatrix}\begin{bmatrix} 1 \\ 0 \end{bmatrix} = \begin{bmatrix} 0 \\ 1 \end{bmatrix}$$

<u>Not of 1</u>

$$not(|1\rangle) = |0\rangle$$

$$not(\begin{bmatrix} 0 \\ 1 \end{bmatrix}) = \begin{bmatrix} 1 \\ 0 \end{bmatrix}$$

$$\begin{bmatrix} 0 & 1 \\ 1 & 0 \end{bmatrix}\begin{bmatrix} 0 \\ 1 \end{bmatrix} = \begin{bmatrix} 1 \\ 0 \end{bmatrix}$$

Figure 27. Performing a NOT operation on a (qu)bit.

In the quantum coin toss example laid out later in this chapter, "heads" and "tails" are expressed within the kets of Dirac notation because they are each one of the possible states of the system, and these possible states are physically represented as vectors. For the purpose of the quantum coin toss example a deeper understanding of the notation is not required. A more detailed description of Dirac notation involves Hilbert spaces (a finite dimensional vector space over complex numbers is an example [32]) and other concepts more related to linear algebra and physics than programming languages[12], which is the focus of this dissertation. Consequently the reader is referred to [12] in particular for a more detailed explanation of Dirac notation, and [24] for linear algebra.

### 2.1.2  Example: Tossing Quantum Coins

Even though nature at the quantum level is random, it is not random in the way most people are familiar with– observation, or lack thereof, plays a key part. A coin toss is a good example to illustrate this random behavior. The coins we are familiar with in everyday life can be referred to as classical coins. Each toss of the coin is independent of all previous tosses. Additionally it does not matter if one observes the result of the toss or not since observation has no effect in a classical toss. In mathematical terms a classical coin is described by the probability in Figure 28.

$$P(heads) = 0.5$$
$$P(tails) = 0.5$$

Figure 28. Probabilities of outcomes of an even coin toss.

---

[12] Granted, physics has only been covered up to this point as a necessary background. For one already familiar with the necessary physics the discussion of quantum computer programming begins in 2.2.

Yet at the quantum level things are not described by simple probabilities as in the classical coin described above. Instead their behavior is described by probability amplitudes, which can be thought of as corresponding to waves. A coin described by probability amplitudes can be a quantum coin. The quantum coin can be specified using Dirac notation as show in Figure 29. In this case the labeling of heads and tables is arbitrary. The complete form is also labeled to show that if it is heads, there is no chance of it being observed as tails.

$$\left|heads\right\rangle = \left|0\right\rangle = 1\left|0\right\rangle + 0\left|1\right\rangle$$

$$\left|tails\right\rangle = \left|1\right\rangle = 0\left|0\right\rangle + 1\left|1\right\rangle$$

Figure 29. Labeling of heads and tails on a qubit.

The coin is then tossed once via a Hadamard operation, which puts the qubit in superposition if the starting state is head or tails. The result of the flip is shown in Figure 30. As the figure shows, this toss occurs from the starting state of heads ($\left|0\right\rangle$) or tails ($\left|1\right\rangle$). Notice that the complex number of $\left|tails\right\rangle$ is negative when flipping from $\left|tails\right\rangle$. As can easily be shown, the results of heads and tails are still equal after this first toss because the probability amplitudes are still even– regardless of the starting state.

$$\left|heads\right\rangle \mapsto \frac{1}{\sqrt{2}}\left|heads\right\rangle + \frac{1}{\sqrt{2}}\left|tails\right\rangle$$

$$\left|tails\right\rangle \mapsto \frac{1}{\sqrt{2}}\left|heads\right\rangle - \frac{1}{\sqrt{2}}\left|tails\right\rangle$$

Figure 30. Quantum coin after one toss.

To get the chances of a particular event happening at any point, we again square the absolute value of the complex number coefficients. As was mentioned earlier $\left(\frac{1}{\sqrt{2}}\right)^2 = 0.5$, so we see that the coin still has a fifty percent chance of being heads or tails after one toss in Figure 30. Thus the quantum coin is not biased– just like a classical coin. It has an equal chance of being heads or tails after one toss. So tossing a coin once and observing the result is the same for a quantum and classical coin: fifty percent chance of heads, fifty percent chance of tails.

The quantum coin is tossed by what is referred to as a Hadamard gate [13]. The primary purpose of the Hadamard gate is to put a qubit in a superposition state as in the example and to take it out of superposition to get the result [21]. As a result it is a quantum gate that is frequently encountered, and will be discussed in more detail later on.

It is when we don't observe the quantum coin after the first toss that things begin to get strange. If we toss a classical coin twice it will still be heads or tails with equal probability. When a quantum coin is tossed once without being observed, then tossed again, the result is always the same after observing the coin after the second toss: in this case heads. This is worked out mathematically in Figure 31 and is the mathematical equivalent of the splitting the beam of photons twice in Figure 8.

After the first coin toss from heads:

$$\left|heads\right\rangle \mapsto \frac{1}{\sqrt{2}}\left|heads\right\rangle + \frac{1}{\sqrt{2}}\left|tails\right\rangle$$

After the second coin toss:

$$\left|heads\right\rangle \mapsto \frac{1}{\sqrt{2}}\left(\frac{1}{\sqrt{2}}\left|heads\right\rangle + \frac{1}{\sqrt{2}}\left|tails\right\rangle\right) + \frac{1}{\sqrt{2}}\left(\frac{1}{\sqrt{2}}\left|heads\right\rangle - \frac{1}{\sqrt{2}}\left|tails\right\rangle\right)$$

$$= \left(\frac{1}{2}\left|heads\right\rangle + \frac{1}{2}\left|tails\right\rangle\right) + \left(\frac{1}{2}\left|heads\right\rangle - \frac{1}{2}\left|tails\right\rangle\right) = 1\left|heads\right\rangle + 0\left|tails\right\rangle = \left|heads\right\rangle$$

Figure 31. Mathematically working out two tosses of a quantum coin.

This problem is symmetric and the labels heads and tails are merely that: labels, and do not have special meaning. The key point is that the quantum coin will only be in one of the two possible states after the second flip. Applying the Hadamard operation once puts the qubit in superposition, but applying it twice returns it to the original state [13].

Results such as this are not possible with a classical coin toss. It is this interference of probability amplitudes that makes these quantum probabilities different from the classical probabilistic behavior most people are familiar with. As pointed out, this also means that the various tosses of a quantum coin are not independent: if the coin is flipped an even number of times without observation it will always be in the same state when observed. This example is just the first example towards showing how quantum computers can efficiently solve problems that are unsolvable on classical computers.

### 2.1.3 Introduction to Quantum Computing and Information

Today's computers and their ancestors are known as "classical" computers. The term "classical" is used frequently throughout this paper to distinguish them from quantum computers. While the speed of modern classical computers has increased in accordance with Moore's Law, fundamentally they operate no differently than Alan Turing originally laid out in the 1930s [9]– they get smaller and faster, but the principles they operate on are the same. The behavior of nature at the quantum scale, as outlined previously, can be employed in order to perform computations on a machine known as a quantum computer[13]. Rieffel and Polak have also put together a good introduction on quantum computing geared towards non-physicists [33], which is a good reference for those who don't wish to read a book.

The various states of a qubit should not be confused with a ternary (3 state, also called trinary) computing system. In a ternary system values can be one of three possible states. When qubits are in a superposition of 0 and 1 they have probabilities of being either 0 or 1 when observed. As an example, a qubit may have a 10% chance of being in a 0 state when observed. This means that on average it will be a 1 state in 9 out of 10 observations, but 0 in the remaining observation. (The state of this qubit would be expressed as $\sqrt{0.1}|0\rangle + \sqrt{0.9}|1\rangle$ in Dirac notation.) Therefore a quantum program may produce different results from execution to execution. A ternary system on the other hand will always produce the same result when given the same input. It is important to make the distinction between a random result selected, in the case of a quantum computer, and that of random input influencing the execution of a classical program, see Figure 32.

---

[13] The question may then arise, if this random quantum behavior forms the basis of matter, how have we created deterministic machines? The answer is in part what makes the physical implementation of quantum computers so difficult: they must be isolated. Interacting with the outside environment acts as a measurement, this measurement collapses the state to one that is not random. This is known as decoherence.

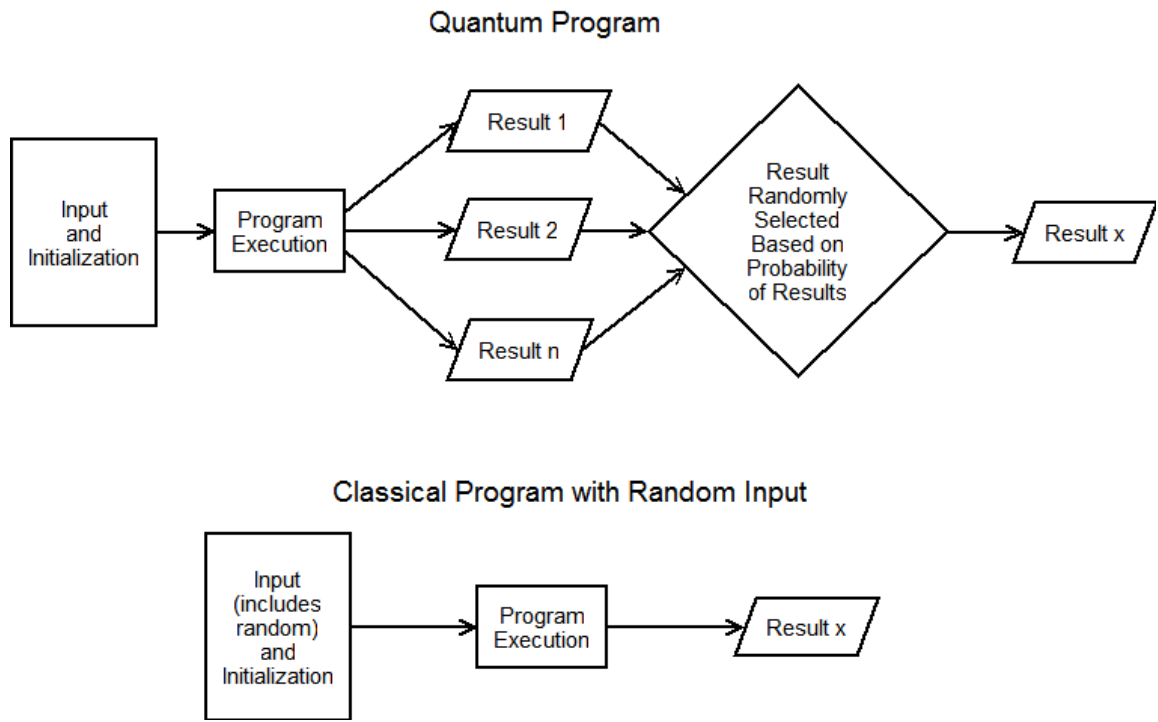

Figure 32. A quantum versus a classical program.

Note that on Figure 32 the quantum computer generates multiple results, but only one of those is selected and becomes the output. Once the classical program is seeded with random input its execution and result are deterministic. As a result, when seeded with the same input a quantum program may generate different results from execution to execution, but a classical program will execute the same given the same input.

How is the quantum computer useful if its result is random? There are two primary reasons why. The first is that the probabilities can be skewed towards the desired result, making it more likely. The second reason is that while performing the computation is hard, checking the answer is easy for some problems. An example would be factoring. If we are trying to factor $N$, where $N = pq$, it is easy to check classically if $p$ and $q$ are factors of *N* if they are the output from the quantum computation. It is the coming up with p and q that is the hard part where a quantum computer is exploited, as will be detailed in 2.1.11.

One might think on first impression that there is no difference between a quantum computer and a probabilistic classical one, but this isn't the case. The primary difference is that the possibilities in a quantum system are allowed to constructively and destructively interfere with each other– something that does not happen in a probabilistic classical system. The quantum coin toss earlier is an illustration of this as the number of tosses before observation matters. Thus the qubit is fundamentally different than any classical unit of information. See Table 1 for a comparison of bits and qubits, which is a subset of the table given in [27].

Table 1. Comparison of bits and qubits.

|  | Bits (classical) | Qubits (quantum) |
|---|---|---|
| Subsets of *n* (qu)bits | Always in absolute states, not influenced by other bits. | Generally have no absolute states, and can be influenced by other qubits when entangled. |
| Can state be learned from (qu)bits? | Yes | No, measurement collapses |
| To get classical information | Look at them | Measure them |
| Information acquired | *X* | $x$ with probability $|\alpha_x|^2$ |
| State after information acquired | Same, *x* | Different, $|x\rangle$, due to collapse |

It should also be noted that qubits are continuous [13], while bits are discrete. This means it takes an infinite amount of classical information to precisely represent the state of an arbitrary qubit [13]. While largely outside the scope of this dissertation, the Bloch sphere in Figure 33 helps to illustrate the difference between quantum and classical units of information. The Bloch sphere is 3 dimensional, while a probabilistic bit is 1 dimensional. Physically, one can think of the projection on the Bloch sphere as the orientation of a particle such as a photon, although this example does not scale beyond a single qubit.

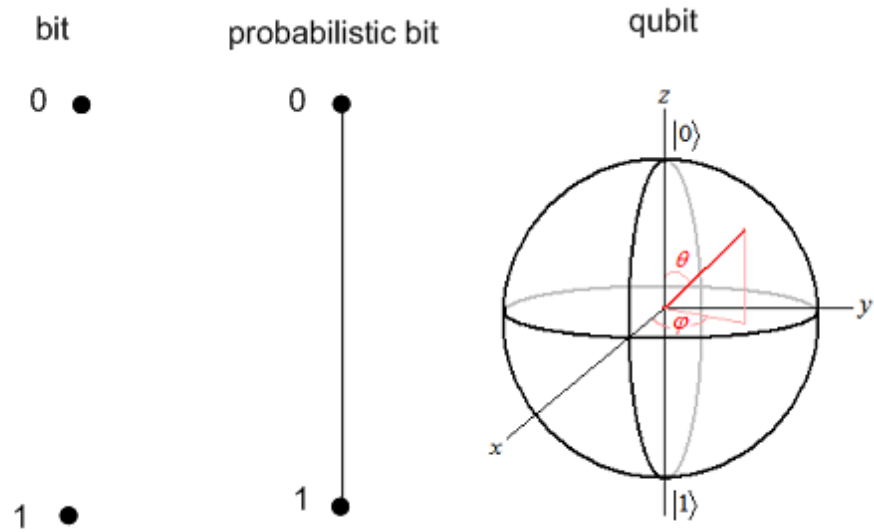

Figure 33. The difference between bits, probabilistic bits, and qubits.

Figure 33 is based on the illustration in [12]. Physically, the Pauli gates (detailed in Quantum Operations, section 3.2.4) are rotations about the three axes in the Bloch sphere– hence the names X gate (not), Y gate, and Z gate. One can see from the illustration how a rotation of 180 degrees, or $\pi$ radians, about the X axis would perform a not operation. Hence "X gate" and "Not gate" are used interchangeably throughout the literature.

There are a few features unique to qubits that should be mentioned. Further details on these topics are beyond the scope of the dissertation, so the reader is referred to the references for more detailed explanations. Nonetheless there are a few examples that help highlight the differences between classical and quantum information:

- No-cloning theorem – an arbitrary quantum state (qubit) cannot be copied [11-13]. This plays into the limitations of quantum computing, which are discussed in section 2.1.6.

- Quantum teleportation – allows for the state of a qubit to be transferred from sender to receiver. The state of the sender's qubit is destroyed, so it does not violate the no-cloning theorem [11-13].

- Superdense coding – sending one qubit transmits two bits of classical information [11-13].

- Entanglement– The measurement on one qubit affects the state of others that are entangled with it [11], regardless of the physical distance between them.

While the focus of this paper is not on quantum computing hardware, it is worth briefly mentioning in order to illustrate the challenges that lie ahead in implementation and why quantum computers are expected to be a number of years away. As of 2006 quantum experimental ion-trap quantum computers have been built with 8 qubits, and nuclear magnetic resonance (NMR) quantum computers have been built with up to 12 qubits [34]. However the scalability of these two approaches is questionable. More promising, and perhaps likely to surpass NMR and ion-trap quantum computers is solid state quantum computing using diamonds [35]. Since there is also considerable infrastructure in place for solid state methods, this may likely be the route to commercial quantum computers [34]. In the summer of 2009 the American NIST demonstrated multiple move and logic operations via the ion-trap approach [36], which is a significant milestone. In 2008 Vuckovic and her team have made progress by developing a solid state method utilizing photons [37].  But these are just a few of the more than a dozen physical implementations being explored [38].

While it may be disappointing to think that systems of only up to a dozen qubits have been built in the 20 years since the idea was introduced, many people feel there are breakthroughs on the horizon [39]. One possible area for improvement is the use of quantum multicomputers: small quantum computers linked together to solve a problem [40]. This avoids the problem of creating a single large quantum computer, but does present problems of its own.

Quantum computing exploits these non-classical features to perform computation. There is another area in computer science that also exploits nature at the quantum level: quantum cryptography. One of the first methods outlined is the Bennett-Brassard scheme which utilizes quantum mechanics to generate a key [28]. This type of method is typically called "quantum key distribution". Unlike many forms of modern cryptography, quantum cryptography is not based on mathematically "hard" problems such as factoring, but on the properties of nature. Zeilinger has distributed quantum keys up to 89 miles and has plans to distribute them across continents using the International Space Station by 2014 [41]. This is important since it shows that researchers are making progress towards being able to use quantum cryptography over distances. Since this dissertation is on quantum computer programming, quantum cryptography will not be explored further. For further details on the subject the reader can consult [28].

### 2.1.4   Example: Entanglement

Through entanglement, changes to a set of qubits may affect the state of another set of qubits. In classical computation, if a single bit is manipulated all the others remain unchanged. This isn't the case in quantum computation: changing the state of one qubit may change the state of other qubits through the principle known as entanglement. In fact when dealing with a set of qubits the user may not even know or care that they are entangled with another set. More formally: entangled states are those that cannot be created by combining (tensoring) smaller ones, that is there is no way to "factor" the state into smaller ones known as separable states [42].

This example covers the simplest case of entanglement: changes to one qubit affecting a second qubit. In order not to confuse them with the X, Y, and Z quantum operations, these qubits will be labeled A and B. Since we are dealing with the qubits as a set, there is a register consisting of A and B. Both qubits are initialized to 0, so the register starts in state $|00\rangle$. The two qubits are then entangled by first applying the Hadamard operation, which was the toss in the quantum coin toss example, to qubit A. After this operation qubit A is in superposition between 0 and 1 and qubit B remains 0. The next step is to apply the controlled Not (CNot) operation to the register, with A being the control qubit. This can be expressed mathematically in Figure 34 or via the quantum circuit diagram in Figure 35. The quantum circuit diagrams and the notations are covered in 4.3, the point of Figure 35 is just to show that entanglement is easily accomplished.

| Action | Matrix Notation | Result in Dirac |
|--------|-----------------|-----------------|
| Initial state | $\begin{bmatrix} 1 \\ 0 \\ 0 \\ 0 \end{bmatrix}$ | $|00\rangle$ |
| Application of Hadamard operation to qubit A | $\left( \frac{1}{\sqrt{2}} \begin{bmatrix} 1 & 1 \\ 1 & -1 \end{bmatrix} \otimes \begin{bmatrix} 1 & 0 \\ 0 & 1 \end{bmatrix} \right) \begin{bmatrix} 1 \\ 0 \\ 0 \\ 0 \end{bmatrix} = \begin{bmatrix} \frac{1}{\sqrt{2}} \\ 0 \\ \frac{1}{\sqrt{2}} \\ 0 \end{bmatrix}$ | $= \frac{1}{\sqrt{2}}|00\rangle + \frac{1}{\sqrt{2}}|10\rangle$ |
| Application of CNot to the register | $\begin{bmatrix} 1 & 0 & 0 & 0 \\ 0 & 1 & 0 & 0 \\ 0 & 0 & 0 & 1 \\ 0 & 0 & 1 & 0 \end{bmatrix} \begin{bmatrix} \frac{1}{\sqrt{2}} \\ 0 \\ \frac{1}{\sqrt{2}} \\ 0 \end{bmatrix} = \begin{bmatrix} \frac{1}{\sqrt{2}} \\ 0 \\ 0 \\ \frac{1}{\sqrt{2}} \end{bmatrix}$ | $= \frac{1}{\sqrt{2}}|00\rangle + \frac{1}{\sqrt{2}}|11\rangle$ |

Figure 34. Working out entanglement mathematically.

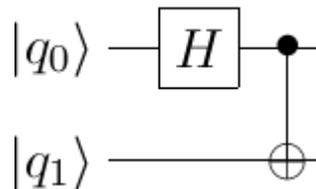

Figure 35. Quantum circuit diagram[14] for entangling two qubits.

After the Hadamard operation, "H" in Figure 35, the register is in the state $\frac{1}{\sqrt{2}}|00\rangle + \frac{1}{\sqrt{2}}|10\rangle$. So the register will collapse to $|00\rangle$ half the time and $|10\rangle$ the remainder of the time. This is just what is expected: qubit A (the first digit in $|00\rangle$ or

---

[14] Quantum circuit diagrams are explained in detail in 4.3, and the notation given in Table 5 through Table 9. "q$n$" is used throughout this dissertation to denote various qubits.

$|10\rangle$) will be 0 or 1 with equal probability and qubit B will still always be 0. At this point measurements on A or B do not influence the other. If we measure B, A is still in super position. If we measure A then B is still 0.

In the next step the CNot operation and the qubits become entangled, in state $\frac{1}{\sqrt{2}}|00\rangle + \frac{1}{\sqrt{2}}|11\rangle$. The register will collapse to $|00\rangle$ or $|11\rangle$ with equal probability when observed. What makes the qubits entangled is that they are no longer independent. For example, let's say we only measure qubit A. Of course there are one of two possibilities: 0 or 1. But we can see from the state of the register that both qubits must take on the same value after measurement. Thus if we observe qubit A and it is 0, then B must also be 0 since the state has collapsed to $|00\rangle$, likewise for observing 1. This may become a little more clear when we expand the state of the register to include the zero probability states. We can see from this expansion in Figure 36 that there is a 0 probability of observing A and B as different values.

$$\frac{1}{\sqrt{2}}|00\rangle + 0|01\rangle + 0|10\rangle + \frac{1}{\sqrt{2}}|11\rangle$$

Figure 36. Entangled 2 qubit register with 0 probabilities included.

The qubits being observed as the same value when they are entangled occurs regardless of the physical distance between the qubits. Distance plays a factor in cases such as the attraction between two bodies, but as can be seen in the mathematics it plays no role in entanglement. There are many other possible entangled states, and they can contain any number of qubits; this example is just a trivial one and the qubits do not all necessarily have to collapse to the same value. As an example, it could be constructed where the two qubits are observed as opposite values instead of the same.

This example of entanglement is a good illustration of how individual qubits may influence one another, which is not possible with classical information. With classical information you can flip bits as much as you like and there is no impact on other bits. With quantum information this isn't true, as entanglement shows that flipping one qubit may impact another. This example also illustrates how the CNot operation plays a role in quantum computation that has no equivalent when used classically[15].

### 2.1.5    *Foundations of Quantum Computing*

This section covers the papers that presented the key ideas within the field of quantum computing as they apply to programming quantum computers. Quantum circuits are not covered here as they lie at a lower level than is generally necessary for quantum computer programming. Nonetheless they are covered in section 4.3 as they are used frequently within the literature. Tangential areas such as the evolution of quantum circuits via genetic programming [21] are also not considered.

In 1982 Feynman implicitly stated that a computing device would need to operate based on quantum mechanics in order to simulate a quantum system efficiently [43]. His proposal was for a "universal quantum simulator" [4, 44]. Although he later expanded his ideas [45], his initial proposal was not a computer as laid out by Turning [11]. The universal quantum simulator Feynman proposed can only be programmed by preparing it in a suitable physical state. One important point Feynman made at this time was that a quantum system cannot be simulated on a classical computer without an exponential slowdown in efficiency [11].

Benioff constructed a model of computation utilizing quantum kinetics and dynamics [46]. Deutsch argued that this approach could be simulated perfectly by a classical Turning machine [4]. As a result Benioff's model of computation is not considered to be the founding work of quantum computing, but rather utilizing quantum mechanics to perform classical computation [46]. A similar idea well covered in the literature is superdense coding– transmitting of a single qubit relays two bits of classical information [11-13].

The field of quantum computing is largely considered to have been founded by David Deutsch in 1985 with his paper <u>Quantum theory, the Church-Turing principle and the universal quantum computer</u> [4]. As previously mentioned, Feynman pointed out that a classical computer could not efficiently simulate a quantum system [44]. However, Feynman did not explicitly state that a quantum computer could efficiently perform classes of computations that are not practical on a classical computer– he merely alluded to it. In his paper Deutsch explains how a quantum computer can perform computations that are inefficient on classical systems– computations that would take so long on classical computers as to be considered impossible with classical technology that will be developed in the future, even if Moore's law continues indefinitely. At the time of his paper, quantum computers were viewed as something of an oddity for two reasons. First, it was unknown (and still unknown according to some) if a practical quantum computer could be built. Secondly, there were no useful algorithms to take advantage of the characteristics of a quantum computer. Although Deutsch laid out an algorithm in his paper where a quantum computer could outperform a classical one, its practical use is limited. The algorithm he laid out is commonly referred to as Deutsch's algorithm.

---

[15] Not meaning there is no classical CNot operation (which there certainly is), meaning that there is no equivalent of creating superposition using it.

This limited practicality of quantum computers later changed when Peter Shor figured out how to efficiently factor integers using a quantum computer [47]. Factoring integers is the basis of many commercial public key (asymmetric) encryption algorithms, so at this point Deutsch's work became more significant because it could be used to tackle what are considered hard problems with classical techniques. Deutsch is a physicist [19], so naturally his paper focuses more on the mathematical and physical aspects of quantum computing. Deutsch also pointed out [4] that a qubit is a closer representation of nature than a bit. As he points out at the end of his paper, "Quantum computers raise interesting problems for the design of programming languages…" [4], primarily due to the fact that the information dealt with does not adhere to classical rules.

Quantum computing is similar to parallel and distributed computing in some aspects. Both attempt to solve a problem by essentially performing multiple solutions in parallel. This is done through use of the extra states in a qubit on a quantum computer. So if there are one thousand possible solutions to try, then in the worst case all one thousand will be attempted. It is true that a parallel computer could carry out all these attempts in parallel, but as will be shown in the next paragraph, a quantum computer quickly scales to a point not achievable by parallel computers.

This isn't the case with a quantum computer– a quantum computer can execute all of those attempts at once. A quantum computer's power increases by a power of two for each qubit added. Furthermore, a quantum computer does not require the communications and partitioning overhead that a parallel or distributed attempt at a solution requires[16]. As an example, take a 37 qubit quantum computer. A quantum computer is able to attempt all the possibilities that those 37 qubits can represent at once: $2^{37}$, or 137,438,953,472 combinations. This number is the same as the total lifetime of the universe in years if the universe is closed [48]. So to achieve the same computational power a classical solution would require some combination of processors, systems, and/or attempts that equals this number. Taking this even further, quantum computers with many qubits have the ability to solve problems that will never be possible classically, no matter what the technology.

A simple illustration of the extra power of quantum computation is Deutsch's algorithm [13]. Given a function, $f(x)$, the goal is to compute the result of $f(0)$ xor $f(1)$. Obviously, on a classical computer $f(x)$ would have to be calculated twice– once to compute $f(0)$, and a second time to compute $f(1)$, after which the result would be obtained. Deutsch's algorithm employs the power of quantum computation to solve the problem while only evaluating $f(x)$ once, a feat that is impossible on a classical computer. Deutsch's algorithm utilizes the quantum mechanical property of interference as illustrated in the coin toss example in Figure 31 order to achieve the result when only querying the function once. A more detailed explanation of Deutsch's algorithm involves quantum circuits, which will be covered in 4.3. For an detailed explanation of Deutsch's algorithm the reader is referred to [13].

---

[16] Admittedly the belief that there will not be the overhead in quantum computers is somewhat speculation since large enough quantum computers do not yet exist.

In 1996 Knill introduced his conventions for quantum pseudocode [3], one of the first efforts to tackle the subject of quantum computer programming [49]. Pseudocode is often used as a concise way to express algorithms within computer science without becoming tied to a particular language syntax or implementation restrictions. As Knill points out in his paper, up until this point there were no conventions for quantum pseudocode. This meant that until this time algorithms were written typically in mathematical notation, which is geared towards mathematicians and physicists, making the algorithms difficult to understand for software developers. Although it is still common to see quantum computer algorithms expressed in mathematical notation, Knill's pseudo code was a good step in moving quantum computing towards mainstream computer science. Quantum circuits are also another common method sometimes used to illustrate quantum algorithms. What is important about Knill's paper is not just the pseudocode convention introduced, but the practical nature of the machine which it will operate on. In the introduction he states:

> It is increasingly clear that practical quantum computing will take place on a classical machine with access to quantum registers. The classical machine performs off-line classical computations and controls the evolution of the quantum registers by initializing them in certain prepareable states, operating on them with elementary unitary operations and measuring them when needed….

Knill calls a machine that behaves in this manner a quantum random access machine (QRAM), see Figure 37. This combination of classical and quantum computing is now generally believed to be how the first commercial quantum computers will appear. The proposed ideas for quantum programming covered in this dissertation utilize an expanded QRAM model, which is detailed in 5.2.

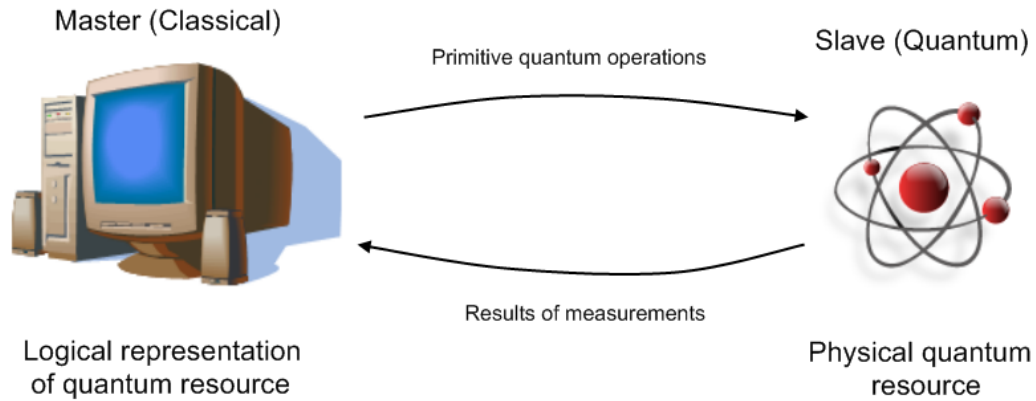

Figure 37. Knill's QRAM model, based on [2].

Even though quantum computing was founded in the 1980's, it largely remained a oddity until a decade later when Shor introduced his factoring algorithm. It is now known that there are several important problems that can be solved more efficiently on a quantum computer. Some examples of these problems include:

- Integer factorization – this forms the basis of many commercial public key (asymmetric) cryptographic algorithms [47].

- Simulation of quantum systems [44] – any simulation of a quantum system will experience an exponential slowdown.

  o Protein folding [50].

  o Reaction dynamics [50].

  o Large molecules, which could lead to new drugs [51].

- Unsorted search [52] (with no key to any of the data)

- Quantum Fourier transform [12] – an algorithm that performs Fourier transformations of quantum mechanical amplitudes [13].

*2.1.6   Limitations of Quantum Computing*

As will be shown in the factoring example (2.1.11), quantum computers are able[17] to carry out computations that cannot efficiently be carried out on classical computers. An example of this is simulation of a quantum system: for all but the smallest simulations the problem is too complex to execute on a classical computer, as there is an exponential slowdown of the simulation [11]. This exponential slowdown makes a simulation of a quantum computer on a classical system impractical for more than a limited number of qubits. Even though quantum computers are able to carry out computations that are not feasible on classical computers, this increase in computing power does come with certain restrictions; those limitations will be covered in this section.

At an abstract level, quantum computers can be thought of as being able to carry out a computation for multiple combinations of inputs at once. It is this capacity for parallel processing that makes quantum computers powerful. There are several restrictions on quantum computers, the most noteworthy ones for classical programmers include: probabilistic output [10], non-observation [11], and reversible computation [53]. Each of these will be discussed in turn.

The first limitation is that the output of a quantum computer is probabilistic–running the same quantum program multiple times may generate different results. The result generated depends on the probability of the possible answers. In the quantum coin toss example covered earlier the result would always be heads. The other possibility, tails, had a zero percent chance of occurring. The coin can also be set up so that tails is always produced. Thus a quantum coin will always have the same output, even given an infinite number of tosses. For more complex algorithms the correct answer doesn't always occur with a probability of one hundred percent, but the probabilities can be skewed towards the correct answer. Expanding on the quantum coin example, this could be thought of as weighting a pair of dice or stacking a deck of cards: while you may not always get the desired result, you've increased the chances that you will.

Although a quantum computer can carry out parallel computation, at the end one of those possibilities is randomly selected based on their probability amplitudes, and returned as the answer. In the quantum coin example, after one toss there is a fifty–fifty chance that it will be heads or tails. There is a one hundred percent chance of heads after the second toss. Once an answer has been obtained from a quantum program, through measurement, the only way to generate another result is to rerun the quantum program. Obtaining the answer through measurement can be thought of as collapsing the system to a classical state. At this point the system can not be put back into the quantum state it was in before the measurement. In effect there is no way to "undo" receiving the answer through observation; the only option is to rerun the program.

A quantum computer may not seem very useful if the answer is essentially random. Even so, there are methods to tilt the probabilities towards the correct answer. In the quantum coin toss example, by flipping the coin twice without observation we can obtain heads every time. This skewing of probabilities towards the correct answer is known as constructive interference, while the minimization of the incorrect ones is destructive interference [32]. These concepts are central to more complex quantum algorithms. In some cases the result returned from the quantum algorithm can be easily and efficiently checked with classical means to determine if it is correct. In the case of factoring, checking the result is efficiently done with classical techniques– factoring $x$ is hard, checking if $y$ and $z$ are factors of $x$ is not.

---

[17] Assuming that they can actually be built of course.

Another limitation on quantum computing is that it cannot be observed while carrying out a computation. Observation (also called measurement) is used in these cases as physicists use it: the quantum computer cannot interact with the outside environment when carrying out its computation. The fact that the quantum computer must be isolated is a large part of what makes constructing quantum computers so hard. When a system is observed, it is said to "collapse the state vector" [11]. All of the parallel computations that were being carried out suddenly collapse into the one randomly selected answer as previously described. Thus a quantum computer can not be queried as to its state while in the middle of the computation without influencing the outcome. This is why in the quantum coin example we must not observe it after the first toss in order to get the result after the second toss to always be heads. If we observe it after the first toss, it will collapse to head or tails, and then be heads or tails after the second toss with equal probability. By not observing the coin after the first toss, constructive interference in the second toss causes the result of heads. Expanding this concept further, it is not possible to determine what the probability amplitudes are for the various possible states before collapse. Nonetheless, many classical simulations of quantum computers allow for this to be obtained to aid students in quantum computing and to help verify programs and the correctness of the simulation[18].

---

[18] This was used extensively during the implementation of the simulation for Cove. Both in unit testing to show that it is functioning correctly at each step, and during debugging to identify and correct errors.

Along the lines of non-observation is what is known as the No-Cloning Theorem. The No-Cloning Theorem states that there is no operation that can produce a copy of an arbitrary quantum state [11, 13]. This prevents one from making a copy of an arbitrary quantum system and observing that copy to get around the problem of observation collapsing the system mentioned in the previous paragraph. Thus observation not only produces an answer, but the only way to obtain another answer is to rerun the computation– due to the No-Cloning Theorem there is no way to "undo" an observation. In essence, the No-Cloning Theorem implies that there is a limited amount of information we can obtain from an arbitrary quantum state [22], in accordance with Heisenberg's uncertainty principle.

Quantum teleportation allows for the state of a system to be sent from a source system to a target system. (Quantum teleportation is a commonly used term, such as in [11, 12].) With quantum teleportation matter is not transferred, merely the information needed to construct the identical state at the target. When this happens the state of the source system is destroyed, thus no-cloning is not violated [22]. One way to think of teleportation is as a move operation, where the state is moved from one system to another while no-cloning prohibits copy (this based on [42]). Consequently teleportation is all about moving information.

A common misconception about quantum teleportation is that it allows for instantaneous communication. This isn't the case: the laws of nature prevent instantaneous communication [22]. So even if two qubits are entangled the observation of one does not communicate information to the other. One may choose to observe one qubit in an entangled pair, and obtain a result, but this does not send any information to the other qubit in the pair. Without going into further details of teleportation, it is this observation that destroys the state that is part of the algorithm for the sender [42].

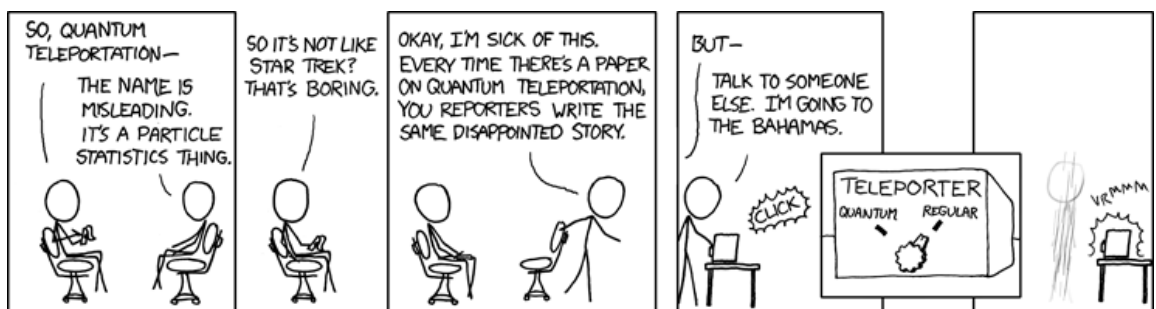

Figure 38. Comic illustrating the common misconception of quantum teleportation, courtesy of XKCD.com [54].

The final limitation of quantum computers that should be mentioned is that the computations they perform must be reversible. Reversible is defined as being able to recover all inputs given any possible output. Essentially the program can be run backwards to recover the input. Many of the operations carried out on a classical computer are not reversible. The AND operation is a simple example: if the result is 0 we do not know what the two inputs were, other than the fact they were not both 1. As an aside, computations that are reversible do not require any energy [55]. Erasing of information is what requires energy and this is known as Landauer's principle [13, 53]. Thus the problem of heat being generated on today's computers partially comes from the fact that information is constantly being erased. So if all computations on today's computers were reversible then the problem of excessive heat would be less of a hindrance.

More formally, the operations on qubits are required to be unitary[19] transformations [27]. Thus any supplied operation can easily be checked to see if it meets the requirement of reversibility. Any operation U is reversible if it satisfies $U^\dagger U = UU^\dagger = I$, where I is the identity matrix [27]. If $U^*$ represents the complex conjugation and $U^T$ represents the transpose operation then $U^\dagger = (U^*)^T$ and $U^\dagger$ is called the conjugate-transpose [13]. Stated another way an operator is considered unitary if $U^\dagger = U^{-1}$ ($U^{-1}$ is the inverse of U) [12].

$U$      any operation

$U^{-1}$      inverse of U

$U^*$      complex conjugation

$U^T$      transpose

$U^\dagger$      conjugate-transpose

Figure 39. Common matrix notations.

The controlled not, or CNot, is a reversible operation, which is one of the reasons it is commonly used as an example of a standard quantum operation since all quantum operations must be reversible. Part of the reason CNots are frequently used as examples is that they can entangle two qubits and are used to create the Bell states (Figure 26). A CNot has two inputs, and the truth table is shown in Table 2. If the first input is 1 then a Not is performed on the second input. There are two outputs– the first input with no changes, and the second input, with potentially a Not performed on it. It is trivial to show that this is a reversible computation. It should be noted that this requirement of reversibility is related to the fact that the quantum computer cannot interact with the outside environment. If a computation is not reversible it will dissipate energy to the environment, which will betray the state of the system [53]. While a Not operation is reversible many other classical operations such as And and Or are not reversible.

---

[19] Unitary operations should not be confused with unary operations, which are operations that take only one input. A unitary operator can have any number of inputs as long as it satisfies the condition $U^\dagger = U^{-1}$.

Table 2. Truth table for the CNot operation.

| Input Control | Input Target | Output Control | Output Target |
|---|---|---|---|
| 0 | 0 | 0 | 0 |
| 0 | 1 | 0 | 1 |
| 1 | 0 | 1 | 1 |
| 1 | 1 | 1 | 0 |

This restriction of reversible computation may initially appear as a huge limitation because many programs and circuits on classical computers are not reversible. Nonetheless, it has been shown that any irreversible computation can be transformed into a reversible one [12]. This can be done by adding extra (ancilla) qubits, as shown in Table 3.

Part of the restriction of reversibility is that in the most general sense quantum gates can only evolve a quantum system using unitary transformations [28], which are reversible [13]. This means that there no quantum operation that can transform the system to a smaller one. An example of something that would evolve the system to a smaller one is the And operation: it has two input bits but only one output bit, meaning it transforms to a smaller state. So not only must the computation be reversible, but it must evolve its inputs into the same number of outputs. Due to the reversibility requirement the quantum

And operation operates on three qubits as shown in Table 3. By flipping the rightmost qubit the additional 4 possibilities are given, for the total of all 8 possibilities, and is given in grey in the table. The key point of the example is that the rightmost qubit flips when both inputs are $|1\rangle$.

Table 3. Classical and quantum And operations.

| Classical And | Quantum And |
|---|---|
| 00 → 0 | $|000\rangle \rightarrow |000\rangle$ |
| 01 → 0 | $|010\rangle \rightarrow |010\rangle$ |
| 10 → 0 | $|100\rangle \rightarrow |100\rangle$ |
| 11 → 1 | $|110\rangle \rightarrow |111\rangle$ |
| 00 → 0 | $|001\rangle \rightarrow |001\rangle$ |
| 01 → 0 | $|011\rangle \rightarrow |011\rangle$ |
| 10 → 0 | $|101\rangle \rightarrow |101\rangle$ |
| 11 → 1 | $|111\rangle \rightarrow |110\rangle$ |

For the quantum And, the And functions on the first two qubits of the input with the result being in the third of the output. (The bits are labeled from left to right.) Note that unlike the classical And, the Quantum And is both reversible and operates on the same system (the output is the same size as the input). While the classical And has a new output of a single bit, the quantum And is different: the same 3 qubits that were the input are the output, only one of them has changed after the quantum And operation is applied. This example in Table 3 is a condensed version of the one given in [28]. However this requirement of unitary transformations is the only limitation for quantum operations. In other words, any unitary transformation is a valid quantum operation [13]. Consequently there are there are an infinite number of valid operations since qubits are described by continuous ranges of probabilities.

### 2.1.7  *Classical Parallel Languages*

Most languages today do not enforce limitations needed for quantum computation. Furthermore, many of today's popular languages were originally used on single processor systems. Consequently not much emphasis was made on parallel programming for these languages. However, Microsoft has been working on a "Parallel FX" library to make better use of parallel features within languages utilizing the .NET framework. Due to the fact that multiprocessor systems are becoming more common, today's popular languages may be replaced by those that have been designed for parallel processing from the start. There are several languages being developed for multiprocessor systems. One or more of the languages may see mainstream use in the near future, and include:

- Sun Microsystems's Fortress [56] – Meant to be a high performance language for the same sort of applications that Fortran has been used for. Fortress is statically typed, allows for component reuse, and supports modular and extensible parsing– which allow for notation to be added to the language. The syntax of Fortress appears similar to Fortran and C derived languages including Java [57] and Adobe's ActionScript.

- Cray's Chapel [58, 59] – The primary reason for the development of Chapel is to introduce high level language concepts for expressing parallelism. The parallel aspects of Chapel are largely based on previous solutions from Cray such as the MTA extensions of C [58, 59]. Chapel's authors claim its prime serial language influences as C derivatives, Ada, and Fortran, which is evident by examining language samples. Unlike Fortress, which is statically typed, Chapel is not. In the spirit of Python this makes it easier for programmers by not forcing them to replicate algorithms for different types.

- IBM's Experimental Concurrent Programming Language (X10) [60] – X10 enforces safety of several different kinds: type, memory, place, and clock.

Unlike Chapel and Fortress, X10 is largely build upon a single language–Java. As such, the current implementation translates X10 to Java and is available on Source Forge at http://sourceforge.net/projects/x10. Furthermore an integrated development environment (IDE) for X10 has been developed for the Eclipse IDE.

- Intel's Ct [61]– Ct is based on C++ and utilizes a nested data parallel model. Importantly, program execution is guaranteed to be the same on any number of cores, which eliminates data races. Data races may occur infrequently and can be difficult to identify and resolve.

- MIT's StreamIt [62, 63] – Unlike the other languages listed here, it is designed as a special purpose language for streaming applications. The argument for this is that streams cannot be expressed naturally or eloquently enough in existing high level languages. While StreamIt may make it easier for programmers to handle streams, this also limits its use. Commercial programmers typically encounter a wide range of problems to be solved, and learning a language to solve only a few of them often isn't worth the time and effort required of the programmer.

Even though languages such as these are intended to be parallel programming languages, they are fundamentally different from quantum languages due to the advantages and limitations of a quantum computer. Even though a quantum computer may operate on a huge number of possible values, in the end only one of the potential results is selected as the output as outlined earlier in this section. Using parallel programming it is possible to achieve all possible outputs. As a result quantum computing can be looked at as a way to more easily operate on multiple inputs; this comes at the expense of only receiving one potential solution.

*2.1.8   D-Wave Systems*

There is a startup company called D-Wave Systems that claimed to demonstrate a 16 qubit quantum computer in February 2007 at the computer history museum in Mountain View, California. D-Wave has not disclosed how their quantum computer works, and rightfully are being viewed very skeptically by those in the field of quantum computing [38, 42, 64, 65]. As the famed quantum computing researcher and author of "Programming the Universe" [66] Seth Lloyd has stated "[D-Wave is] certainly not the kind of company I'd invest my money in" [67]. Furthermore their quantum computer was not physically present at the demonstration– it was being accessed remotely. This lack of evidence is hardly what one would expect about a commercial device that can break encryption commonly used on the Internet as it scales up. As the late astronomer Carl Sagan has said, "Extraordinary claims require extraordinary evidence." [68]. To date D-Wave's evidence is lacking.

Quantum computers have the ability to carry out computations that are impractical on classical computers due to their ability to operate on multiple inputs at the same time. It is the three limitations: probabilistic output, non-observation, and reversible computation, which contribute to making quantum computers difficult to implement and program. Due primarily to the limitation of non-observation[20], it is generally believed that we will not see commercial quantum computers until about a decade from now.

### 2.1.9   Limitations of Simulating a Quantum Computer on a Classical Computer

It has been estimated that practical quantum computers will not appear for another 10 to 20 years or so [39]. Given that there will be no quantum computers in the near future, we need a way to test various software techniques. In absence of an actual quantum computer, the only way to do so is to simulate a quantum computer on an existing classical computer. There will still be the exponential slowdown on the classical system [11], but a simulator does allow programmers to write limited quantum programs and learn methods of developing quantum programs. Being able to learn how to write good quantum programs before the introduction of quantum computers may reduce the learning curve necessary to utilize quantum computers once they become a reality. Furthermore students of quantum computing can use simulated quantum computers to understand and develop new algorithms. It is also likely that quantum computers will be expensive and their use limited when first realized, much like classical computers were at the time of their introduction. In this case it becomes expensive to utilize actual quantum resources for learning quantum programming when simulated ones suffice. Aside from the exponential slowdown, there are other limitations one needs to be aware of when simulating a quantum computer on a classical one.

Quantum operations and registers can be represented classically using matrices, as already discussed. The size of these matrices increases exponentially with the number of qubits and is the reason for the exponential slowdown when simulating quantum computers. More efficient approaches to simulating quantum computers have been proposed, by means such as Deutsch and Jozsa [69]; but these approaches still experience an exponential slowdown in the worst case. These types of techniques for more efficient simulation of quantum computers shall not be explored further since the focus of this dissertation is on usable programming techniques as opposed to efficient simulation.

One area to be concerned about in simulating quantum computers is the round off errors that may occur. Numbers on a classical machine are discrete, and often of a fixed precision. The state of a quantum system can at times not be exactly represented using common classical data types. An example would be the representation of the square root of two, which is an irrational number. One needs to be aware of the possible rounding errors that could accumulate and make the classical simulation not reflect the reality that would occur on a quantum computer. Some existing approaches, such as Spector's automatic quantum computer programming approach, enforce limitations to minimize these errors [21]. One method to limit these errors is to limit the number of manipulations on a register so that the errors do not accumulate. An alternate approach is to use floating point data types more accurate than the standard 32 or 64 bit floating point numbers. In other words the accuracy of standard floating point numbers is limited much more than if we were to use more bits to express the number. A qubit is represented by values in a continuous range, meaning that no finite discrete system such as a classical computer can precisely represent an arbitrary qubit [13]. Thus quantum computers are proof that there are limitations to classical computation.

---

[20] A quantum system interacting with the outside environment and thus causing an observation is known as decoherence. Hence for quantum computing the system must be isolated during computation. This is one of the fundamental challenges of physically implementing a quantum computer.

Classical computers face an exponential slowdown when simulating quantum systems because a quantum system can be in multiple states at once (superposition). For example if there is a register of 8 qubits then 256 ($2^8$) probability amplitudes must be kept track of for the register. As a reminder these amplitudes are the entries in the matrix representing the register, or the numbers in front of each value in Dirac notation: $x_0 |00000000\rangle + x_1 |00000001\rangle + ... + x_{255} |11111111\rangle$. Additionally, this represents the quantum system only at a particular instant in time– these amplitudes also need to change as the system evolves by applying operations. Adding an additional qubit to a total of 9 means that the set can now be in 512 ($2^9$) states, and so on exponentially as we increase the number of qubits. For a more detailed explanation of this example the reader is referred to pages 45-50 in Johnson [43].

This exponential slowdown was recognized by Feynman, who in 1982 wondered what would happen if a computing device operated based on quantum instead of classical mechanics [43]. Deutsch expanded on this idea in 1985 to suggest quantum computers, as covered in section 2.1.5.

A quantum computer essentially carries out computations on an array of possibilities, yet only one of those possibilities is returned as a result– the one returned is randomly selected based on the probabilities of the possible outcomes. Consequently, repeated runs of the same quantum program may return different results [21] as already mentioned. It is impossible to see what the other potential answers are unless the program is run enough times to gather all possibilities. While it is impossible to examine the system to gauge the probability of all potential answers, these probabilities can be determined by repeatedly running the program. The idea is similar to a pair of dice: if the dice are thrown an infinite number of times then we can determine the probabilities of all possibilities.

Even though there are limitations simulating a quantum computer on a classical one, there is one area where there is an advantage. When the simulation is carried out it is possible for the user to examine what the possible answers are and their corresponding probabilities [21], violating the limitation of probabilistic output. This eliminates the need for repeated runs of the program. Carrying out a simulation on a classical computer also allows users to do another thing that isn't possible on an actual quantum computer: peek at the state of the system in the middle of computation, violating the limitation of no-observation[21]. On a quantum computer these observations would collapse the system to one result, making it impossible to resume the program without starting over. Since the limitation of no-observation isn't a physical one in a simulation, it can be broken. This ability to peek at the state of a system is obviously useful in testing to confirm that the simulator functions correctly.

Both of these features, seeing all possible results and examining the state of the system during execution, may be useful to students of quantum computer programming in order to better understand the computation. Even though they may be useful, their use should be discouraged in all but the most elementary exercises since they are impossible on an actual quantum computer. The prime reason for discouraging these behaviors is that students and practitioners of quantum programming should not become accustomed to features that are impossible to implement on working quantum computers[22]. Furthermore the users may reach false conclusions when it comes to quantum computing. Aside from the slow down of a simulation, the user of a particular quantum programming method would ideally not know if they were writing against a simulation or actual quantum computer. Nonetheless, existing simulations often provide these methods to peek at the state of a system– primarily for learning and debugging purposes.

### 2.1.10  Quantum Algorithms

Quantum computer algorithms are important to the study of quantum computer programming techniques, as they are typically demonstrated by implementing some of these algorithms. In this section some of the more widely known algorithms are briefly covered. Some of these do little more than illustrate basic advantages of quantum computing over classical computing, while others have real world applications. Since this work is for quantum computer programming and not algorithms, only a brief introduction into the most frequently covered algorithms is necessary to better understand the code examples that follow.

---

[21] Which lends challenges to debugging quantum programs as well since the state cannot be observed during execution.

Until the mid 1990s there were no known algorithms that used quantum computers to solve useful problems. The quantum algorithms introduced earlier by people such as Deutsch and Jozsa illustrated the power of quantum computers [70], but didn't utilize it for problems that had applications in the real world. Consequently quantum computing was viewed as something of a novelty. After a decade since Deutsch's proposal for quantum computers Peter Shor published his paper Polynomial-Time Algorithms for Prime Factorization and Discrete Logarithms on a Quantum Computer [47] in the mid 1990s. In this paper Shor outlines algorithms for quantum computers that allow factoring and discrete logarithm problems to be solved in a polynomial number of steps, based on input size. The discrete logarithm and prime factorization problems are generally considered to be hard on classical computers, with no known efficient algorithms. As a result these two problems form the basis of many modern cryptographic systems, especially the integer factorization problem. Integer factorization forms the basis of the RSA public key (asymmetric) cryptographic

---

[22] By being able to swap implementations, a classical simulation could be utilized for debugging. This would allow one to watch the state of the system as it evolves. Due to the exponential slow down this would only work for small numbers of qubits.

algorithm. RSA forms the basis of many commercial communication security algorithms [71].

If quantum computers become a reality, then Shor's algorithm makes many modern commercial encryption systems obsolete. Due to their widespread use this has profound implications, not only for ecommerce, but perhaps for national security as well, because the systems based on these algorithms would be rendered obsolete. Their wide spread use would also make them difficult to replace, especially for legacy systems. Due to these consequences, a branch of cryptography called quantum cryptography has arisen in recent years. Quantum cryptography attempts to create cryptographic systems that base their security on the laws of nature as opposed to problems that are hard to solve, as most modern cryptographic systems such as RSA do [28].

Lov Grover also introduced an important algorithm for quantum computers in the mid 1990s, his algorithm for fast database search [52] is commonly known as Grover's algorithm. Shor's integer factorization and Grover's fast database search are largely considered to be the most important quantum algorithms to date (2009). One or both of these algorithms are frequently covered in modern quantum computing texts [10, 12, 13, 28, 32]. On a classical computer, searching through an unordered list of objects requires $O(n)$ time. What Grover's algorithm does is allow for this list to be searched in $O(\sqrt{n})$ time. Furthermore, Grover's algorithm is the fastest possible quantum algorithm for this problem. In his paper Grover also points out that this algorithm is likely to be easier to implement than other quantum mechanical algorithms [52].

While Shor's and Grover's algorithms are the most written about algorithms due to their practical nature, several others are frequently encountered in quantum computing literature. The reader should consult the references for further details on these algorithms. Some of these other algorithms and a short description of what they accomplish include:

- Deutsch's algorithm [4] – Allows for `f(0) xor f(1)` to be determined with only one query to `f(x)` using the concept of interference covered in the quantum coin toss example in 2.1.1. A classical algorithm would have to query `f(x)` twice.

- Deutsch-Jozsa [70] – A more generalized version of Deutsch's problem that solves `f(x)` for *n* bits instead of 1 as in Deutsch's algorithm.

- Quantum Fourier transform [13] – The quantum version of the Fourier Transform.

- Generalized Simon's algorithm [72] – An approach for finding a hidden sub group.

While the number of known algorithms that take advantage of the power of quantum computers is limited, the development of genetic programming techniques for quantum programming holds promise [73]. In particular, the evolutionary approach combined with the power of a quantum computer to carry out parallel computations could allow for quantum computation to solve hard problems for which a quantum algorithm does not yet exist.

While there have been several important quantum algorithm developed to date, development of more algorithms continues to be an area of intense research within quantum computing– this is due in part to the limited number of existing algorithms. The algorithms presented in this section are often implemented in proposed quantum programming techniques and can give good insight into how practical and readable the proposed techniques are. The sections that follow cover various quantum programming languages, which are illustrated using these algorithms.

### 2.1.11 Example: Factoring (Shor's Algorithm)

Shor's algorithm to perform factoring is perhaps the most famous example of a quantum algorithm. As such it is often used as a real world example when demonstrating quantum programming techniques. In this section the workings of the algorithm will be detailed, how specific quantum operations are detailed in 4.3.3, while how factoring is accomplished in Cove is covered in section 5.4.3. This is a probabilistic algorithm, so several runs may be required to obtain the factors.

Shor's algorithm finds two prime numbers, $p$ and $q$, where N = $pq$. While it is very easy to compute N given $pq$, it is generally accepted that it is hard to find $pq$ given N. This assumption forms the basis of many codes, including the commonly used RSA algorithm [23] for asymmetric encryption[23]. Without going into detail on the workings of RSA, it suffices to say that N forms the basis of the public key while $pq$ form the basis of the private key.

---

[23] Recall asymmetric encryption is encryption where there are two keys, typically referred to as the public and private key. Messages encoded with one key can be decoded with the other. In typical practice the public key can be distributed publically, hence the name. If Alice wants to send Bob a message that only he can read, Alice then encrypts it with Bob's public key, sends it to Bob, who then decrypts it with his private key (that only he has).

| Symbol | Description |
|--------|-------------|
| N | The number to factor. |
| *p, q* | The factors of N. |
| P | The period. |
| *m* | A distinct random number in each iteration. |
| *N* | The number of (qu)bits needed to express N. |
| *F* | A function of integers that is periodic under addition. |

Table 4. Description of symbols used in Shor's.

There is a classical and quantum part of the algorithm. The key part of the algorithm where a quantum computer is exploited is finding the period P of a function *f*. *f* is a function on integers that is periodic under addition, where $f(x) = f(y)$ for a distinct *x* and *y*. In this case *x* and *y* differ by an integral multiple of P, in other words $f(x+P) = f(x)$ for every *x*.

Figure 40 provides a high level view of Shor's algorithm and is based the description in [23]. As the illustration shows, step 2 is the only part where a quantum computer is required in order to efficiently factor. The details of each step are covered in the text following the figure.

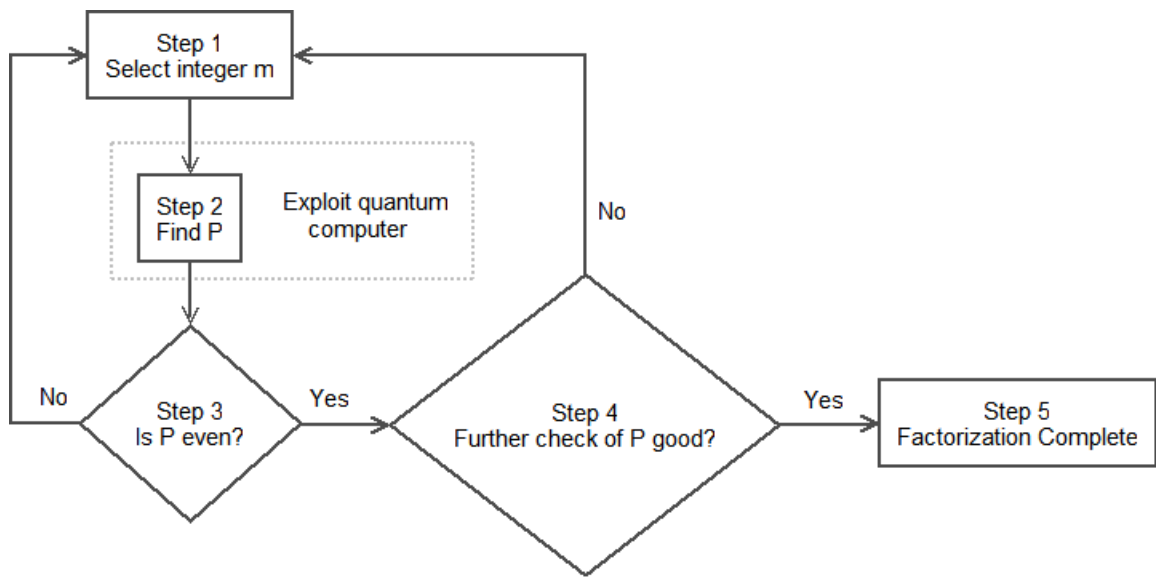

Figure 40. High level flow chart of factoring.

The first step of the algorithm is selection of an integer, *m*. *m* is selected randomly and is less than N (the number we are trying to factor). If the algorithm starts over due to the checks in step 3 or 4 then a value for *m* that we have not tried yet is selected. Once *m* is selected the greatest common divisor of N and *m* is calculated. This can be done via the Euclidean algorithm, which is an efficient algorithm on classical computers [28]. If the greatest common divisor is not 1 then m is *p* or *q*. In this case we have just randomly selected one of the factors through pure luck, and the algorithm is finished. Obviously there is nothing special about luckily selecting *p* or *q*− with a quantum or classical computer.

Step 2 is the only part of the algorithm that requires a quantum computer to make factoring efficient. For step 2 we are trying to find the period *P*, which may allow a factor to be obtained as will be shown. The function $f(x) = m^x \bmod N$ has a period *P* where $P \le N$. The period *P* is the smallest integer where $f(x + P) = f(x)$ for every *x* [28].

When calculating $f(x)$ enough values of x are used so the period can be obtained. Doing step 2 with a quantum computer is covered in detail shortly, as not to interrupt the discussion of the entire factoring process.

When using nontrivial examples the $m^x$ in $f(x)$ will quickly overflow in many languages that do not support arbitrary sized integers. Nonetheless there is as easy way to get around this problem, as shown in Figure 41. Thus to calculate the next $f(x)$ we use the result of the previous evaluation. For a more detailed description of obtaining Figure 41 the reader is referred to [42].

$$m^x \bmod N = ((m^{x-1} \bmod N)m) \bmod N$$

Figure 41. Obtaining *f(x)* while avoiding overflow [42].

For step 3 we first check to see if $P$ is even. If it is odd then we start the algorithm over while selecting a different *m*. These first steps continue until an even $P$ is obtained.

For step 4 we must check[24] $P$ to see if $m^{P/2} + 1 \equiv 0 \bmod N$, given the values of P and *N*. If this true then we start the algorithm over again, as $\gcd(m^{P/2} - 1, N) = 1$ (no factor found). As when starting over from step 3, we make sure to choose a value for *m* that we have not yet tried. If $m^{P/2} + 1 \not\equiv 0 \bmod N$ then we continue onto step 5.

Step 5 is the conclusion of the algorithm. Either $\gcd(m^{P/2} - 1, N)$ or $\gcd(m^{P/2} + 1, N)$ is a factor of N that is greater than 1 (possibly both). Thus $\gcd(m^{P/2} - 1, N)$ and/or $\gcd(m^{P/2} + 1, N)$ is one of the factors, either *p* or *q*. Given one of the two factors it is trivial to obtain the other, as $q = \dfrac{N}{p}$. We have thus efficiently found the factors of N by exploiting a quantum computer in step 2.

### 2.1.12  Replacing Step 2 with a Quantum Computer

What follows is a description of utilizing a quantum computer to obtain the period P. In this section we will not be concerned with how the necessary operations are constructed from elementary quantum operations, we will just assume that they are already defined unitary operators. The building of these operators from elementary quantum operations is detailed in 4.3.3.

For the quantum version of step 2, there are two registers, which we will refer to as Register 1 and Register 2: $\left| REG1 \right\rangle$ and $\left| REG2 \right\rangle$. $\left| REG1 \right\rangle$ will hold the possible values of $x$, while $\left| REG2 \right\rangle$ will hold the result of $f(x)$. $n$ is then the number of (qu)bits needed to express the number to be factored, $n = \log_2 N$ [42]. We need to express all possible values of x $(0 \le x \le N^2)$ in $\left| REG1 \right\rangle$, thus $\left| REG1 \right\rangle$ must consist of $2n$ qubits. By doubling the number of qubits in $\left| REG1 \right\rangle$ there will be at least N periods of $f(x)$ [27]. The result of $f(x)$ will always be $< N$, so $n$ qubits are required for $\left| REG2 \right\rangle$. In total $3n$ qubits are required [42].

In many explanations of Shor's algorithm, including [23], it is not often clear that these two registers are logical subsets of a single register, since entanglement between the two is utilized. In other words: even though we have two registers they need to be operated on as a whole for some parts of the algorithm. These registers are initialized to all zeros: $\left|0...0\right\rangle$. For the remainder of this example the state of qubits will be expressed in strings of binary digits that represent an unsigned integer. Example: $\left|0101\right\rangle$ means a register of 4 qubits is representing the decimal number 5, $0(2^3) + 1(2^2) + 0(2^1) + 1(2^0) = 4 + 1 = 5$. (No number in front of the state means it is representing the decimal number 5 with 100% probability when observed.)

---

[24] $x \equiv y (\mathrm{mod}\, z)$ means that $x - y$ is divisible by $z$.

Now that the quantum registers have been initialized we apply the Hadamard operation to all the qubits in $\left|REG1\right\rangle$. After the application of the Hadamard operations $\left|REG1\right\rangle$ is in a superposition of all values $0 \le x \le (N^2 - 1)$. In Dirac notation the state is expressed as shown in Figure 42.

$$\frac{\left|0...00\right\rangle + \left|0...01\right\rangle + \left|0...10\right\rangle + \left|0...11\right\rangle ... + \left|1...11\right\rangle}{2^n}$$

Figure 42. State of $\left|REG1\right\rangle$ after application of Hadamard operation.

Next we define an operation $U_f \left|REG1\right\rangle \left|REG2\right\rangle = \left|REG1\right\rangle \left|f(\left|REG1\right\rangle)\right\rangle$ and apply it ($f(x)$ as defined in the previous paragraph). Construction of $U_f$ is detailed later in 4.3.3, as it deals with constructing a quantum circuit from simple gates. Put plainly, we have created an operation $U_f$ that takes the values of $x$ from $\left|REG1\right\rangle$ and places all the results of $f(x)$ in $\left|REG2\right\rangle$. After $U_f$ is applied $\left|REG1\right\rangle$ and $\left|REG2\right\rangle$ are entangled. At this point one can think of $\left|REG1\right\rangle$ containing all possible $x$'s and $\left|REG2\right\rangle$ containing the

result of $f(x)$ for each of those values of $x$. Herein lies the power of a quantum computer: *all* of the values of x, and *all* of the results of $f(x)$ are now held in the quantum computer!

After this entanglement $\left| REG2 \right\rangle$ is measured, which collapses it to a single classical value. This collapse of $\left| REG2 \right\rangle$ also changes the state of $\left| REG1 \right\rangle$ because they are entangled. Due to this measurement of $\left| REG2 \right\rangle$ $\left| REG1 \right\rangle$ now only represents the $x$'s that resulted in the collapsed value in $\left| REG2 \right\rangle$.

Next the Quantum Fourier Transformation is applied to $\left| REG1 \right\rangle$, then $\left| REG1 \right\rangle$ is measured. This result in $\left| REG1 \right\rangle$ is called P and is the period, where $P$ is a natural number ( $P \in \square$ ) and $m^P \equiv 1 \bmod N$. Once $P$ is obtained the quantum part of the algorithm is finished and we can move onto step 3 and continue the algorithm with a classical computer. [23] describes an alternate implementation of using a quantum computer that utilizes fewer operations but is more difficult mathematically.

### 2.1.13  Factoring 15

Shor's algorithm can be a little abstract when read, so it will be demonstrated by detailing the following simple example of factoring 15 into 3 and 5. This example also works through all the states during the algorithm, but again the detailed construction of the operations used in the algorithm is covered in 4.3.3.

First we must randomly choose a value $m$, where $1 < m < N$. For this example we'll first select a value of 8 for $m$. This meets the constraint of $1 < m < N$, explicitly for this example $1 < 8 < 15$. Next we check $\gcd(8,15)$, which is $1-$ so m is not a factor of N. This concludes step 1.

We then calculate the values of $f(x)$ (for $0 \le x \le (15^2 - 1)$). In practice it is a little easier to understand and costs us nothing extra if we do the values $0 \le x \le (n^2 - 1)$, or $0 - 255$, instead of $0 \le x \le (N^2 - 1)$, or $0 - 225$ ($15^2$). The reason is that we are effectively putting the qubits in equal probabilities of all possible states (0 - 255) for the given number of qubits instead of the explicitly needed subset ($0 - 225$). Constructing 0 - 255 is easier because we simply have to apply the Hadamard operation to each qubit. If we were to do $0 - 225$ then the operations to apply would not be so simple. Figure 43 shows the results of the first few calculations of $f(x)$, while Figure 44 shows this as a graph with each period boxed in red. In actuality we are going up to x = 255, as is shown in Figure 45.

$$f(0) = 8^0 \bmod 15 = 1$$

$$f(1) = 8^1 \bmod 15 = 8$$

$$f(2) = 8^2 \bmod 15 = 4$$

$$f(3) = 8^3 \bmod 15 = 2$$

$$f(4) = 8^4 \bmod 15 = 1$$

$$f(5) = 8^5 \bmod 15 = 8$$

$$f(6) = 8^6 \bmod 15 = 4$$

$$f(7) = 8^7 \bmod 15 = 2$$

$$f(8) = 8^8 \bmod 15 = 1$$

$$f(9) = 8^9 \bmod 15 = 8$$

$$f(10) = 8^{10} \bmod 15 = 4$$

$$f(11) = 8^{11} \bmod 15 = 2$$

$$f(12) = 8^{12} \bmod 15 = 1$$

$$f(13) = 8^{13} \bmod 15 = 8$$

$$f(14) = 8^{14} \bmod 15 = 4$$

$$f(15) = 8^{15} \bmod 15 = 2$$

...

Figure 43. Finding the period, calculation of *f(x)* when *m*=8 and N=15.

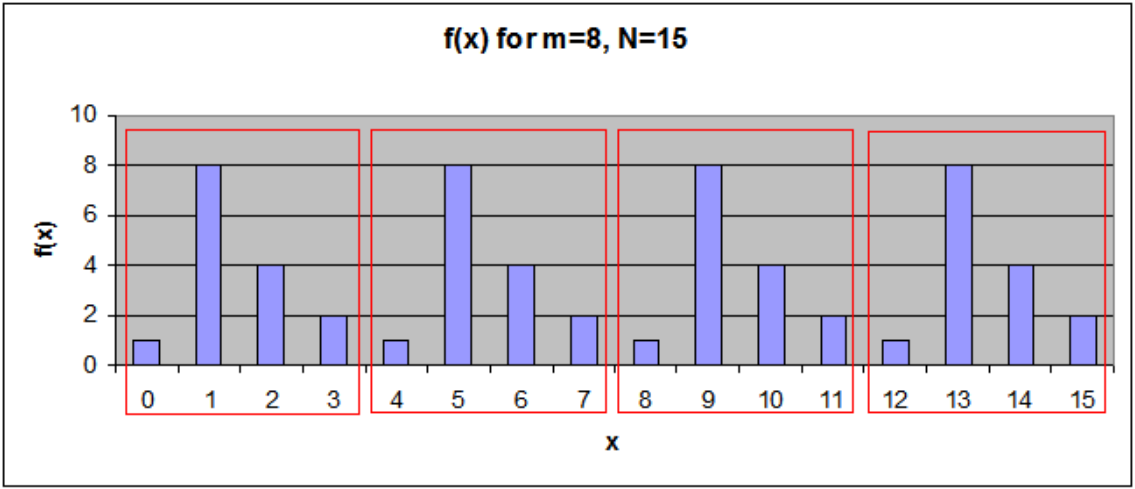

Figure 44. Finding the period as a graph, *f(x)* when *m*=8 and N=15.

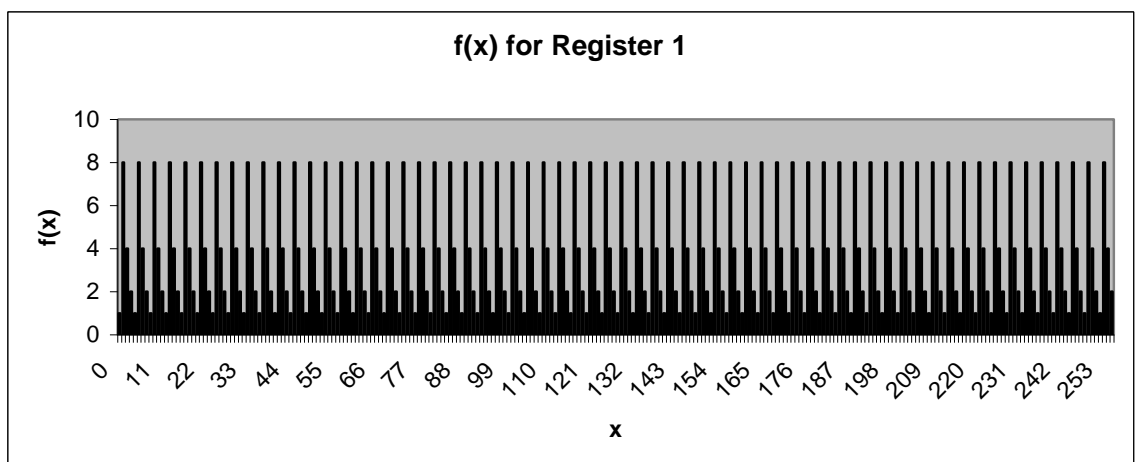

Figure 45. Calculation of *f(x)* for all values of *x*.

As can easily be seen from Figure 43 and Figure 44, the period *P* is 4 since the pattern repeats every 4 numbers: 1, 8, 4, 2, … Finding the period is step 2, so we have completed the second step.

When this part (finding the period) of the algorithm is replaced by the quantum part, one can think of the quantum computer representing all values of *x*. These values of *x* are then used to calculate $f(x)$, from which the period is found.

4 qubits are required to express the decimal integer 15, so *n* = 4. (Recall *n* is the number of (qu)bits to express the number being factored.)  Hence we allocate a register of 12 qubits (3*n*): 8 (2*n*) qubits for $\left| REG1 \right\rangle$ and 4 (*n*) for $\left| REG2 \right\rangle$. All qubits in the registers are initialized to 0, so $\left| REG1 \right\rangle = \left| 00000000 \right\rangle$ and $\left| REG2 \right\rangle = \left| 0000 \right\rangle$. Next we put the first register in superposition of all possible states by applying the Hadamard operation to all qubits, leaving it in the state shown in Figure 46– evenly representing the values 0-255 at once. Recall that since $\left| REG1 \right\rangle$ is in superposition we cannot actually observe this state in Figure 46; if we try to observe the state it will collapse randomly to one of the possible values, $0 - 255$.

$$\frac{\left|00000000\right\rangle + \left|00000001\right\rangle + \left|00000010\right\rangle + \left|00000011\right\rangle + ... + \left|11111111\right\rangle}{\sqrt{256}}$$

Figure 46. State of $\left|REG1\right\rangle$ after application of Hadamard to all qubits.

Next $U_f$ is applied (see 4.3.3 for details on constructing the operator), which leaves $\left|REG2\right\rangle$ representing $f(x)$ for all the values in $\left|REG1\right\rangle$. $f(x)$ will range from 0 to 15. Thus we can think of $\left|REG2\right\rangle$ as keeping a tally of these possible values as we evaluate $f(x)$ for each $x$. Explicitly, $\left|REG2\right\rangle$ is now in the state shown in Figure 47 in Dirac notation and Figure 48 graphically. As can be seen $f(x)$ only evaluates to 1, 2, 4, and 8 for this case. $\left|REG2\right\rangle$ also has an equal probability (25%) of collapsing to one of these four values if observed.

$$\frac{1}{\sqrt{16}}\left|0001\right\rangle + \frac{1}{\sqrt{16}}\left|1000\right\rangle + \frac{1}{\sqrt{16}}\left|0100\right\rangle + \frac{1}{\sqrt{16}}\left|0010\right\rangle + \frac{1}{\sqrt{16}}\left|0001\right\rangle + \frac{1}{\sqrt{16}}\left|1000\right\rangle + \frac{1}{\sqrt{16}}\left|0100\right\rangle + \frac{1}{\sqrt{16}}\left|0010\right\rangle$$

$$+ \frac{1}{\sqrt{16}}\left|0001\right\rangle + \frac{1}{\sqrt{16}}\left|1000\right\rangle + \frac{1}{\sqrt{16}}\left|0100\right\rangle + \frac{1}{\sqrt{16}}\left|0010\right\rangle + \frac{1}{\sqrt{16}}\left|0001\right\rangle + \frac{1}{\sqrt{16}}\left|1000\right\rangle + \frac{1}{\sqrt{16}}\left|0100\right\rangle + \frac{1}{\sqrt{16}}\left|0010\right\rangle$$

$$= \frac{1}{\sqrt{4}}\left|0001\right\rangle + \frac{1}{\sqrt{4}}\left|1000\right\rangle + \frac{1}{\sqrt{4}}\left|0100\right\rangle + \frac{1}{\sqrt{4}}\left|0010\right\rangle$$

Figure 47. Register 2 after application of Uf (Dirac Notation).

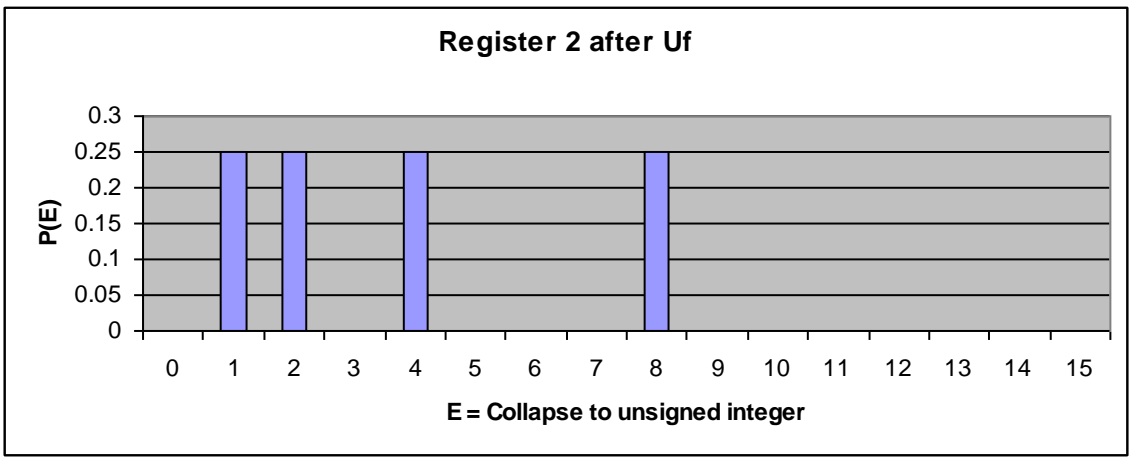

Figure 48. Register 2 after application of Uf (Graphically).

The next step in the quantum part of the computation is the measurement of $|REG2\rangle$. This collapses $|REG2\rangle$ to 1, 2, 4, or 8 with a probability of 0.25 for each as shown in Figure 48. For this example let's assume that $|REG2\rangle$ collapses to 4, leaving it in the state $|0100\rangle$. However since the two registers are entangled $|REG1\rangle$ is also effected. In this case every x that resulted in 4 remains, leaving it in the state shown in Figure 49 (in a superposition of 2, 6, 10, 14, and so on while $< 256$) and graphically in Figure 50.

$$\frac{|00000010\rangle + |00000110\rangle + |00001010\rangle + |00001110\rangle + ...}{\sqrt{64}}$$

Figure 49. Register 1 after the application of U*f*.

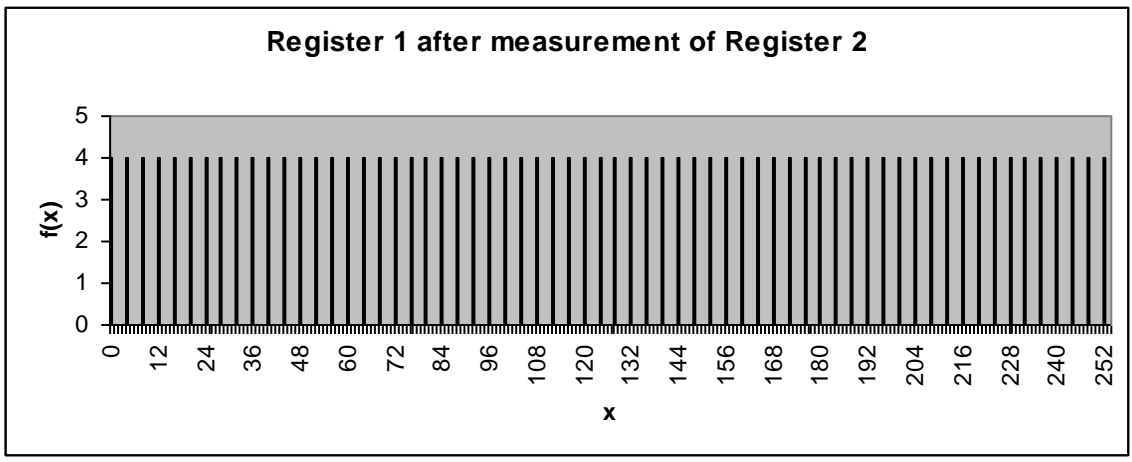

Figure 50. Graphical representation of register 1 after measurement of register 2.

The final step is to find the period from $\left| REG1 \right\rangle$. To do this we utilize the quantum Fourier transformation, followed by a measurement. This essentially returns the period from the superposition in $\left| REG1 \right\rangle$. What is happening in this final step of utilizing the quantum computer is illustrated in Figure 51. Construction of the quantum Fourier transformation is detailed in Figure 104.

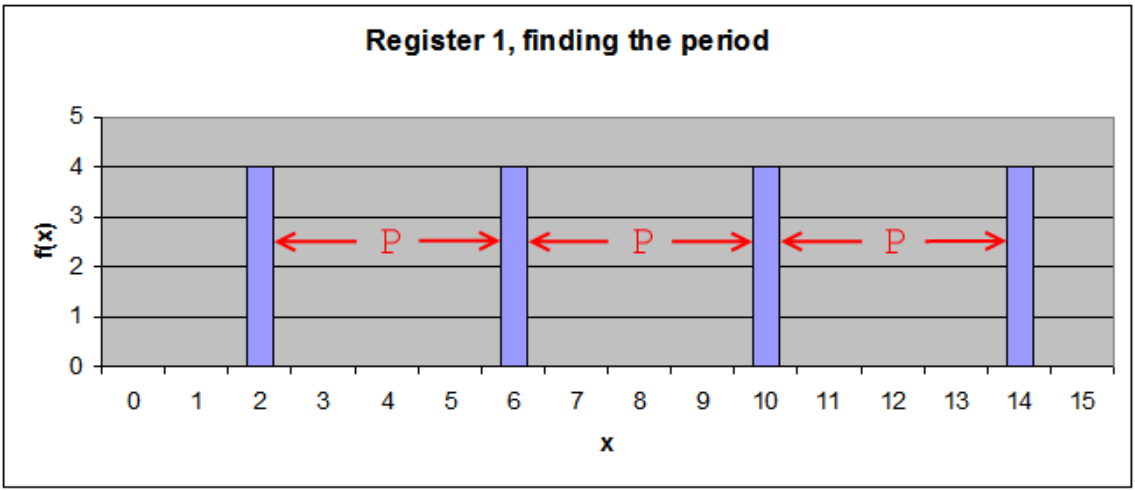

Figure 51. Finding the period (P) from Register 1.

We now have the period, 4, so we can continue the rest of the steps in the algorithm classically. This period of 4 is even, which passes the check for step 3. Continuing onto step 4 we check $8^{4/2} + 1 = 8^2 + 1 = 65$ and $8^{4/2} - 1 = 8^2 - 1 = 63$. We then find the greatest common denominator of these two against the number we are factoring: $\gcd(65,15) = 5$ and $\gcd(63,15) = 3$. In this case we have reached step 5 and found the factors of 15: 5 and 3. One can easily walk through repeating the early steps of the algorithm for different values of *m*.

### 2.1.14 Further Scaling

Factoring 15 is a trivial example, so this section discusses how factoring larger numbers really takes advantage of a quantum computer. As an example let's consider a 128 bit number, which is a more realistic factoring example than 15. For the quantum part of the factoring we will need 384 qubits (3 * 128).

The important part is that the quantum operations are run once, just on more qubits. A classical equivalent of scaling up is addition: the procedure is the same, it just operates on more bits. For factoring we are calculating $f(x)$ for a huge number of values: 0 to $2^{(2*128)}$. This can be done all at once with the quantum computer, but is too many calculations to do on a classical computer. Consequently it isn't that calculating $f(x)$ is hard– it is the large number of times that we have to do it. Not only do we have to do all these calculations, but we have to store the results somewhere too in order to find the period.

So we confronted with a choice. The period can be found classically by performing such a large number of operations that we won't be able to do them in the foreseeable future. The alternate is to use just a few hundred qubits and easily perform the needed calculations by performing the quantum just once for each iteration of step 2 in the algorithm. It is this massively parallel nature of quantum computing that allows for feats that may never be possible on classical computers.

## 2.2    Quantum Computer Programming

The programming languages examined in this section are divided into two categories: procedural and object oriented languages are covered in 2.2.1, while all others are covered in 2.2.2. Due to their popularity in commercial environments, procedural and object oriented languages are examined separately since Cove is an object oriented approach. Procedural languages are examined along with object oriented languages because they have many traits in common, and in instances such as C/C++/C# the object oriented language has evolved out of the procedural one. Other techniques are covered for comparison purposes and to identify common themes throughout a majority of quantum programming techniques.

### 2.2.1    Survey of Procedural and Object Oriented Quantum Computer Programming Techniques

In the academic world a variety of languages are studied and used. But with the exception of a few domains, most commercial applications use object oriented languages. A partial list of these languages includes many that would be familiar to any commercial developer: Visual Basic, C#, Java, Python, Fortran, Cobol, and so on. For the power of a quantum computer to be utilized economically in commercial applications, the programming must be easy for existing commercial developers to learn and employ. This is best done by piggy backing off of the languages and techniques they are already familiar with– this means that successful quantum languages for existing commercial developers will likely be related to one of more of these languages, or quantum frameworks (libraries) for these languages. The popularity of languages changes with time, so as new languages come into popularity their potential for quantum computing also needs to be kept in mind. Many of today's popular languages were not designed to easily take advantages of multiple cores or processors. Consequently it is quite feasible that other languages that take advantage of these parallel processing capabilities will rise in popularity in the near future and be excellent candidates for extending to carry out quantum computing. Of course, the alternate is that existing languages will be extended.

The structure of quantum programming languages differ from existing classical languages in that the limitation outlined in section 2.1.6 must be enforced. Depending on the proposed approach, violation of these limitations may be caught at compile time or at run time. The quantum languages typically include statements for the 3 core tasks needed for quantum computing:

1. Initializing the quantum state of the system.

2. Manipulating it through (unitary) operations.

3. Measurement, which results in classical data.

When languages are developed for quantum computing additional statements are included to carry out classical computation, since something must be done with the result. When frameworks or APIs for classical languages are developed the focus of those can be on quantum computation instead of also having to provide classical capabilities as a language must do.

As mentioned in section 2.1.5, Knill has introduced pseudocode conventions [3]. His pseudocode is based on imperative program techniques, as it utilizes variables and flow control statements based on that methodology. Within his paper he also provides several elementary examples of the use of his proposed pseudocode; one example is shown in Figure 52. As mentioned previously, the importance of Knill's paper lies not necessarily in the proposed pseudocode conventions, but in the use of his quantum random access machine model (QRAM). While Knill's work is an important step forward, pseudocode has little use for writing actual applications.

$a \leftarrow \text{MEASUREDFOURIER}(\underline{a}, d)$

**Input:** A quantum register $\underline{a}$ with $d$ qubits. The most significant qubit has index $d - 1$.

**Output:** The amplitudes of $\underline{a}$ are Fourier transformed over $\mathbb{Z}_{2^d}$, and then measured. The most significant bit in the output has index 0, that is the ordering is reversed. The input quantum register is returned to a classical state in the process.

$\omega \leftarrow e^{i2\pi/2^d}$
$\phi \leftarrow 0$
**for** $i = d - 1$ **to** $i = 0$
$\quad \mathcal{R}_\phi(\underline{a_i})$
$\quad \mathcal{H}(\underline{a_i})$
$\quad a_i \leftarrow \underline{a_i}$
$\quad \phi \leftarrow (\phi + a_i \pi)/2$
$\qquad$ **C:** *The expression on the right of this assignment statement requires $a_i$ to be in a classical state as it involves operations not allowed for quantum registers.*

Figure 52. Measured Fourier transform utilizing Knill's pseudo code [3][25]

---

[25] Each programming method discussed in this chapter is intentionally not covered in great detail for the sake of being concise. The code snippets are included so the reader can get a brief feel of the particular method. For more details, the reader should consult the appropriate reference.

Sanders and Zuliani developed the programming language qGCL as a means to express quantum algorithms through "rigorous semantics and associated refinement calculus" [74]. The refinement calculus allows for the specification and code to be combined, thus making the language more formal. The authors state the primary purpose of the language is for program derivation, proof of correctness, and teaching. As the authors point out, qGCL does not aim to do numerical simulations of quantum algorithms like Omer's QCL, which will be covered later. Within the paper they first describe a probabilistic extension to Dijktra's guarded command language (GCL) [75], which they appropriately call pGCL. They then extend pGCL to invoke quantum procedures and call the resulting language qGCL. Thus qGCL is like many other proposed quantum programming techniques where the computation is controlled by a classical computer utilizing a quantum sub system. The three quantum procedures they outline and place emphasis on are fundamental to any system carrying out quantum computation: initialization, evolution, and finalization (or observation). They also provide implementations of several quantum algorithms, including Shor's [47] and Grover's [52]. Since GCL was proposed in 1975, and qGCL is an augmentation to it, qGCL may be too limited and dated to construct commercial applications. Like Knill's pseudo code, qGCL also suffers from a very mathematical syntax– something that is harder for commercial programmers to understand and even type. As the authors point out though, this simplicity makes it an effective tool for teaching the basics of quantum programming.

$$\textbf{var } t : \mathbb{B}, \ a, d, p : 0 \mathbin{..} (n+1) \ \bullet$$
$$\quad t := 0 \ \mathring{,}$$
$$\textbf{do } \neg t \ \longrightarrow$$
$$\qquad a :\in 2 \mathbin{..} n \ \mathring{,}$$
$$\qquad d := gcd(a, n) \ \mathring{,}$$
$$\qquad \textbf{if } d \neq 1 \ \longrightarrow \ t := 1$$
$$\qquad [] \ d = 1 \ \longrightarrow \ Q(a, n; p) \ \mathring{,}$$
$$\qquad\qquad\qquad \textbf{if } p \text{ odd} \ \longrightarrow \ t := 1$$
$$\qquad\qquad\qquad [] \ p \text{ even} \ \longrightarrow$$
$$\qquad\qquad\qquad d := gcd(a^{p/2}{-}1, n) \sqcup gcd(a^{p/2}{+}1, n) \ \mathring{,}$$
$$\qquad\qquad\qquad t := (d \neq 1)$$
$$\qquad\qquad\qquad \textbf{fi}$$
$$\qquad \textbf{fi}$$
$$\textbf{od}$$

Figure 53. Shor's algorithm in Sanders and Zuliani's qGCL [74].

Bettelli, Calarco, and Serafini have developed a preliminary extension for C++, in the form of a library, for quantum computer programming [2], which evolved into Bettelli's Ph.D. thesis [29]. These seem to be referred to as Bettelli's C++ extensions, but the author sometimes refers to it as the Q language; the former will be used throughout this dissertation. This library exposes several classes that can be utilized for quantum computation. The use of classes provides the important benefit of encapsulating the workings of the library and hiding the implementation from users. Furthermore, unlike some procedural implementations, rules can be better enforced and valid states maintained through the use of classes[26]. Bettelli's implementation also generates quantum operations, and these byte codes could be piped to an actual quantum subsystem or a simulator.

---

[26] This does not mean to imply that classes are required.

While the library is in a preliminary form, Bettelli's paper also contains a list of features desirable for a scalable quantum programming language. One of the most important of these points is that a quantum programming language should be an extension of a classical language. Extensions can take a variety of forms: class libraries and dynamically linked libraries to name a few. These don't add new features to the language itself, but provide functionality not included "out of the box". Not only does extending a classical language make it easier for existing programmers to utilize quantum features, but it also helps to keep the library applicable to new classical techniques as the language surrounding it evolves to tackle classical problems. Thus the author of the quantum extension can focus on tackling only those issues that apply to quantum computing instead of also tackling classical computation as must be done with a proprietary language. Bettelli's work is the only programming proposal examined that focuses on a practical approach, but there is still room for improvement. His approach also focuses on byte codes and error corrections[27], which widens the focus of his work to include not just the C++ extensions, but implementation issues.

It appears that Bettelli halted work on his C++ extensions in 2003. At this point all of his basic classes seem to have been implemented, but his list of improvements and to-dos (places marked in the code where work needs to be done) in the most recent version [76] outlines some areas that new quantum programming proposals should take into account:

- A quantum simulator to carry out tests of the proposal.

---

[27] One source of errors, and the primary challenge in building quantum computers, is decoherence: unintended interaction with the outside environment which collapses all or part of the system.

- Use of namespaces.

- Tests of the proposal to strengthen claims of correctness and completeness.

- Multithreading is something to consider, both for thread safe (or not) access to objects and for speeding up any simulation.

In light of Bettelli's work, it is important to note that some languages, such as Python, are evolving iteratively through open source methods [77] as opposed to large standards developed over a period of years as is the case with C and C++[28]. C++ was developed in 1984 [57], but the standard was not approved until 1998 [78]– enough time for processors to double in speed seven times in accordance with Moore's law. Additionally, there have been over 8,500 programming languages developed [1], yet only a select few of these are actually used in industry– further strengthening the argument for creating extensions of existing languages instead of new languages. Bettelli's work is the most useful to existing programmers because C++ is a widely used language and an ancestor of many others. Additionally, only the library needs to be learned, not an entire new language. As new languages are developed and speed and efficiency of a language are not as important due to increased computing power, C++ seems to be declining in popularity. Thus as we move forward a higher level language such as Java or C# would be perhaps more useful, and also avoids the development expense of memory management.

---

[28] Although the languages were created and in use well before the standard.

```
Qbitset run_Grover(bool(*f)(int), int n) {
    int repetitions = sqrt(pow(2.0,n));
    Qop phase_oracle(f,n);
    Qop invert_zero(f_0,n);
    Qop mixer = QHadamard(n);
    Qop invert_mean = mixer & invert_zero & mixer;
    Qop grover_step = phase_oracle & invert_mean;
    Qreg input(n);
    mixer(input);
    for (int i=0; i<repetitions; ++i) grover_step(input);
    return input.measure();
}
```

Figure 54. Grover's algorithm in Bettelli's C++ extension [2]

Over a period of six years, 1998 – 2004, Omer has developed what is arguably the most complete quantum programming language to date: Quantum Computation Language, or QCL [30, 79-81]. This claim of completeness is reinforced by the fact that QCL has been used for an introduction to the subject of quantum computing [82]. QCL is a language that has a structure similar to C, making it easy to learn for many programmers because C and its descendants such as C++, C#, and Java are popular languages [57]. However this strength of basing QCL on C is also part of its downfall. C is still used for low level functions such as drivers, but not often for new applications. As a result, QCL does not have many of the features available in modern languages. By being a proprietary language QCL would be difficult to adopt in the real world for many programmers writing applications since it does not have the power and libraries available in modern languages for classical computation. Omer has also created a complete simulator for QCL programs, including an interpreter. Having an interpreter for QCL allows for students of the language to create and see how code behaves in real time. In a benefit to all studying quantum computing, Omer has also made the source code of the interpreter available [83]. While the inclusion of the interpreter and source code makes QCL useful, especially for those studying quantum programming proposals, the fact that it is a new language does present an obstacle to those wishing to learn quantum computer programming. While learning a new language is not always a large learning curve, reproducing existing classical features already present in other languages would be a significant expense. As with all new languages, it also makes it harder to integrate quantum algorithms into existing code bases.

```
/* Define Oracle */

const coin1=(random()>=0.5);        // Define two random boolean
const coin2=(random()>=0.5);        //   constants

boolean g(boolean x) {              // Oracle function g
  if coin1 {                        // coin1=true  -> g is constant
    return coin2;
  } else {                          // coin1=false -> g is balanced
    return x xor coin2;
  }
}
qufunct G(quconst x,quvoid y) {     // Construct oracle op. G from g
  if g(false) xor g(true) { CNot(y,x); }
  if g(false) { Not(y); }
}
/* Deutsch's Algorithm */

operator U(qureg x,qureg y) {       // Bundle all unitary operations
  H(x);                             //   of the algorithm into one
  G(x,y);                           //   operator U
  H(x & y);
}
procedure deutsch() {               // Classical control structure
  qureg x[1];                       // allocate 2 qubits
  qureg y[1];
  int m;
  {                                 // evaluation loop
    reset;                          //   initialize machine state
    U(x,y);                         //   do unitary computation
    measure y,m;                    //   measure 2nd register
  } until m==1;                     // value in 1st register valid?
  measure x,m;                      // measure 1st register which
  print "g(0) xor g(1) =",m;        //   contains g(0) xor g(1)
  reset;                            // clean up
}
```

Figure 55. Deutsch's algorithm expressed in Omer's QCL [30]

Blaha has introduced a quantum assembly language and quantum C language [84]. In his two language proposals the languages themselves are algebraic in nature, which he argues allows for better understanding of the language and proof of correctness if necessary. Within Blaha's work however, less than one page is dedicated to his quantum C language, and most of that involves an explanation of pointers in C. So while he proposes a quantum C language, there isn't much of an explanation of how it works other than defining the algebraic representation of the pointer operations. It is also interesting to note that Blaha was able to obtain trademarks for what would seem to be generic terms in the field of quantum computing, including "Probabilistic Grammar", "Quantum Grammar", and "Quantum Assembly Language". Like Bettelli's work, Blaha's use of C makes the approach viable. However, without further details it is hard to gauge how easy it is to actually use.

Markus has devised a method to simulate quantum computing using Fortran [85]. While not a true language or framework in itself, it is worth noting because it is an example of how such a library would work. Currently any quantum computing language or library must simulate the quantum system since quantum computers are currently unavailable for use in programming. Many languages are derived from Fortran [57][29] [30], so Markus's paper gives a good insight on how to actually accomplish that for a variety of languages. Included in the paper is the full source code listing for the simulation, along with debugging statements. It is also notable that Fortran has been used as a parallel programming language in the Fortran-K implementation, which is based on Fortran-90

---

[29] Some of lauages that Sebesta includes as being desecended, sometimes several generations, from Fortran include: Basic, Algol, Modula-2, Ada, and C.

[86]. Nonetheless, more modern languages such as Fortress [56] could also be used to simulate quantum computing and be more accessible. Providing the source code is invaluable for others developing quantum libraries as it provides a source of solutions for problems that may arise during implementation, and this is a benefit of the work Markus has done.

Carini has developed a method to simulate qubits using the programming language Ruby [87]. Like Markus's Fortran simulation [85], even though it is not a language or framework[31] it is noteworthy due to the implementation techniques. Carini's implementation involves simulating the states of a qubit on separate threads, although she admittedly ran into some scheduling issues. This is another important insight for the simulator of any proposed language or framework– the simulation should take advantage of today's multiprocessor systems. Doing so increases efficiency of the simulation, but presents challenges of its own through the need to implement parallel processing techniques.

In particular this presents a problem for any framework or language built upon the Python programming language due to the global interpreter lock. While Python is a concise and easy to program in language, only one thread within a process can access Python objects at a time [88]. This means that even with a multiprocessor system, multithreaded Python programs cannot take full advantage of it as they effectively use one processor. The work around for this is to implement multiple processes within Python instead of multiple threads. Even with this difficulty Python is still a good candidate for building a quantum computing framework on. Python is an interpreted language, which allows for one to dynamically interact with the program much like Omer's QCL [79].

---

[30] Algol lead to more structure, and as a descendent of Fortran we can say that a lot of structure in future languages is thus indirectly descended from Fortran.

Svore and colleagues have developed a suite of tools for use in quantum computation [89]. These tools include a language, compiler, optimizer, simulator, and layout tools. A key feature to the language, as others have pointed out as necessary, is that it is machine independent. For practical purposes quantum computers are not yet a reality, so any proposal for programming them must be independent of whatever physical solution is used to realize them. Within their paper they also propose translating their high level language into a quantum intermediate language (QIR) which then gets translated into a quantum assembly language (QASM), and finally a physical language (QCPOL). This is approach is the similar to many modern day classical languages. As with many other quantum programming proposals, this one also makes use of Knill's QRAM model [3]. Another key to the proposal is that quantum error correction[32] would be implemented on a lower level and not within the higher level language itself. This higher level abstraction is akin to how modern day programmers are not concerned with error correction within RAM or through a network connection, except when fatal errors occur such as a network connection that cannot be made or a file not found. While the purpose of the various languages and transitions between them are described, the work does not actually include specifications for the languages themselves. As such, the languages themselves remain an open problem as is pointed out at the end of the paper as an important challenge.

---

[31] Although a simulation still has to be setup somehow, and depending on the simulation that might be considered a graphical or visual programming method.
[32] Quantum error correction essentially allows the state of the qubits to be reliably maintained.

Tucci has developed quantum compiler that compiles steps of an algorithm into a sequence of elementary operations [90-92]. The implementation of his compiler proposal is called "Qubiter", for which he has made the source code in C++ freely available. While still in a basic state as he admits and lacking a GUI[33], it is still a valuable resource for those developing quantum programming methods, again because the source code is available. Being able to examine the source can lead to insights for solutions to implementation problems. Notable about his compiler is that it will also perform optimizations. Clearly this work on optimization would be useful for any other quantum programming system in order to increase efficiency. Tucci also received a patent for the ideas that Qubiter represent in 2002 [90]. Figure 56 shows the output of Tucci's Qubitter for the 4 qubit Hadamard matrix [92], which is also known as Hardamard-Walsh transform and will put 4 qubits into equal superposition of 0 and 1 when starting from state where each qubit is $|0\rangle$ or $|1\rangle$. Recall that the Hadamard operation is used to toss the coin in the quantum coin toss example.

---

[33] A GUI isn't necessarily required, but is helpful. One can certainly program most text based languages using a simple text editor, but this isn't nearly as productive.

```
ROTY 3 45.0000000
ROTY 2 45.0000000
ROTY 1 45.0000000
CPHA 1 T 180.000000
CPHA 2 T 180.000000
CPHA 3 T 180.000000
ROTY 0 45.0000000
CPHA 0 T 180.000000
```

Figure 56. Output of Tucci's Qubitter for the input 4 bit Hadamard matrix[92].

While there has been a small variety of quantum computing programming proposals utilizing the imperative or object oriented approach, none of them is equivalent to or utilizes the more widespread modern programming languages such as C#, Visual Basic, Java, or Python. The lack of a quantum computing framework for any of these languages makes quantum computer programming less accessible to the average commercial developer. Just as important, usability has also been neglected. So while the languages and libraries presented could be used, the fact that they are not similar to or use modern languages represents a significant hurdle to their use by practicing commercial developers. The fact that modern languages are not utilized for quantum computer programming and usability has been largely ignored represents an excellent candidate for work in the field of quantum computer programming.

### 2.2.2   *Brief Overview of Select Other Quantum Programming Techniques*

Now that procedural and object oriented methods for programming quantum computers have been covered, some alternate methods will be briefly explored. Many of the techniques utilized by these solutions are not further explored because they are too foreign to many commercial programmers for a variety of reasons, predominantly their techniques and syntax. It could be argued that most undergraduates would have some exposure to concepts such as functional or logical programming, making them valid candidates. Even if this is the case in undergraduate education, according to the United Engineering Foundation only 40% of practicing developers hold a Bachelors degree in software related disciplines [5]. In the United States 50,000 software development jobs are created each year, but only 35,000 computer science related degrees are awarded [5]– so this trend of many developers not having a strong background is likely to continue. Consequently a majority of existing commercial programmers have no exposure to these methods. Nonetheless, these approaches need to be explored on a limited basis in order to identify commonality with imperative approaches and to point out alternate techniques.

In 1996 Baker introduced QGOL, which is a system for simulating quantum computers using a functional approach [93]. QGOL remains an incomplete work, even though he admits rewriting it several times. He partially blames this on standard object oriented design techniques and says there would not be a better way to partition the problem. Almost always there are many different ways to design a system using object oriented techniques, so without further details on his design and implementation this assessment needs to be viewed critically. What is most useful about this paper is not the solution that he came up with, but the implementation issues that were encountered. Some of these issues include: detecting unitary operations, partitioning the problem, use of inheritance for operations, and so on. This knowledge is invaluable for anyone looking to implement a quantum computer language and/or simulator on a classical computer.

In 2003 Sabry proposed a method of programming quantum computers utilizing Haskell [94]. Haskell is a functional language, which contains no imperative constructs and has no side effects [57]. Due to these characteristics, Haskell is a good language to model quantum computing in because it deals well with the limitations on quantum computing outlined earlier in section 2.1.6.

$$deutsch \; :: \; (Bool \; \rightarrow \; Bool) \; \rightarrow \; IO \, ()$$
$$deutsch \; f \; =$$
$$\textbf{do} \; inpr \; \leftarrow \; mkQR \, (qFalse \; \&* \; qTrue)$$
$$\textbf{let} \; both \; = \; virtFromR \; inpr$$
$$top \; = \; virtFromV \; both \; ad\_pair_1$$
$$bot \; = \; virtFromV \; both \; ad\_pair_2$$
$$uf \; = \; cop \; f \; qnot_{op}$$
$$app1 \; hadamard_{op} \; top$$
$$app1 \; hadamard_{op} \; bot$$
$$app1 \; uf \; both$$
$$app1 \; hadamard_{op} \; top$$
$$topV \; \leftarrow \; observeVV \; top$$
$$putStr \; (if \; topV \; then \; "Balanced" \; else \; "Constant")$$

Figure 57. Deutsch's oracle for his algorithm in Sabry's Haskell approach [94].

Karczmarczuk has also developed an approach using Haskell around the same time as Sabry, and acknowledges Sabry's work. Karczmarczuk calls his approach a "framework for representing quantum entities in Haskell" [95]. Within this approach both quantum states and operators are functions. Karczmarczuk also emphasizes that the level of abstraction in the approach is high. This can be considered an advantage from a programming standpoint since the approach is not strongly tied to a physical implementation or problems that could potentially be overcome. One important advantage of this framework is that it is difficult to perform operations that are illegal from the quantum perspective, which Karczmarczuk identifies as good for the discipline of the programmer.

```
circuit f =
  let in1 =
      (boost2 ax_had ax_had) (ket B0 <*> ket B1)
    in  boost2 ax_had id (fcnot f in1)
xout = circuit id
yout = circuit (const B0)
```

Figure 58. The oracle for Deutsch's algorithm in Karczmarczuk Haskell framework [95].

Grattage has developed a functional language called QML [96]. A compiler for QML has also been created which takes QML programs as input and creates Haskell data types from them. QML is quite mature in the sense that the compiler is complete, so working examples can be constructed and compiled. Using Haskell also highlights the popularity of the language within the community of those researching functional quantum programming techniques.

$Had, Qnot, Meas \in Q2 \multimap Q2$
$Had\ b\ = \mathbf{if}^{\circ}\ b\ \ \mathbf{then}\ \mathbf{qfalse} + -\ \mathbf{qtrue}$      -- The Hadamard operation
          $\mathbf{else}\ \ \mathbf{qfalse} + \mathbf{qtrue}$

$Qnot\ b = \mathbf{if}^{\circ}\ b\ \mathbf{then}\ \mathbf{qfalse}\ \mathbf{else}\ \mathbf{qtrue}$      -- The Not operation (Pauli X)
$Meas\ b = \mathbf{if}\ b\ \mathbf{then}\ \mathbf{qtrue}\ \mathbf{else}\ \mathbf{qfalse}$      -- A measurement operator, using $\mathbf{if}$
$CNot\ \ \ \in Q2 \multimap Q2 \multimap Q2 \otimes Q2$      -- A quantum CNOT operation
$CNot\ s\ t = \mathbf{if}^{\circ}\ s\ \mathbf{then}\ (\mathbf{qtrue}, Qnot\ t)$
           $\mathbf{else}\ \ (\mathbf{qfalse}, t)$

$Epr \in Q2 \otimes Q2$      -- The EPR pair, $|00\rangle + |11\rangle$
$Epr = (\mathbf{qtrue}, \mathbf{qtrue}) + (\mathbf{qfalse}, \mathbf{qfalse})$
$Bmeas\ \ \ \in Q2 \multimap Q2 \multimap Q2 \otimes Q2$      -- The Bell-measurement operation
$Bmeas\ x\ y = \mathbf{let}\ (x', y') = CNot\ x\ y$
          $\mathbf{in}\ (Meas\ (Had\ x'), Meas\ y')$
$U\ \ \ \in Q2 \multimap Q2 \otimes Q2 \multimap Q2$      -- The unitary correction operations
$U\ q\ xy = \mathbf{let}\ (x, y) = xy\ \mathbf{in}\ \mathbf{if}\ x\ \mathbf{then}\ \ \ (\mathbf{if}\ y\ \mathbf{then}\ U_{11}\ q\ \mathbf{else}\ U_{10}\ q)$
                             $\mathbf{else}\ \ \ (\mathbf{if}\ y\ \mathbf{then}\ U_{01}\ q\ \mathbf{else}\ q)$

$U_{01}, U_{10}, U_{11} \in Q2 \multimap Q2$
$U_{01}\ x = \mathbf{if}^{\circ}\ x\ \mathbf{then}\ \ \ \mathbf{qfalse}\ \mathbf{else}\ \mathbf{qtrue}$
$U_{10}\ x = \mathbf{if}^{\circ}\ x\ \mathbf{then}\ -\ \mathbf{qtrue}\ \mathbf{else}\ \mathbf{qfalse}$
$U_{11}\ x = \mathbf{if}^{\circ}\ x\ \mathbf{then}\ -\ \mathbf{qfalse}\ \mathbf{else}\ \mathbf{qtrue}$

   -- The quantum teleportation algorithm
$Tele\ \ \ \in Q2 \multimap Q2$
$Tele\ q\ = \mathbf{let}\ (a, b)\ \ \ \ = Epr$      -- $a$ is given to Alice, $b$ to Bob
           $f\ \ \ \ \ \ = Bmeas\ q\ a$    -- Result of Alice's Bell-measurement is classical data
         $\mathbf{in}\ U\ b\ f$      -- Bob applies $U$ to his qubit, using classical data $f$

Figure 59. Teleportation in Grattage's QML.

In April 2007 Danos, Kashefi, and Panangaden introduced their measurement calculus [97]. What is notable about their approach is that they utilize a measurement based computation model as opposed to the traditional circuit model utilized by other approaches. In essence, their measurement calculus provides a mathematical model for expressing programs or "patterns" as the authors call them. Since the model is extremely mathematical, its structure would be very unfamiliar to most commercial programmers.

$$(\mathcal{I}(1) \otimes \langle (3,4) \rangle) \wedge \mathcal{Z}(1,3)(\mathcal{I}(1) \otimes \langle (2,3) \rangle) = X_4^{s_3} M_3^x E_{34} E_{13} X_3^{s_2} M_2^x E_{23}.$$

By standardizing:

$$X_4^{s_3} M_3^x E_{34} \boxed{E_{13} X_3^{s_2}} M_2^x E_{23} \qquad \Rightarrow_{EX}$$

$$X_4^{s_3} Z_1^{s_2} M_3^x \boxed{E_{34} X_3^{s_2}} M_2^x E_{13} E_{23} \qquad \Rightarrow_{EX}$$

$$X_4^{s_3} Z_4^{s_2} Z_1^{s_2} \boxed{M_3^x X_3^{s_2}} M_2^x E_{13} E_{23} E_{34} \Rightarrow_{MX}$$

$$X_4^{s_3} Z_4^{s_2} Z_1^{s_2} \overline{M_3^x M_2^x E} {}_{13} E_{23} E_{34}.$$

Figure 60. The CNOT operation in Danos, Kashefi, and Panangaden's measurement calculus [97].

Selinger has introduced a statically typed, functional language, which the author describes as "quantum data, classical control" [98]. The language Selinger introduces is called "Quantum Flow Charts", or QFC. As the name implies, the language is based largely on flow charts, except that they are functional in nature: that is, they transform specific inputs into matching outputs as opposed to updating a state as in an imperative approach. Based on QFC, Selinger also introduces an actual programming language, QPL, which is largely based on the principles laid out in QFC. While the syntax of QPL looks imperative at a glance, it is still functional. Selinger also proposes an alternate of QPL called "Block QPL". While this is a very detailed and thorough paper, it still suffers from the deficiencies previously outlined– its functional nature is foreign to many practicing developers, and the introduction of a new language introduces a significant hurdle for its use. Selinger has also done some work developing a high order quantum language based on linear typed lambda calculus [99]. It is worth mentioning that Selinger has also written a brief survey of quantum programming languages [100], in which he notes the practicality of Bettelli's C++ extensions.

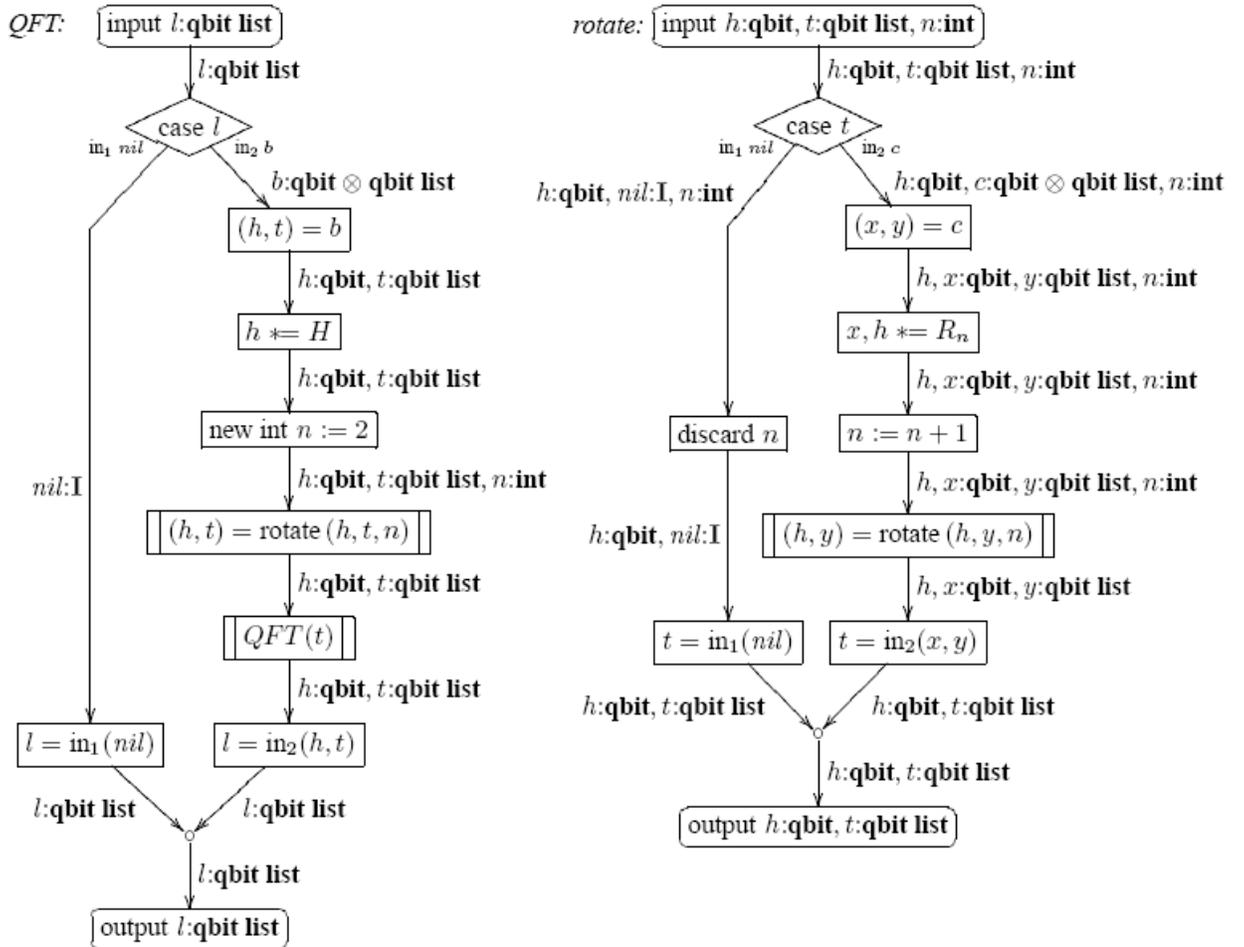

Figure 61. The Quantum Fourier Transform[34] in Selinger's QFC [98].

Tafliovich has developed a method of programming quantum computers based on an extension of probabilistic predicative programming [101, 102]. Hehner, who was Tafliovich's advisor, defines predicative programming as a method in which each step of the program is proven as it is made [103]. This is both the strength and weakness of the programming style. While it results in software that is formally proved, this approach is likely to be too time consuming and foreign for many commercial programmers. For non-critical systems, this cost of formally proving the software is likely to be too high as certain defects do not render the software un-shippable. This follows the philosophy of "Good Enough" software, where the product does not have to be defect free or meet a very high "good enough" standard– simply free of enough and the right kind of defects to be "good enough" [104]. Additionally the cost of formally proving software is impractical for prototyping; where prototyping is defined as quickly creating partially working programs to illustrate certain pieces of the end result. It has also been noted that the most efficient software organizations are those that do not remove all of the defects [5]. The obvious exception to this "good enough" approach is mission or life critical software, which has to go through a more rigorous testing cycle. As with many other methods of programming quantum computers, Tafliovich assumes that the programs will be executed on a classical machine with a quantum subsystem as proposed by Knill [3]– further enforcing this theme across different proposed programming methodologies for quantum computers.

---

[34] According to Hirvensalo on page 58 [11], "[QFT] can be done in time O($m^2$), which is exponentially separate from the classical counterparts of Fourier transform." (For $m$ qubits.)

$$x' = f0 \oplus f1 \wedge t' = t + 1$$

$$= |(((-1)^{f0}/2 + (-1)^{f1}/2) \times |0\rangle + ((-1)^{f0}/2 - (-1)^{f1}/2) \times |1\rangle)\ x'|^2 \times$$

$$(t' = t + 1)$$

$$= |H(U_f(H|0\rangle))\ x'|^2 \times (t' = t + 1)$$

$$= \psi := |0\rangle;\ \psi := H\psi;\ t := t + 1;\ \psi := U_f\psi;\ \psi := H\psi;$$

$$|\psi\ x'|^2 \times (t' = t)$$

$$= \psi := |0\rangle;\ \psi := H\psi;\ t := t + 1;\ \psi := U_f\psi;\ \psi := H\psi;\ \textbf{measure}\ \psi\ x$$

Figure 62. Deutsch's algorithm utilizing Tafliovich's approach [101].

Aggour, Guhde, Simmons, and Simon have developed a method of simulating quantum computation called "Quantum Express" [105]. This is essentially a quantum computer simulator that is written in Java. They state the objective is to test and quantify the performance of quantum algorithms, which is a worthwhile goal. This simulator is worth noting because it takes two XML files as input: a state file and an algorithm file–effectively allowing basic programming via XML. They have also developed a graphical user interface to create these files, meaning that it can be considered programmable by this GUI. This GUI allows for the creation of programs by apparently following the low level quantum circuit model. (Section 4.2.1 details quantum circuits.) Figure 63 shows an example of using the Quantum Express GUI for a 3 qubit register.

There is a very importation observation to make here: They allow the "CCPhaseShift" operator to apply to any 2 qubits, which do not have to be "next" to each other as the middle "CCPhaseShift" on qubits 1 and 3 shows in Figure 63. In the literature the operations are usually specified as applying in a specific order to a register only containing the same number of qubits. As an example a CNot operation applies to a pair of qubits, so the matrix specified operates only on two qubits. The matrix representing the operation is different if one applies it to a pair of qubits in a register of more than two qubits. This is expanded on in section 5.5.5, and there is an unpublished paper on the web site that goes into this in greater detail (see Appendix A).

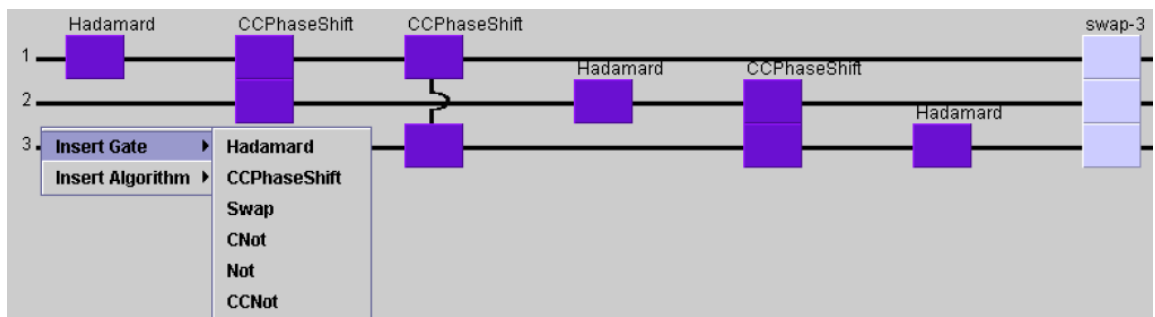

Figure 63. Constructing the 3 qubit Quantum Fourier Transform algorithm using the GUI of Quantum Express, from [105].

CCPhaseShift in Figure 63 is assumed to mean "control-phase shift. However Nielsen and Chuang list the first CCPhaseShift in this diagram (3 qubit QFT) as a control-S, the second a control-T, and third a control-S [13]. The S and T gates are simply phase shifts by different amounts. Thus it is unclear from this diagram how the S and T gates are distinguished.

Most of the paper on Quantum Express discusses the implementation and reasoning, such as using an array of doubles instead of creating a complex number class [105]. Quantum Express can also handle qutrits and higher base quantum units of information. A qutrit is the quantum equivalent of the classical 3 state unit of information, the trit: thus it can be in a super position of $|0\rangle$, $|1\rangle$, and $|2\rangle$ instead of just $|0\rangle$ and $|1\rangle$ as in a qubit. Just as three state units of information are not encountered much in classical information, qutrits do not seem to be encountered much in the quantum computer programming literature.

While programming techniques that are not procedural or object oriented are alien to many commercial programmers, examining them reveals implementation hurdles and a better idea of what might work or not. Furthermore, Knill's QRAM model is a common theme across many programming proposals. Quantum computing itself relates closely to functional programming, so it is not surprising that a variety of the quantum computer programming proposals utilize this approach. Additionally, more mathematical and formal methods have been selectively examined. These techniques are helpful for anyone trying to get a lower level view of quantum computer programming, such as a physicist who might work on the implementation of quantum computers. While the programming of quantum computers may be considered a challenge to computer scientists, the implementation of quantum computers is largely a challenge to physicists.

## 2.3    Framework Design

A software framework, or framework for short, is a group of cooperating classes that form a skeleton to address a specific problem domain[35] [106]. Frameworks typically contain abstract base classes or interfaces. Additionally there are often concrete subclasses of these abstract base classes or interfaces, especially for black box frameworks. These class implementations allow for the framework to be used in the development of working software. The use of the abstract components also means that the implementations can be swapped out with ideally no impact[36] on software that has been developed to use the framework.

This dissertation is about the design and implementation of a quantum framework, named Cove. Cove is considered a framework and not a library since the framework can be extended by users. A framework is also easier to use in a practical sense since a user only has to search through several methods on a class in a framework once they find the appropriate class as opposed several thousand methods in a more traditional API which may only provide a collection of method calls. In light of this, a simulation of a quantum computer could be a part of the present implementation, but this could be switched out with an implementation that runs on actual quantum computers once they become viable. This swapping of implementations is shown in Figure 64. While there is an exponential slow down when classically simulating quantum computation as outlined in section 2.1 it does allow for users to write code that can be executed.

---

[35] That problem domain may be general data structures, .NET, MFC, and Java libraries being examples.
[36] Obviously the interface itself does not specify exact behavior, although some work has been made in this sense as will be detailed. One way behavior can helped to be enforced is by testing. As an example: if I push an item onto a stack with $n$ elements then there should be $n + 1$ elements, with the added element being at the top of the stack.

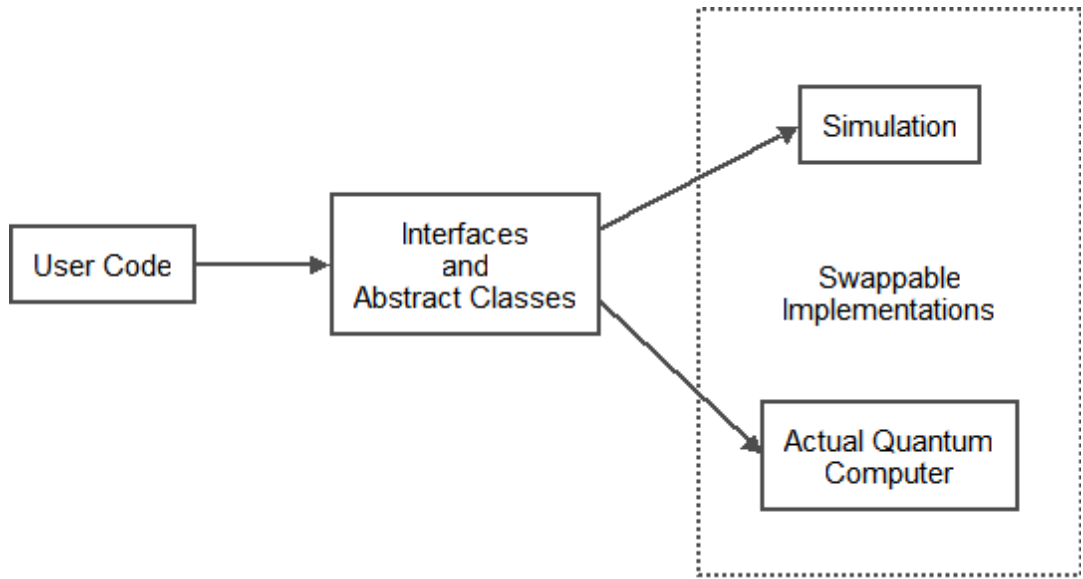

Figure 64. Swappable implementations when users write against interfaces and abstract classes.

To define frameworks another way, frameworks are partially complete solutions to a set of problems [6] and therefore allow for the reuse of both design and code [107]. There is a fundamental difference between traditional libraries and frameworks[37]: libraries are meant to be utilized by the application while frameworks provide important parts of the application architecture [6]. Design of programming frameworks is generally considered to be harder than application development, primarily due to the fact that frameworks need to be more flexible [108]. For this reason, it is important to examine the literature and established techniques before carrying out a project in framework design. In short, "framework design is all about developing the right abstractions" [108], although modeling behavior plays a role as well. Consequently designing frameworks is a balancing act between providing simplicity and power [109].

---

[37] Although this is not a clear division between the two; there is some grey area. Frameworks can just be collections of libraries in some cases.

The first framework to see widespread use was the Smalltalk-80 user interface framework called Model/View/Controller (MVC), and was developed in the late 1970s [109]. Some other frameworks that have seen extensive use include MacApp, zApp, Microsoft Foundation Classes (MFC), and the Microsoft .NET Framework (as of 2009 at version 3.5). Regardless of the framework, complete and accurate documentation has proven critical to the adoption of a framework [110, 111]. In the same spirit the classes, methods, and parameters must also have descriptive names so that users can easily deduce their purpose. Abbreviations and nondescript names mean that the programmer has to stop programming and reference the documentation frequently.

Exceptions are commonly used within frameworks, as they are considered to be a more structured way of error handling [57, 112]. Implementation of languages typically provide an exception base class, which all other exceptions are derived from– C# being a good example. This allows for specific exceptions to be caught and dealt with. Alternately more general (higher up the inheritance tree) can be caught to allow for generic error handling. Furthermore, there can be a base exception class that frameworks can derive their own framework specific exceptions from, allowing for framework specific exceptions to be caught and handled. Exceptions also allow for libraries or programs to easily handle errors not thrown by their code [57]. Thus exception handlers can be used to handle a variety of errors if needed. Exceptions are also typically thrown at or close to the source of the error[38].

---

[38] It is generally considered good practice to throw the exception as close to the source of the error as possible to aid in debugging.

Johnson has made contributions to framework design [107]. His 1997 paper "Components, Frameworks, and Patterns" [109] makes the distinction between the three and provides a brief history of frameworks up until 1997. One crucial thing he points out in this paper is that the most important part of a framework is typically its abstract classes. It is the abstract classes that lay out the design for how all the pieces fit together. Frameworks also typically come with component libraries, which are libraries that provide usable implementations of the abstract ones. Johnson also points out that one benefit of frameworks is that they let users create new components from existing ones, such as a user interface out of widgets [109]. Furthermore, frameworks also provide templates on how to create new components. He also advocates iteration in the framework design process, as do many others.

Two types of frameworks are commonly encountered: white-box frameworks and black-box frameworks. The difference in the two is primarily whether or not the user of the framework needs to be familiar with the internal workings of the framework. A white-box framework is typically used by deriving classes from the framework, and the user needs to be aware of the workings of the base classes. A black-box framework is one in which a user typically instantiates framework classes and sends them messages [113]. Microsoft's .NET framework could generally be considered a black-box framework because users typically instantiate its classes– therefore they do not need to be concerned with the inner workings of the framework. Using black-box frameworks typically results in software more loosely coupled to the framework when compared to using white-box frameworks. Loose coupling is considered to be a good trait in software [106]. While black-box frameworks are more usefully coupled, white-box frameworks can be more powerful and thus potentially more promising.

Within frameworks there is the concept of frozen spots and hot spots [108]. Frozen spots are those components of the framework that are fixed. They are called frozen spots because they are not changed by users of the framework. An example of a frozen spot would be a static method, a method on a class that can be called without instantiating that class. Another example of a frozen spot would be a sealed class– a class that cannot be derived from or changed [114]. Hot spots are those that vary in a framework; pure virtual methods are an example. Thus there lies a distinction between prepackaged parts of the framework (frozen spots) and those where the implementation may change (hot spots). Obviously when designing a framework special care needs to be taken in order not to place components in the wrong category. An example of a correctly classified hot spot might be a method to get the area of a shape- the implementations (triangles, circles, squares, and so on) define how to get the area.

Roberts and Johnson have established a series of patterns that they intend to be applied in a particular order in order to develop frameworks [115]. They also expect the process to be carried out in an iterative fashion. Essentially what they propose is to come up with several examples of how the framework will be used, and then derive a black-box framework through a series of steps. Their purpose in starting with examples is because they believe people generally develop abstractions, and thus frameworks, from concrete examples. They have developed a complete process for designing frameworks as Figure 65 illustrates. In this figure the arrows represent the flow of the framework development, starting with the three examples. However the later steps, the visual builder and language tools, are not applicable to the framework in this dissertation since the framework is not necessarily a complete commercial grade framework.

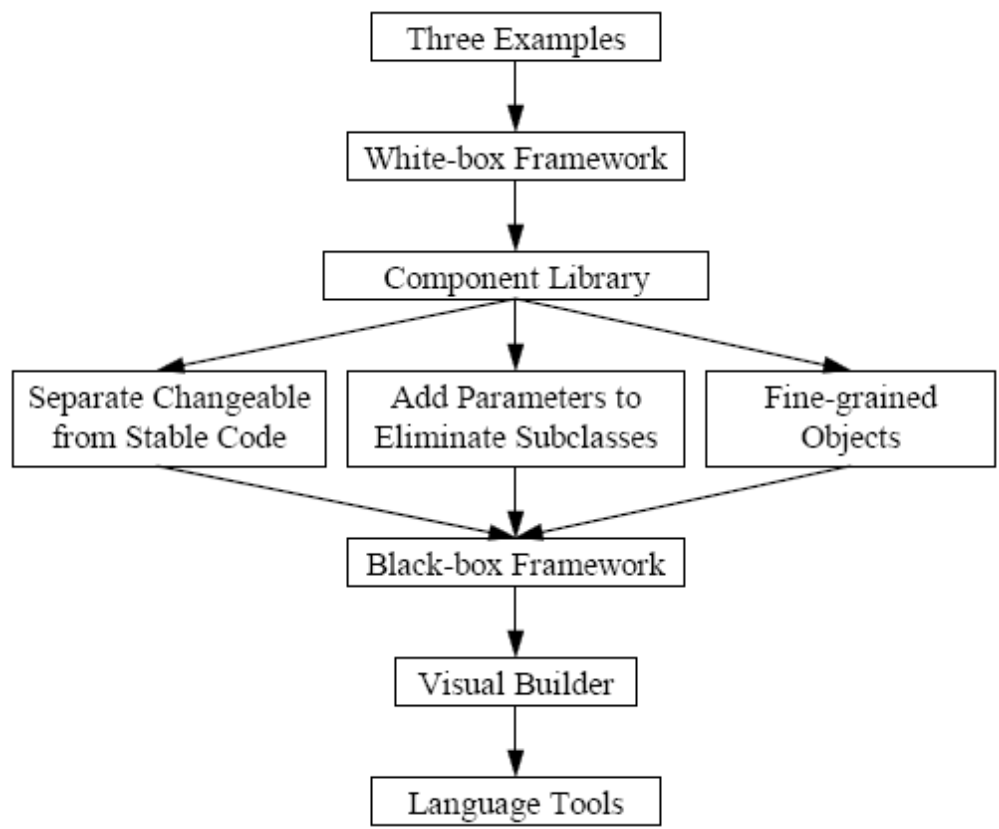

Figure 65. Roberts and Johnson's method to develop a framework, as illustrated in [115].

Throughout their paper Roberts and Johnson make several important points [115]:

- They advocate the use of inheritance[39] within the framework and for white box frameworks.

- Frameworks that are usable out of the box are much easier to use.

- Separate changeable code from stable code (hot spots and frozen spots).

---

[39] Cove uses single inheritance (by the nature of C#) but classes implement multiple interfaces.

- More objects increase the complexity of the framework, so there is a fine line between creating trivial subclasses and making methods take additional parameters.

Imaz and Benyon advocate a similar approach to Roberts and Johnson. Imaz and Benyon encourage using what they call "blends", in simple terms taking multiple input spaces and mapping them onto a more generic space which is the blend [116]. The purpose of the blend is to get a new structure not provided by the input spaces. When their approach is applied to framework design, examples are first gathered. These examples are then used to establish the classes and various entities, which are continuously refined as analysis continues [116].

Hou and Hoover have developed a method for formally specifying the constraints of a framework [117]. Their method allows for the constraints of a framework to be specified and violations identified. When violations are identified they also show the relevant documentation on why the constraint is needed. Within their paper [117] they also give an example of how their method, FCL, can be used to specify constraints in the Microsoft Foundation Classes (MFC), the precursor to the .NET framework. Figure 66 shows example classes that derive from ATM (called specialATM) that can either call debit on ATM or on Account, but not both (xor)– effectively keeping the two from both being debited on one call to the debit method.

```
specialATM : subclass(ATM)
//forall specialATM
{
    method debit(Account &o)
    =>
    {
      calls(ATM::debit(o)) xor calls(o.debit())
    }
}
```

Figure 66. An example of Hou and Hoover's FCL.

Constraints of a framework are often informal and based on the designer's intentions on how it should be used. Further complicating matters can be the lack of documentation of the constraints and intentions. An example would be an abstract vehicle class provided by a framework. A constraint intended by the framework designers would be to decrease the speed of the vehicle when braking occurs. This constraint is not enforced, so a user can misuse the vehicle class and derive a class where the vehicle's speed increases when braking occurs. FCL provides a formal way to specify and enforce constraints such as this. While FCL does not see much use, the fact that it exists is important. It illustrates through concrete examples that the constraints of the framework should be documented at a bare minimum, and better, enforced as much as possible. However, a framework designer must be careful not to make the constraints too limiting so that they unnecessarily reduce the applications the framework can be used for, as there may be uses unanticipated by the designer of the framework.

While not specific to framework design, Bruce Eckel has laid out some programming guidelines in his popular Java text <u>Thinking In Java</u>. These selected guidelines [118] from his text also apply to frameworks:

- Separate the class creator from the class user. That is, the user of a class should not be concerned with how it works[40].

- Make classes as atomic as possible. Atomic is used in the sense that classes should exhibit high cohesion: handle a few things and not become bloated "mega" classes.

- Watch for long argument lists.

- Use exception hierarchies.

- Keep things as private as possible. (As in the access levels of private, protected, and public.)

- Avoid using magic numbers; constants should be used instead.

- Choose interfaces over abstract classes. This is especially important when multiple inheritance is not allowed in the target language.

- In constructors, do only what is necessary to set the object into the proper state.

- Remember that code is read much more than it is written.

Krzysztof Cwalina and Brad Abrams have outlined several qualities of well designed frameworks in [119]. While their book focuses on designing frameworks based on .NET, these guidelines are general and apply to any framework. These qualities of a well designed framework are laid out in the first chapter as follows:

1. Is simple

2. Is expensive to design, as it takes a lot of care to design well

3.  Is full of trade-offs

4. Borrows from the past

5. Is designed to evolve

6. Is integrated

---

[40] An exception would be when one may need to know the basics of how it works for performance or reliability considerations.

7.  Is consistent

Abi-Antoun has discussed his experience in the design of a large framework [110]. Most of his discussion focuses on challenges an organization faces when developing a framework. There are several more general points that seem logical to follow in framework design such as in this proposal:

- There should be a small number of interfaces that are used consistently throughout the framework.

- Design for future changes wherever possible

- The framework should be developed iteratively, including liberal refactorization where deemed appropriate.

- Prove by demonstration. This will be extensively used in this dissertation.

Most of the literature on frameworks does not concentrate on the design, but rather how to document it well [6]. While the literature is sparse on framework design, there are some various themes that run common throughout the literature:

- Repeat patterns of engagement [6].

- Develop the framework through multiple iterations.

- Design the framework with future changes in mind.

Frameworks are an important topic in software engineering, in which the literature is surprisingly light. It is worth examining in the context of this project since the goal is to develop a framework.

CHAPTER III

HYPOTHESIS AND PROOF CRITERIA

This chapter details the hypothesis and proof criteria of this dissertation. The hypothesis is presented in section 3.1, along with details further explaining and elaborating on it. Section 3.2 presents the list of proof criteria. These are the criteria that will be used to judge that the research is complete.

## 3.1 Hypothesis

"A practical framework for quantum computing can be designed for existing modern object oriented classical languages that gives conventional programmers a toolkit to carry out quantum computation."

With the exception of developments such as quantum computing, all software has been written to run on classical[41] Turing machines. Therefore it should not be surprising that nearly all software has been written utilizing classical languages. Much in the same way that classical computers may be extended to utilize quantum computers at the hardware level in accordance with Knill's QRAM model [3], the software programming techniques should also follow this approach of extending what is already present.

At first quantum computation may seem like a trivial programming challenge given that only three things need to be done: initialization, manipulation through unitary operations, and measurement. Nonetheless the numerous flaws in existing programming proposals in chapter 2 show that this isn't so. With further consideration perhaps this isn't too surprising: after all one could view classical programming techniques as little more than the manipulation of bits; to program quantum computers programming techniques must be expanded to manipulate qubits.

The list of commonly encountered flaws in the introduction of this dissertation suggests that a list of usability criteria should be established for quantum computing, although it is admitted that parts of this list are somewhat subjective. Additionally no list of functional properties for a quantum computer programming has been encountered in the literature. Without this functional list, it is possible that proposed programming methods may be incomplete– something that may easily be overlooked without close inspection.

While a bit early in this dissertation since the framework itself is not being covered yet, some mention of how the framework is partitioned is necessary in order to have a clearer picture for the discussions that follow. The framework is broken into two distinct parts: the base library and any number of implementations. The base library defines the minimum functionality each implementation must support through a collection of interfaces. Hence the base library defines what classes and calls must be supported and the implementations provide how to carry them out. Not only does this draw a clear line between what and how, but if users program to the interfaces of the base library then implementations can be swapped out with little work as will be detailed later.

---

[41] Throughout this dissertation the term "classical" is used to distinguish existing computers that operate on bits from quantum computers that operate on qubits.

This research seeks to combine quantum computer programming and usable framework design in order to create a usable quantum computer programming framework. Hence, the goal is a prototype[42] of a practical framework that is suitable for use in a commercial software development environment much in the same way that there has been a focus in making parallel programming practical.

## 3.2    Proof Criteria

The proof criteria in this section have been established to judge the completeness of this research work. The proof criteria are functional properties that Cove must satisfy. The non-functional requirements, usability in particular, are too subjective to be included in the proof criteria, but are detailed in 5.6. If these functional properties are not met then Cove does not provide a complete tool kit for quantum computation. These functional properties specify *what* the framework must be able to do, while the usability properties (section 5.6) specify *how* the framework should accomplish them.

Instead of later specifying how each of these properties is satisfied, which is extremely repetitive and detailed, the framework will be judged to be complete for the purpose of this dissertation when the pieces needed to carry out the examples are implemented. The three examples for this will be the quantum coin toss (described in section 2.1.2), entanglement (described in 2.1.4), and factoring (described in 2.1.11). The first two examples are simple and illustrate that the framework can do a few basic quantum computing tasks. The factoring example is representative of a complex real world task that a quantum computer would carry out.

---

[42] As of May 2009 the framework consists of a few dozen interfaces and classes.

There are a wide variety of functional properties that techniques for programming quantum computers must have. Some of these apply to the specific implementations (such as a simulation), others to the more general interfaces. What follows is a list of those properties, under general categories.

### 3.2.1 Quantum Computing Components

The following components are required to carry out quantum computation:

- Quantum data types: qubits and quantum registers[43], which are collections of qubits.

- (Unitary) Operations – Unitary Operations that manipulate qubit(s). Requiring operations to be unitary implies that the computation is reversible [12]. Irreversible operations require energy to erase information due to Landauer's principle, and this input of energy[44] would act as a measurement and collapse the system [13].

- Classical conversion – To allow qubit(s) to be initialized based on classical types[45] such as an unsigned integer. Measurement also collapses qubit(s) back to states that can be represented classically. Like any classical data type, these initializations and measurement results are merely a series of bits, and those bits need to be easily converted to types more commonly used in high level languages.

The correctness of the above will be shown via automated testing[46]. Additionally one must allow for the fact that the quantum resources may be remote in accordance with the QRAM model.

---

[43] The standard within existing programming techniques is that registers are initialized to $|0...0\rangle$. While not a requirement, this seems to be standard practice, much like how an integer is typically initialized as 0 in a classical language.

[44] Put another way, energy is expelled from the system to the environment. The key point is that the system must interact with the outside environment, causing a measurement.

[45] A register cannot be initialized to an arbitrary quantum state because we could then be copying the state of the initialization register to the new register – which would be a violation of the No-cloning theorem.

[46] While testing does not *prove* correctness, it certainly shows correctness under some cases and is better than no testing at all.

*3.2.2   Limitations of Quantum Computing*

The limitations of quantum computing need to be enforced by any quantum programming technique. Specifically, the following limitations need to be enforced:

1. Unlike classical information, qubits cannot be copied [13, 27]. This is due to what is known as the no-cloning theorem [13].
2. Measurements cannot be undone, as they collapse the system to a specific state.
3. Operations must be reversible.

*3.2.3   Limitations of the Local Simulation*

In addition to limitations for general quantum computing, there will also be some limitations for the local simulation. The local simulation is a prototype implementation of the interfaces in Cove that runs on the local computer. These limitations of the local simulation include:

1. Since the provided simulation will be a local classical simulation, the location of the qubit cannot be changed to something other than the localhost. This is not a limitation of quantum computing, but a limitation of the simulation.
2. The user should not be able to peek at the state of a qubit[47] through methods in the interfaces.
3. A hard limit on the number of qubits allowed in the register. Due to the exponential increase in storage size there will be some limit in the size of simulated registers due to factors such as maximum array size. If there is a hard limit then more graceful errors can be given to the user.

*3.2.4   Quantum Operations*

There are several common operations, or gates, that need be applied to quantum registers and any programming method must support a universal set at minimum[48] [49]. These gates can be written as two by two matrices of complex numbers for single qubits, as shown in Figure 67. In general an operation on $n$ qubits is represented by a $2^n$ x $2^n$ matrix [99][50]. Thus it follows that operations on two qubits are represented by a four by four matrix.

---

[47] However peeking will be allowed for unit testing to verify the correctness of the simulation.
[48] The operations are included in the proof criteria since quantum programming methods must support them.
[49] A universal set allows for any other operations to be constructed in the same way classical NAND gates can be used to build up operations on a classical computer.

$$|0\rangle \mapsto a|0\rangle + b|1\rangle$$
$$|1\rangle \mapsto c|0\rangle + d|1\rangle$$

$$\begin{bmatrix} a & c \\ b & d \end{bmatrix}$$

Figure 67. Abstract representation of an operation that operates on a single qubit, based on [11].

Put another way, an operation on a single qubit changes the state as shown in Figure 68. The operation (the 2 x 2 matrix) can be anything, as long as it is unitary. All of the numbers in the figure are complex numbers. The same principle applies as we scale up to greater number of qubits: the operation must transform the input into the desired output.

$$\begin{bmatrix} a & b \\ c & d \end{bmatrix} \begin{bmatrix} x \\ y \end{bmatrix} = \begin{bmatrix} (ax)+(by) \\ (cx)+(dy) \end{bmatrix}$$

Initial state: $x|0\rangle + y|1\rangle$

After operation: $((ax)+(by))|0\rangle + ((cx)+(dy))|1\rangle$

Figure 68. How the state of a single qubit is altered by an operation.

---

[50] The local simulation has a limit of 62 qubits meaning that a $2^{62}$ x $2^{62}$ matrix would be required for an operation!

The ordering of the qubits in all the operations is arbitrary and the matrix representation listed is the commonly encountered one. This is something to take into account: users may wish to alter the conventional ordering of qubits in operations.

One fact that allows the following definitions of common operators to be expressed as complex number in the form $a + bi$ is that $e^{\varphi + i\theta} = e^{\varphi}\cos\theta + ie^{\varphi}\sin\theta$ and $e^{i\theta} = \cos\theta + i\sin\theta$. When visualized on the Bloch sphere many of these single qubit operations (Table 5) are nothing more than rotations about the X, Y, or Z axis as shown in Figure 69:

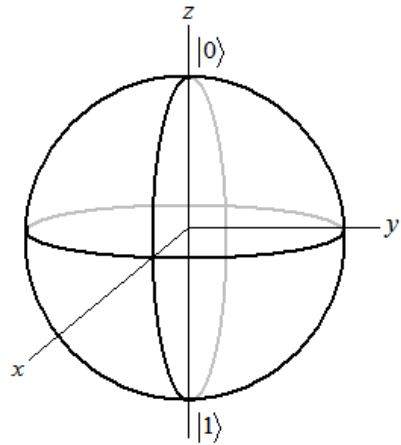

Figure 69. Bloch sphere.

Table 5. Common operations on a single qubit.

| Name | Matrix representation | Quantum circuit representation[51] |
|---|---|---|
| Hadamard (H, Hadamard-Walsh, or square root of not) [10, 12, 13, 27, 28, 32, 55, 120] | $\frac{1}{\sqrt{2}}\begin{bmatrix} 1 & 1 \\ 1 & -1 \end{bmatrix} = \begin{bmatrix} \frac{1}{\sqrt{2}} & \frac{1}{\sqrt{2}} \\ \frac{1}{\sqrt{2}} & -\frac{1}{\sqrt{2}} \end{bmatrix}$ | $\lvert q_0 \rangle$ — $\boxed{H}$ — |
| Not (bit flip, X, or Pauli X gate) [11-13, 27, 28, 120] | $\begin{bmatrix} 0 & 1 \\ 1 & 0 \end{bmatrix}$ | $\lvert q_0 \rangle$ — $\boxed{X}$ — |
| Y (Pauli Y gate) [12, 13, 27, 120] | $\begin{bmatrix} 0 & -i \\ i & 0 \end{bmatrix}$ | $\lvert q_0 \rangle$ — $\boxed{Y}$ — |
| Z (phase flip or Pauli Z gate) [11-13, 27, 28, 120] Leaves $\lvert 0 \rangle$ unchanged and flips the sign of $\lvert 1 \rangle$ | $\begin{bmatrix} 1 & 0 \\ 0 & -1 \end{bmatrix}$ | $\lvert q_0 \rangle$ — $\boxed{Z}$ — |
| S (phase gate) [13, 120] (Also referred to as K in mathematical texts such as [55]) | $\begin{bmatrix} 1 & 0 \\ 0 & i \end{bmatrix}$ | $\lvert q_0 \rangle$ — $\boxed{S}$ — |
| T ($\frac{\pi}{8}$ phase gate) [12, 13, 28, 120] | $\begin{bmatrix} 1 & 0 \\ 0 & e^{i\frac{\pi}{4}} \end{bmatrix}$ | $\lvert q_0 \rangle$ — $\boxed{T}$ — |
| $R_k$ (Used in the quantum Fourier transform) [13, 21] | $\begin{bmatrix} 1 & 0 \\ 0 & e^{2\pi i/2^k} \end{bmatrix}$ | $\lvert q_0 \rangle$ — $\boxed{R_k}$ — |
| Arbitrary real rotation [95] | $\begin{bmatrix} \cos(\theta) & -\sin(\theta) \\ \sin(\theta) & \cos(\theta) \end{bmatrix}$ | N/A |

---

[51] Quantum circuits are covered in 4.3.

| $R_x(\theta)$ (Arbitrary rotation about the x axis) [13, 42] | $\begin{bmatrix} \cos(\frac{\theta}{2}) & -i\sin(\frac{\theta}{2}) \\ -i\sin(\frac{\theta}{2}) & \cos(\frac{\theta}{2}) \end{bmatrix}$ | 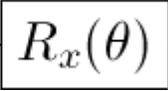 (Not established in the literature) |
|---|---|---|
| $R_y(\theta)$ (Arbitrary rotation about the y axis) [13, 42] | $\begin{bmatrix} \cos(\frac{\theta}{2}) & -\sin(\frac{\theta}{2}) \\ \sin(\frac{\theta}{2}) & \cos(\frac{\theta}{2}) \end{bmatrix}$ | 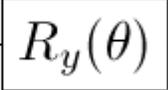 (Not established in the literature) |
| $R_z(\theta)$ (Arbitrary rotation about the z axis) [13, 42] | $\begin{bmatrix} e^{-i\theta/2} & 0 \\ 0 & e^{i\theta/2} \end{bmatrix}$ | 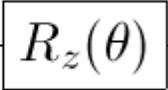 (Not established in the literature) |
| Identity [12, 13, 27] | $\begin{bmatrix} 1 & 0 \\ 0 & 1 \end{bmatrix}$ | 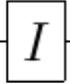 |
| Measurement [13][52] (on a single qubit) | N/A | 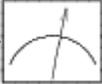 |

Table 6. Common operations on two qubits.

| Name | Matrix representation | Quantum circuit representation |
|---|---|---|
| Controlled Not (CNot) [10-13, 28, 99, 120] If control qubit is set to $|0\rangle$ then target qubit is left alone; else the target is flipped. | $\begin{bmatrix} 1 & 0 & 0 & 0 \\ 0 & 1 & 0 & 0 \\ 0 & 0 & 0 & 1 \\ 0 & 0 & 1 & 0 \end{bmatrix}$ | 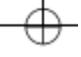 |
| Swap [82, 120] | $\begin{bmatrix} 1 & 0 & 0 & 0 \\ 0 & 0 & 1 & 0 \\ 0 & 1 & 0 & 0 \\ 0 & 0 & 0 & 1 \end{bmatrix}$ | 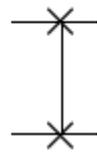 |

[52] The double wires after the measurement gate denote that the information is classical at that point. A single wire represents quantum information.

For the CNot listed in Table 6 the first qubit is the control and the second qubit is the target. Since the ordering is arbitrary, reversing the control and target is possible [27]. When reversed it is still a unitary transformation, thus still meeting the requirement of reversibility that all quantum operations must obey. Figure 70 is the reversed CNot operation when the first qubit is the target and the second is control [27],

$$\begin{bmatrix} 1 & 0 & 0 & 0 \\ 0 & 0 & 0 & 1 \\ 0 & 0 & 1 & 0 \\ 0 & 1 & 0 & 0 \end{bmatrix}$$

Figure 70. Matrix form of the reversed CNOT operation.

Table 7. Common operations on three qubits.

| Name | Matrix representation | Quantum circuit representation |
|---|---|---|
| Toffoli (double controlled not or controlled controlled not) [10, 13, 27, 28, 120]<br><br>If the two control qubits are $\lvert 1 \rangle$ then the single target qubit is flipped, else nothing happens.[53] | $\begin{bmatrix} 1 & 0 & 0 & 0 & 0 & 0 & 0 & 0 \\ 0 & 1 & 0 & 0 & 0 & 0 & 0 & 0 \\ 0 & 0 & 1 & 0 & 0 & 0 & 0 & 0 \\ 0 & 0 & 0 & 1 & 0 & 0 & 0 & 0 \\ 0 & 0 & 0 & 0 & 1 & 0 & 0 & 0 \\ 0 & 0 & 0 & 0 & 0 & 1 & 0 & 0 \\ 0 & 0 & 0 & 0 & 0 & 0 & 0 & 1 \\ 0 & 0 & 0 & 0 & 0 & 0 & 1 & 0 \end{bmatrix}$<br><br>$= \begin{bmatrix} I_2 & . & . & 0 \\ . & I_2 & . & . \\ . & . & I_2 & . \\ 0 & . & . & Not \end{bmatrix}$ | 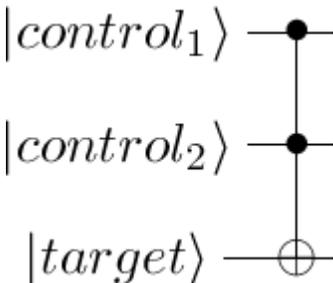 |
| Fredkin (controlled swap) [13, 27, 120]<br><br>The number of $\lvert 1 \rangle$s in the input matches the number of ones in the output [13]. | $\begin{bmatrix} 1 & 0 & 0 & 0 & 0 & 0 & 0 & 0 \\ 0 & 1 & 0 & 0 & 0 & 0 & 0 & 0 \\ 0 & 0 & 1 & 0 & 0 & 0 & 0 & 0 \\ 0 & 0 & 0 & 1 & 0 & 0 & 0 & 0 \\ 0 & 0 & 0 & 0 & 1 & 0 & 0 & 0 \\ 0 & 0 & 0 & 0 & 0 & 0 & 1 & 0 \\ 0 & 0 & 0 & 0 & 0 & 1 & 0 & 0 \\ 0 & 0 & 0 & 0 & 0 & 0 & 0 & 1 \end{bmatrix}$ | 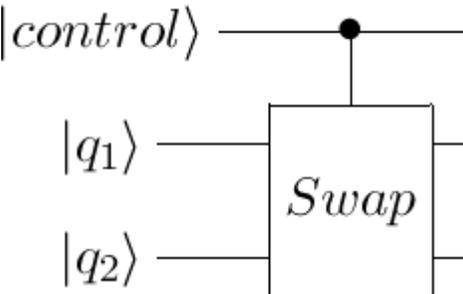 |

---

[53] $I_2$ is the 2 x 2 identity matrix and Not is the 2 x 2 Pauli X gate matrix (Not).

Table 8. Using the Fredkin gate to produce classical operations.

| Name | Inputs and Outputs on Fredkin Gate | | |
|------|------|------|------|
| And [13] | | Inputs | Outputs |
| | a | 0 | xy |
| | b | y | !xy |
| | control | x | x |
| Not or Fanout [13] | | Inputs | Outputs |
| | a | 1 | !x |
| | b | 0 | X |
| | control | x | X |
| Crossover (or swap) [13] | | Inputs | Outputs |
| | a | x | Y |
| | b | y | X |
| | control | 1 | 1 |

The operations in Table 8 are just a sampling of how any classical operation can be made reversible by adding extra (qu)bits.

Table 9. Common operations on *n* qubits.

| Name | Matrix representation | Quantum circuit representation |
|---|---|---|
| Walsh Transformation [23] (Applies the Hadamard operation over *n* qubits: tensor Hadamard *n* times) | $\begin{bmatrix} \frac{1}{\sqrt{2}} & \frac{1}{\sqrt{2}} \\ \frac{1}{\sqrt{2}} & -\frac{1}{\sqrt{2}} \end{bmatrix} \otimes ... \otimes \begin{bmatrix} \frac{1}{\sqrt{2}} & \frac{1}{\sqrt{2}} \\ \frac{1}{\sqrt{2}} & -\frac{1}{\sqrt{2}} \end{bmatrix}$ | 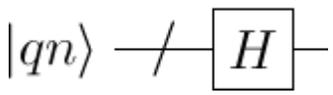 |
| Controlled-U with 1 control qubit and *n*-1 target qubits[54] [10, 13]<br><br>If the single control qubit is $\lvert 0 \rangle$ then nothing happens. If the single control qubit is $\lvert 1 \rangle$ then the single qubit gate U is applied to each of the *n*-1 target qubits. | $\begin{bmatrix} I_2 & 0 \\ 0 & U \end{bmatrix}$ | 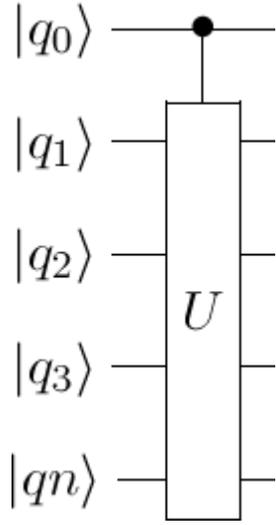 |
| Controled-U, with U = Not (X gate) Over 2 qubits, so this is the controlled not (CNot) operation | $\begin{bmatrix} 1 & 0 & 0 & 0 \\ 0 & 1 & 0 & 0 \\ 0 & 0 & 0 & 1 \\ 0 & 0 & 1 & 0 \end{bmatrix}$ | 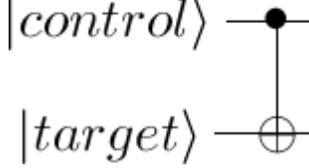 |

---

[54] I(*n*) denotes identity over *n* qubits, so in this case it represents identity over a single qubit which is defined in Table 5. (Identity of 1 is a 2x2 matrix since a single qubit is represented by a matrix of two elements.)

| Controlled-U with *n*-1 control qubits and 1 target qubit [23]<br><br>If all the *n*-1 control qubits are $\lvert 1 \rangle$ then the operation U happens on the target qubit. | Varies with operation | 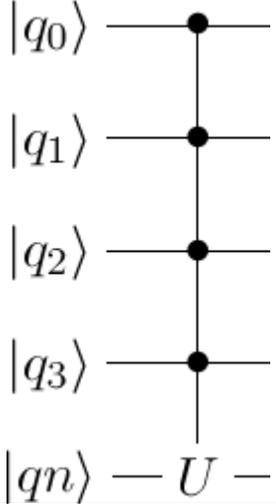 |
| --- | --- | --- |
| Quantum Fourier Transformation (QFT) [10]<br>$\omega$ is the $(2^n)^{th}$ root of unity, $\exp(2\pi i/2^n)$ | $\frac{1}{\sqrt{2^n}}\begin{bmatrix} \omega^1 & \omega^2 & \omega^3 & \omega^4 & ... \\ \omega^1 & \omega^2 & \omega^3 & \omega^4 & ... \\ \omega^1 & \omega^2 & \omega^3 & \omega^4 & ... \\ \omega^1 & \omega^2 & \omega^3 & \omega^4 & ... \\ . & . & . & . & ... \end{bmatrix}$ | 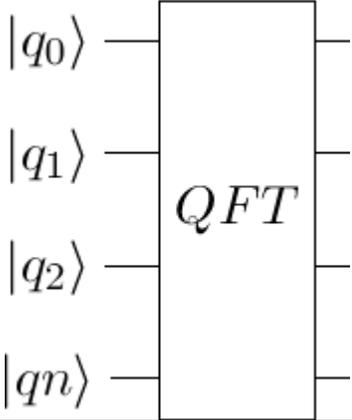 |
| Inverse Quantum Fourier Transform (QFT$^\dagger$)[55] [42] | Inverse of QFT | 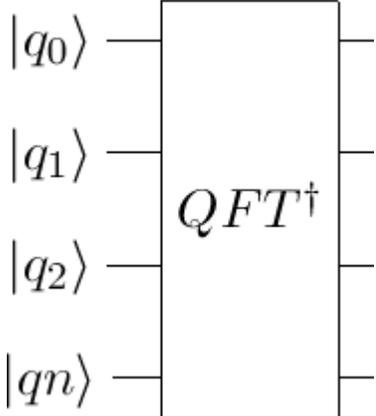 |
| Reverse order of qubits.<br>[23, 27] | Swap over the desired set | 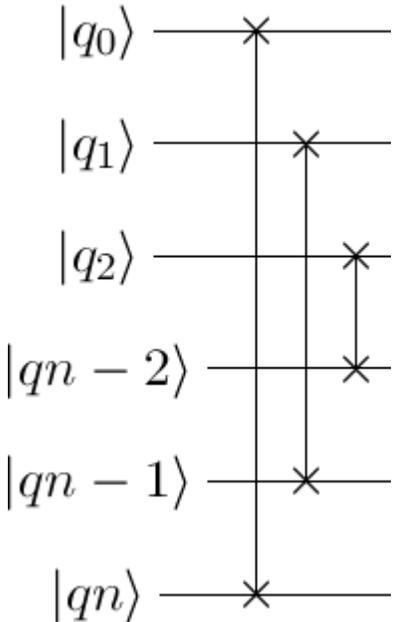 |
| Identity on *n* qubits<br>[24] | $\begin{bmatrix} 1 & 0 & . & . & 0 & 0 & 0 \\ 0 & 1 & & & & 0 & 0 \\ . & & . & & & & 0 \\ . & & & . & & & . \\ 0 & & & & . & & . \\ 0 & 0 & & & & 1 & 0 \\ 0 & 0 & 0. & . & . & 0 & 1 \end{bmatrix}$ | 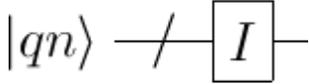 |

Much like classical computation, there are sets of universal gates. Universal is defined as any set of gates that all other gates can be constructed from. For quantum computation the gates that form a universal set are CNot, Hadamard, and T [12].

### 3.2.5   Quantum States

In addition to common quantum operations, there are also several common register states that are detailed within the literature as shown in Table 10. Unfortunately the abbreviated names of the Bell states seems to vary depending on the resource consulted. Since a framework for quantum computing should provide common operations, it should also provide the ability to easily create registers of these states.

Table 10. Common states of quantum registers.

| Name | State |
|------|-------|
| EPR pair (Einstein, Podolsky, Rosen) [31]<br>$\left\| \beta_{00} \right\rangle$ [13]<br>$\left\| \Phi^+ \right\rangle$ (Phi plus) [23]<br>$\left\| \chi \right\rangle$ (chi) [95]<br>$\left\| \psi_B \right\rangle$ [121]<br>(Bell state) [11, 13, 23] | $\frac{1}{\sqrt{2}} \left\| 00 \right\rangle + \left\| 11 \right\rangle$ |
| $\left\| \beta_{10} \right\rangle$ [13]<br>$\left\| \Phi^- \right\rangle$ (Phi minus) [23]<br>(Bell state) | $\frac{1}{\sqrt{2}} \left\| 00 \right\rangle - \left\| 11 \right\rangle$ |
| $\left\| \beta_{01} \right\rangle$ [13]<br>$\left\| \Psi^+ \right\rangle$ (Psi plus) [23]<br>(Bell state) | $\frac{1}{\sqrt{2}} \left\| 01 \right\rangle + \left\| 10 \right\rangle$ |
| $\left\| \beta_{11} \right\rangle$ [13]<br>$\left\| \Psi^- \right\rangle$ (Psi minus) [23]<br>(Bell state) | $\frac{1}{\sqrt{2}} \left\| 01 \right\rangle - \left\| 10 \right\rangle$ |
| Greenberger-Horne-Zeilinger state (GHZ state) [23] | $\frac{1}{\sqrt{2}} \left\| 000 \right\rangle + \left\| 111 \right\rangle$ |
| W state [23] | $\frac{1}{\sqrt{3}} \left\| 100 \right\rangle + \left\| 010 \right\rangle + \left\| 001 \right\rangle$ |

### 3.2.6   Measurement of Quantum Data Types

It is not often emphasized in the literature [27], but measurement plays an important role in quantum computation. First of all the act of measurement can be used for state preparation– that is, to place qubits in an initial state [27]. The second, more obvious, role of measurement is that it is used to extract classical information from qubits. The act of measurement collapses a qubit into $|0\rangle$ or $|1\rangle$. This act of measurement cannot be undone, so it isn't possible to "peek" at a qubit.

Consequently, measurement results in classical data. So there is a need to convert this classical result into classical types in the target language. The measurement concept can also be expanded to include the initialization of qubit(s) to particular states. As an example a register of 8 qubits may be initialized to or measured as the bits 10100110 (0xA6). This equates to the unsigned integer of 166 (decimal). Hence an unsigned integer could, and often is in similar programming proposals, used as the classical input and output of a quantum computation.

### 3.2.7   Potential Measurement Pitfall in Simulations: Entanglement

With a simulation of a quantum system[56] there is a chance for the functional requirement of measurement to be violated when qubits are entangled. As stated previously, when multiple qubits are entangled the measurement of one affects the others. In an actual quantum computer this would be enforced by the laws of nature. In a simulation of a quantum computer there is the possibility that this would not be preserved.  A simple example of a violation would be "peeking" at all of the probability amplitudes in a simulation, since a simulation has to keep track of them all. Violations of the limitations of quantum computing by users should not be allowed and should be handled within the implementation, which in this example would be the simulation[57]. Thus anything that violates what can be done on an actual quantum computer should be prohibited for end users of the framework[58].

A simple case is the entanglement of two qubits into the state $\frac{1}{\sqrt{2}}\left(|00\rangle+|11\rangle\right)$, which is an EPR pair [12]. With this EPR pair, observing one of the two qubits results in it collapsing to 0 or 1, with equal probability. However once the first qubit has collapsed the second will have also collapsed to the same value, meaning that the qubits cannot have different values [11]. This is fairly straightforward if the two qubits are handled as a pair by the framework. For ease of use quantum programming techniques should allow for subsets of qubits to be worked with, as classical techniques allow users to work with subsets of memory. With this entanglement example if a subset of one of the entangled qubits is collapsed, this collapse has to be reflected on the second qubit– even if the user of the first or second qubit isn't aware or cares about the entanglement. This presents some implementation challenges as various logical registers may impact the state of each other by sharing the sets of the same actual qubits. As an example: if there are two entangled qubits the collapse of one will alter the other, even if those qubits are in completely different registers.

---

[56] None of the examined quantum programming proposals are known to run on actual quantum computers yet– hence they all run as simulations.

[57] Of course if it is an actual quantum computer than violations of physical limitations are impossible.

[58] Again, violations can be allowed for testing in order to help verify the correctness of a classical simulation.

The phrase "actual qubits" is used in the preceding paragraph because there will likely be multiple physical qubits for error correction (similar to how the state of classical bits must be preserved) that are presented to the programming interface as a logical qubit [28]. These could then be shared between various registers. This is illustrated by Figure 71: physical qubits are red, logical qubits green, and registers blue. As shown, five physical qubits are needed for error correction via a 5 qubit error correction code [28] and are grouped together as a single logical qubit for the quantum computer. Thus this example contains four logical qubits available for use by the quantum computer. These four logical qubits can be grouped into registers in any manner, including being shared between registers. In this case the upper left logical qubit is shared between register y and register x, so operations on one register may affect the other because they share a qubit.

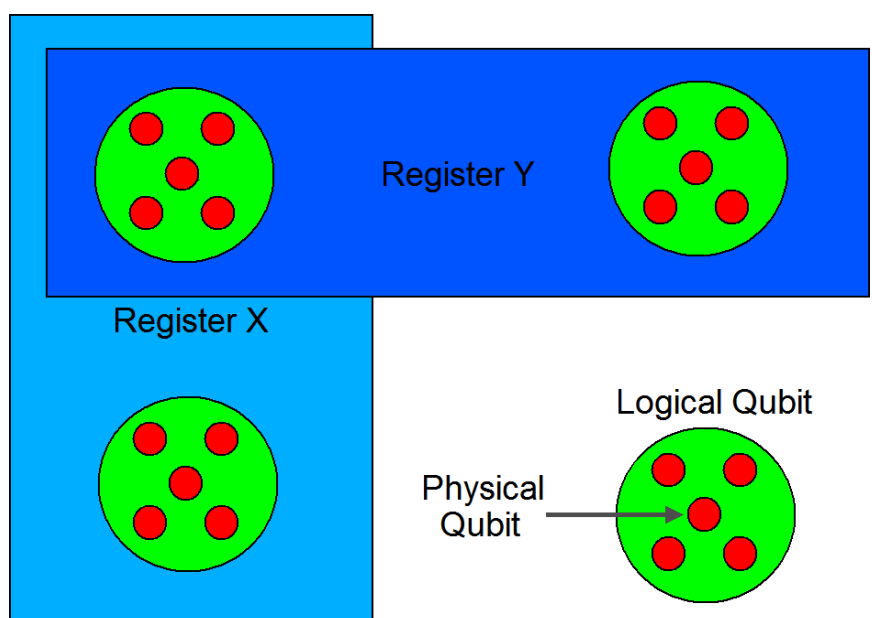

Figure 71. Relation of physical qubits, logical qubits, and registers.

### 3.2.8   Conversion of Qubits to Classical Types

The measurement of a quantum system collapses it to one that can be represented classically. Thus the measurements of qubits need to be converted to bits. Quantum systems can also be initialized based on classical data, so a quantum system needs to be able to be initialized based on bits too[59]. Thus the input and output to a quantum computation is classical data. It is a functional property that this conversion can take place, while making the conversion more flexible is a usability property. One can think of the classical output from a measurement as just a series of bits. As an example, it is often useful to transform that series of bits to an unsigned integer.

### 3.2.9   Accessing a Remote Quantum Resource

Initial implementations of quantum computers will likely be expensive, and thus used as remote resources in accordance with Knill's QRAM model. One can easily imagine a quantum computer being a shared resource between multiple classical computers.

Surprisingly the flexibility to access a remote quantum resource is something that has not been considered by many other quantum programming proposals. While most proposals do so through use of Knill's QRAM model, they don't specify how you would programmatically interact with it. Much as one has to specify the location of a file before using it, one should also have to specify the location of the quantum resource.

In the end, any implementation must be able to work with actual or simulated quantum resources, both local and remote:

---

[59] This can be done by just performing Not on the appropriate qubits if all are initialized to 0 by default.

- Classical simulation, accessed as a local resource. This would be limited in size by the resources of the computer it is run on.

- Classical simulation, accessed as a remote resource, perhaps through web services.

- Quantum implementation, accessed as a local resource.

- Quantum implementation, accessed as a remote resource, perhaps also through web services.

An initial thought is to add an "end point" property to all components. This could be set to "localhost" for local implementations and to a user defined URL for remote ones.

Furthermore one should not be able to entangle qubits that reside on different remote resources[60] with the present implementation. While the general approach of this dissertation has not been to place restrictions based on physical implementations of quantum computers, this one seems reasonable given the current difficulty of even entangling qubits within a single quantum computer. While not addressed in this dissertation, if this becomes practical in the future it could be addressed at that time much as distributed memory has in classical systems. Additionally, it is up to each implementation to enforce that components are located in the same runnable place for that implementation, so an implementation that operates across different quantum resources could still be built within this framework.

This chapter outlined the hypothesis for the research in addition to the proof criteria. These proof criteria will be used to judge that the work has been completed in an

---

[60] One may want to do this as a type of distributed quantum computing where one may need to pass qubits between nodes.

acceptable manner: that it is complete and correct. Explicitly, the proof criteria is the following:

1. The design and construction of a set of interfaces for a quantum computer, which may be accessed remotely.

2. The design and construction of a prototype[61] local simulation implementation that provides the components necessary to execute the following examples: tossing quantum coins, entanglement, and factoring.

3. A set of automated tests that can be used to help show the correctness of an implementation.

4. Documentation detailing each class, interface, method, parameters, return types, and possible exceptions.

Once these items can be accomplished the work can be considered correct and complete. Failure of any of these means that the work is incomplete or incorrect.

---

[61] By using the term prototype it is meant that there may be methods that are not fully implemented, although the examples will still be possible.

# CHAPTER IV

## METHODOLOGY

Cove has been developed using a transparent process, with entire history and stumbling blocks documented on the website at https://cove.purkeypile.com/, see Appendix A for further details and access instructions. This follows the "Science 2.0" approach[62] advocated and used by some within the scientific community [122]. Section 4.1 provides some details about the implementation supplied. Section 4.2 discusses some of the methods and procedures used to develop Cove. Finally 4.3 introduces the quantum circuit diagram notation, which is currently the most frequently used form of expressing quantum computation in the literature. This section also details the quantum circuit diagrams used for the three examples in this dissertation: tossing quantum coins, entanglement, and factoring. The circuit diagrams follow the methods and procedures, as how to implement these circuits in Cove played a central role in its design.

---

[62] The "Science 2.0" approach is essentially open disclosure of the research process via the Internet.

## 4.1     Proof-of-concept Prototype

Cove is based on two types of requirements: functional and usability. Evolved from these are the design, base classes, local simulation implementation, tests, and documentation. The implementation is *how* Cove carries out quantum computation[63], the base classes define *what features* any implementation must supply. The idea is that implementations should be more or less interchangeable. Since no commercial quantum computers exist that could be currently used by Cove, the prototype consists of only a local simulation for an implementation. This local simulation simulates quantum computation on the classical computer on which it executes. While this simulation experiences an exponential slowdown that any classical simulation of quantum computers will exhibit, this does allow for user code to be executed and produce results, albeit for a limited number of qubits. It is anticipated that this initial prototype simulation would be replaced with more sophisticated simulations or implementations that run on an actual quantum computer. Examples of problems that can be solved with the simulation is factoring of small numbers and anything else that requires only a few handfuls of qubits[64].

## 4.2     Methods and Procedures

The design and implementation of Cove has followed accepted software development techniques by utilizing an iterative approach. This iterative approach for frameworks is also advocated by others such as Abi-Antoun [110]. In addition to an iterative approach, unit testing has also been implemented to help ensure the correctness of the simulation as elaborated on in 4.2.4. At present there are hundreds of unit tests against the base interfaces (so they apply to any implementation) in addition to tests specifically for the local simulation[65].

---

[63] Currently Cove carries out quantum computation through a simulated quantum computer.

[64] For the simulation of $n$ qubits $2^n$ complex numbers are needed to represent the state and $(2^n)(2^n)$ complex numbers are needed for the operations. So the simulation is limited based on the amount of memory to hold these data types and how efficiently they can be performed. (Efficient simulation is not a goal of the prototype, so techniques to improve the efficiency of general simulations are generally not utilized.)

### 4.2.1   Gathering Examples

As outlined in 2.3, use cases are the examples that frameworks can be designed and verified against. As a result it is necessary to gather several examples before even starting the project, as these examples influence everything from the architecture to specific methods supplied. One obvious selection of examples is the commonly used quantum algorithms such as those of Deutsch, Shor, and Grover. Analysis of examples coded in other quantum programming approaches may also give insight into the necessary properties. In addition to the automated tests, the correctness of Cove is shown through the implementation of three use cases: the quantum coin toss, entanglement, and Shor's algorithm (factoring)[66].

### 4.2.2   Design of Interfaces

Interfaces are functionally similar to abstract base classes that consist of only pure virtual functions. That is, interfaces can define signatures of methods, delegates[67], and events or messages [123] but no implementation. Any implementer of that interface must provide an implementation for the signatures outlined in the interface. So in a sense an interface can be viewed as a programming contract. Like classes, interfaces can also be inherited from other interfaces.

---

[65] Since the tests can "peek" at the state of a simulation as it evolves to ensure not just that the end result is correct, but that each step to reach it is also correct.

[66] Many of the components for factoring exist at the completion of this dissertation (September 2009), but a complete working example of factoring has not yet been implemented and will be the immediate focus of future work.

In languages such as C# and Visual Basic multiple inheritance (deriving from more than one class) is not allowed. Even though multiple inheritance is disallowed, these languages allow for multiple interfaces to be implemented by a class. Thus any class can only have at most one base class, but implement any number of interfaces. Other languages such as Python and C++ allow multiple inheritance. Single inheritance needs to be targeted to make the design applicable to a wider variety of classical languages. Gosling and McGilton state that multiple inheritance is discarded from Java because it causes too many problems [124]. Instead they claim the attractive features of multiple inheritance are supported through the use of interfaces.

Thus single inheritance is a requirement of Cove, but not a functional one. That is to say, that single inheritance is selected because of the language used (C#) and for the design to be applicable to as large of number of languages as possible. If multiple inheritance were used then the design could not be ported to other languages unless they supported it or significant restructuring took place.

It could be argued that implementing multiple interfaces suffers some of the same flaws as multiple inheritance, even if implementing multiple interfaces is allowed in a language that allows only single inheritance. Nonetheless being able to implement multiple interfaces is a necessity to avoid interfaces becoming bloated[68]. Hence the cost of cohesion of interfaces is the added complexity in the ancestors of the class. As an example even the seemingly simple String class within the .NET framework implements 7 interfaces: `IComparable`, `ICloneable`, `IConvertible`, `IComparable(Of String)`, `Enumerable(Of Char)`, `IEnumerable`, and `IEquatable(Of String)`.

---

[67] Similar to function pointers in C++, but type safe.
[68] This can be worked around by having an interface serve up other interfaces.

The advantage of interfaces is that code can be written against the interface with the expectation that it will be carried out appropriately[69] by any implementer since it is a contract. In true object oriented fashion, the user of an interface is only concerned with what the interface does and typically not how it does it. Since the code is written against the interface any implementations of that interface should be interchangeable as far as the particular interface is concerned. While there is no restriction in place that forces interfaces to describe behavior and implementers to carry it out, that is largely their purpose.

The following C# example illustrates how implementations of interfaces are interchangeable through basic polymorphism:

```csharp
using System;

namespace InterfaceImplementationExample
{
    //this program is used to show that the implementations
    //of the interface are interchangable
    class TestProgram
    {
        static void Main(string[] args)   //entry point of the program
        {
            //iterate through an array of objects that implement
            //GenericInterface.
            foreach (GenericInterface cCurImplementor
            in new GenericInterface[] { new FirstImplementer(),
            new SecondImplementer() })
```

---

[69] Granted, the interfaces do not enforce behavior. This problem can be reduced through examples, complete documentation, and automated tests against implementers of an interface.

```csharp
            {
                Console.WriteLine(
                cCurImplementor.PrintWelcomeMessage("TestProgram"));
            }

            Console.ReadKey();
        }
    }

    //a simple example of an interface
    public interface GenericInterface
    {
        string PrintWelcomeMessage(string SomeParameter);
    }

    //example of a class that implements the interface
    public class FirstImplementer : GenericInterface
    {
        public string PrintWelcomeMessage(string SomeParameter)
        {
            return ("Bye " + SomeParameter + " from FirstImplementor");
        }
    }

    //another example of a class that implements the interface
    public class SecondImplementer : GenericInterface
    {
        public string PrintWelcomeMessage(string SomeParameter)
        {
            return ("Hi " + SomeParameter + " from SecondImplementor");
        }
    }

}                                                   //end of namespace
```

Figure 72**.** Example of how interface implementers are interchangeable in C# via polymorphism.

```
Bye TestProgram from FirstImplementor
Hi TestProgram from SecondImplementor
```

Figure 73. Output from the code in Figure 72.

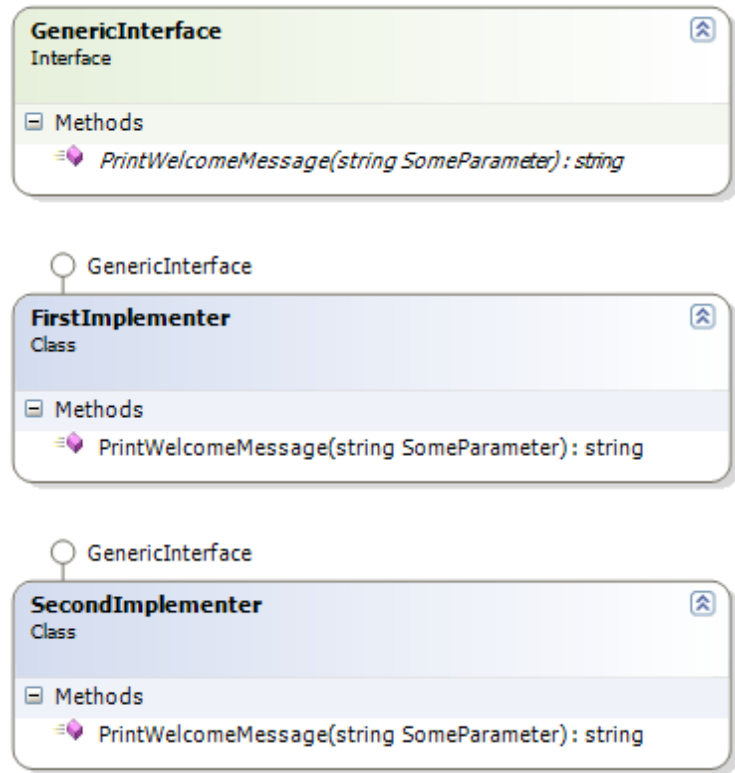

Figure 74. UML diagram of the example.

The purpose of the functional properties is to outline *what* the framework must do and provide. The interfaces concretely specify how the framework would be used, but not how those tasks will be carried out. Since interface implementers are interchangeable, a classical simulation of a quantum computer could be swapped out with one that runs on or against an actual quantum computer.

Implementers of interfaces are intended to carry out certain behavior, but there are no constraints within the language itself for enforcing this. Take a simple `Equals()` method in an interface which is supposed to check for equality. There is no constraint that keeps the implementer from modifying the state, printing a string, or generally having the method do anything they want as long as the signature matches. Hence the problem of incorrect implementations isn't limited to Cove, but any interface– although errors could occur at run time. To help prevent this problem adequate documentation of the interfaces is essential to the point it should be considered a requirement for any interface supplied[70].

The interfaces in Cove are designed as programming language independent as possible, while targeting object oriented languages. Consequently the interfaces have been designed using the unified modeling language (UML). The actual code of these interfaces is written in C# utilizing Microsoft's .NET Framework, version 3.5. Frameworks that are usable by multiple languages, such as Microsoft's .NET framework, see widespread use. Part of the reason for their popularity is that a user can switch between languages without having to learn a new framework or API, which was one of the goals in the development of the .NET framework [125]. By utilizing this platform, Cove will also be usable by any .NET language, allowing users to write code against it in a language they are familiar with instead of the particular one the framework is written in. At the time of this writing these languages include Visual Basic, C#, IRON Python, and F# (still under development) [126].

### 4.2.3   Implementation of Interfaces via a Classical Simulation

While the interfaces allow for one to see how code for quantum computers could be written, interfaces alone do not allow for working code to be executed. In order to allow for executable code to be written, a prototype implementation is provided. This implementation is a locally run simulation of a quantum system, as opposed to a remote simulation of a quantum computer. The primary reason for this is that the local simulation is easier to implement and the focus is on easy to use interfaces that working code can be written against– not an elaborate simulation. In any case, the interfaces have been designed with the ability to access remote resources in mind.

---

[70] And hence documentation as one of the proof criteria for this work.

### 4.2.4   *Automated Unit Testing*

Automated testing is an important part of the implementation. Perhaps all of the existing quantum programming proposals are carried out by no more than a few people. This leaves little resources for software quality assurance (QA). Unit testing does not guarantee the correctness of the software, but does provide an easy way to verify that common test cases function correctly[71]. As with any software that increases in size, the possibility of introducing new defects when making seemingly minor changes increases.

Cove has been written partially with a test driven development approach– that is the automated tests are written before the implementation of these interfaces are written. Newkirk, who administers and contributes the NUnit[72] unit testing project, believes that unit testing is beneficial since it also helps the software lifecycle move from requirements to running code [127]. The test libraries have been divided into ones specific to the local simulation implementation and those that apply to any implementation. The idea is that the later could be run against any implementer of the interfaces and thus be used to test any implementation, not just the local simulation initially developed. The reason for implementation specific tests is to test specific behaviors of that implementation. For the local simulation implementation the state can be peeked at before measurement to ensure it is correct. This isn't possible on an actual quantum computer, hence the reason for the test being specific to the implementation.

---

[71] Since the author has written both the test cases and implementation, where possible the tests are based on examples in other works. A simple example is the tensoring of matrices- the tests are based on examples from a linear algebra text.

[72] http://www.nunit.org/index.php

The unit tests also have an added benefit: they force code to be written that actually uses the framework. In fact the construction of the unit tests for Cove lead to numerous changes and additions of the framework. Some things seem fine when constructing the framework itself, but once you start writing code for the framework it becomes obvious that there are better ways to do things. Unit testing is advocated by Henning, who also points out that they can help improve documentation as well [128].

The unit tests developed will also help to ensure that this implementation functions as an actual quantum computer will, aside from the exponential slowdown in performance[73]. While these tests do not guarantee that the implementation is correct, they do provide a higher confidence that it is correct. The tests also have the added benefit of being able to gather performance metrics. This may be valuable for being able to compare performance of different implementations.

### 4.2.5 Other Potential Implementations

Cove currently consists of a base library and local simulation. It is anticipated that other implementations could be developed as well. This encapsulates Cove into the following components:

- Primary library: the base library. This contains all interface definitions and any (base) classes that are implementation independent.

---

[73] This slow down became noticeable around 8 qubits, and quickly worsened from there as detailed in Figure 113.

- Secondary library: classical simulation accessed locally, which has been developed. This allows for programs to be written and run on existing classical computers. This is the only secondary library developed to date.

- Secondary library: classical simulation accessed remotely. A good area for future work in order to identify and resolve problems with accessing resources remotely. A remote simulation could also employ distributed computing to increase the speed of the simulation. This distributed computing would be transparent to the framework user.

- Secondary library: quantum implementation accessed remotely. This is likely how the first commercial quantum computers would be utilized.

- Secondary library: quantum implementation accessed locally. Likely the last method to arrive for quantum computing, as it would require a likely expensive quantum computer to be built into the system being used.

- Secondary library: a library that works with any remote quantum resource through the expanded QRAM model (section 5.2).

The relation of components is shown in Figure 75. The arrows point to what that component is based on. Thus the base libraries are the core piece since both applications and implementations are built on them

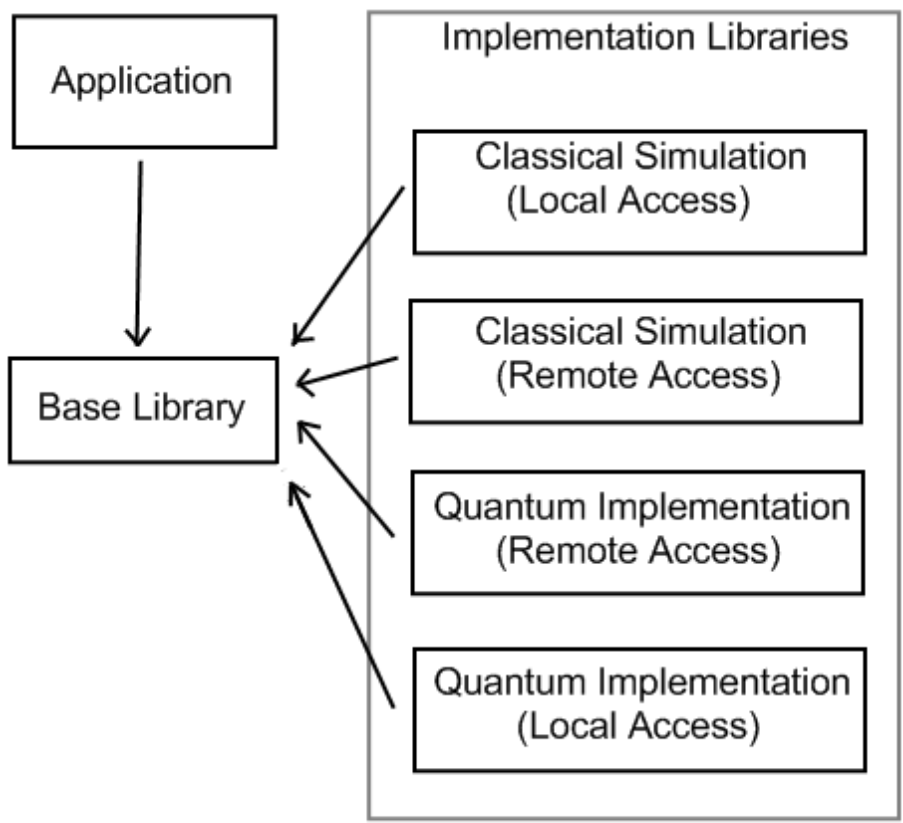

Figure 75. Implementation graph.

One of the goals of Cove is for the classical simulation and quantum implementation to be as interchangeable as possible. Code can then be developed using the simulation and subsequently deployed to actual quantum computers, helping to reduce development costs since the cost of quantum resources would be reduced during the development phase.

There are also some specific things that need be taken into account with the implementation that don't fall under functional properties (mostly in section 3.2) or usability properties (section 5.6), but are specific to the local simulation:

1. Seeding the random number generator– just cannot create a random number generator seeded with the timer for each qubit. If reseeding a bunch of times in quick succession the results will be anything but random because they all have the same seed or a related set of seeds.

2. Specific exceptions for the implementation should be developed.

3. Interface, class, and method documentation should be based on source code comments for ease of maintenance [129].

Cove has been developed using C# and the .NET 3.5 framework. Nonetheless there are several classical features missing that have been developed in order to carry out the simulation:

1. Complex number class

2. Complex matrix class, the matrix class in the .NET 3.5 framework is only for a 3 by 3 matrix [130] and is sealed (cannot be used as a base class).

### 4.2.6   Documentation

Documentation is essential to the success of any framework or API. No matter how good they are, they will see limited or no use without documentation unless there is no choice but to use the framework. The documentation must be clear and easily searchable so that users can quickly look up topics when needed. The documentation should also contain plenty of examples. These examples can be cut, pasted, and modified by users. The goal of any programming approach should be to allow users to express their thoughts as quickly in code as possible. While the syntax of a particular language helps a great deal, the documentation is crucial for when the user must pause to look up a certain topic. With poor documentation the user is forced to consume time researching in detail how to express their idea, which detracts from the primary goal of expressing it with the framework.

The primary documentation for Cove is available through several means since it is automatically generated based on comments in the code. This documentation is also viewable in three languages: C#, Visual Basic, and Visual C++. The documentation is accessible in the following ways:

- Within the source code of the framework itself (Figure 76).

- Via intellisense documentation while writing the code when using Microsoft Visual Studio (Figure 77).

- A local help (chm) file. This allows for the user to consult a local copy of the documentation (Figure 78).

- Through the Cove website (Figure 79).

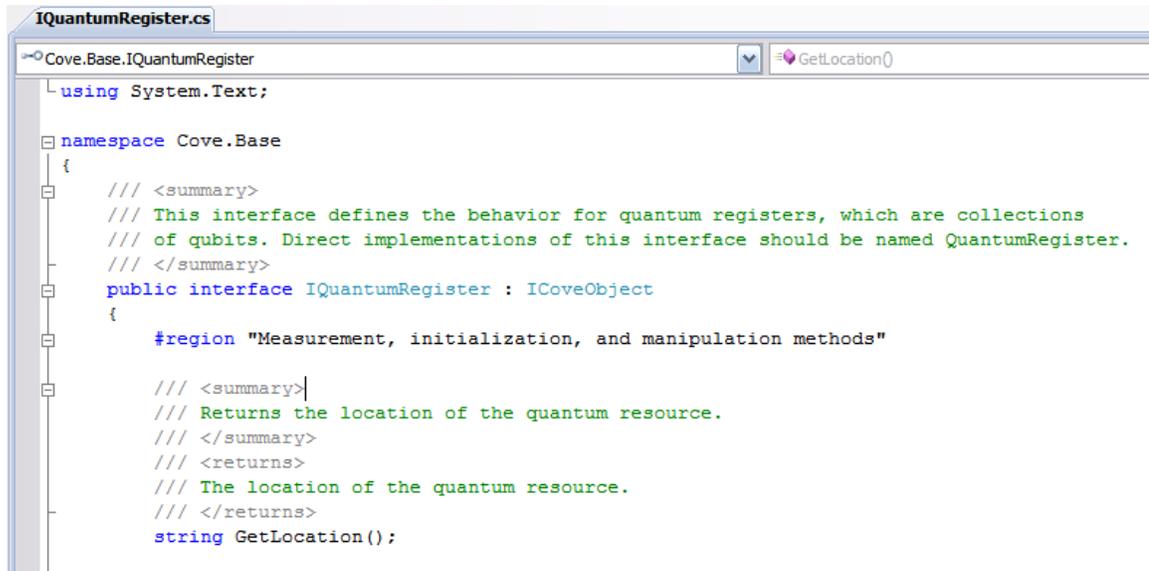

Figure 76. Viewing documentation within the code.

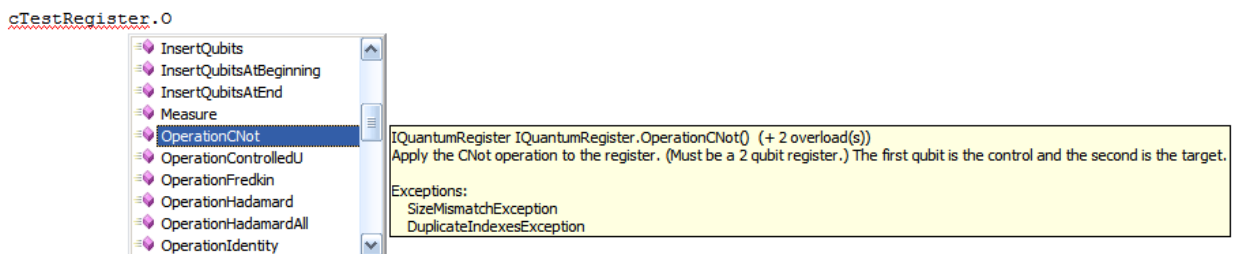

Figure 77. Intellisense documentation while writing code.

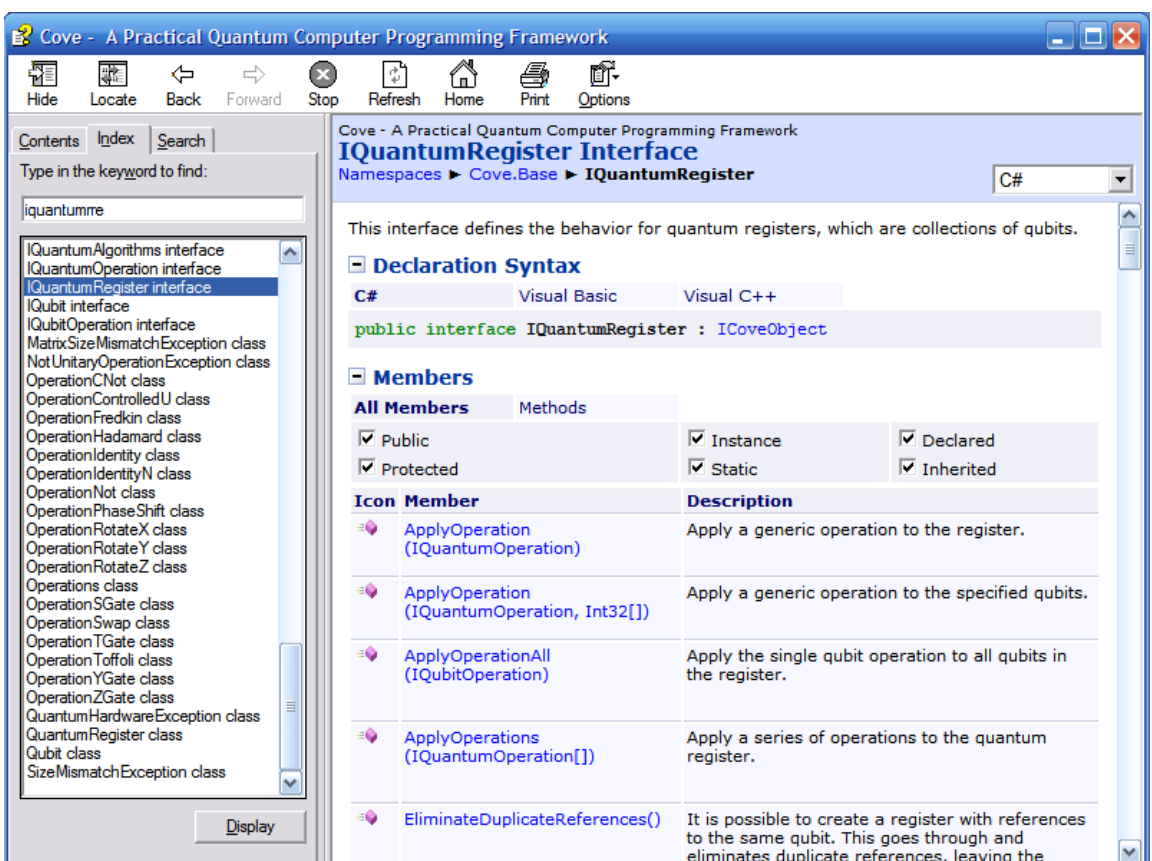

Figure 78. Viewing documentation through the local help file.

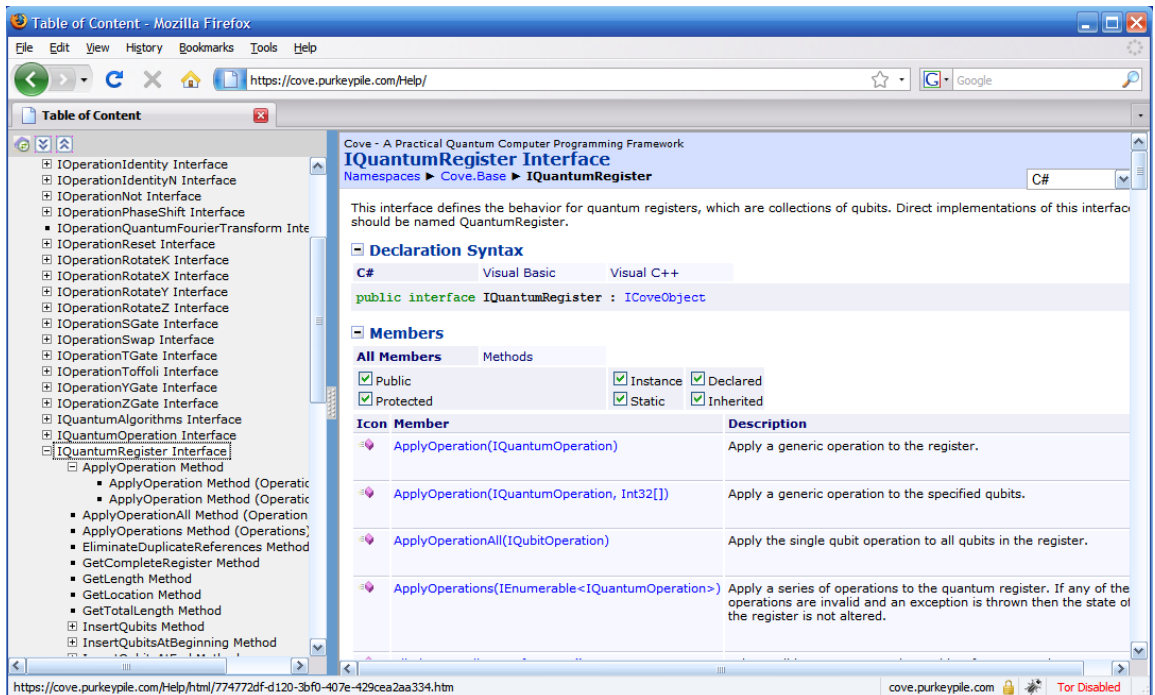

Figure 79. Viewing documentation online.

The intellisense documentation is an important tool as it allows for users to see brief documentation within the IDE. This allows for users to see applicable documentation while writing code as shown in Figure 77 and Figure 80. Figure 80 shows how the various overloaded versions of the method can be cycled through– allowing the user to select the desired one and see the details of all parameters:

```
//create a register of two qubits and apply the CNot operation
IQuantumRegister SampleRegister = new QuantumRegister(2);
SampleRegister.OperationCNot(
▲ 3 of 3 ▼  void IQuantumRegister.OperationCNot (int ControlIndex, int TargetIndex)
ControlIndex: Index of the qubit which will be the control.
```

Figure 80. Example of inline intellisense documentation.

### 4.3 Quantum Circuit Diagrams

Quantum circuit diagrams are frequently used in the literature to detail quantum computing. While Knill's pseudo code [3] is a more concise representation, circuit diagrams seem to be the preferred method of illustrating quantum computing within the literature. Like flow charts, quantum circuit diagrams are only useful for small sets of operations. As with flow charts, beyond small sizes they become unreadable– which is one of the usability flaws for visual programming methods.

In quantum circuit diagrams qubits are represented by "wires", which are horizontal lines. Each line represents a qubit. Operations are represented by marks on the lines, typically a box for a single qubit operation. The diagram is read from left to right, and represents the progress of time. Figure 81 illustrates a quantum circuit diagram of three qubits and three operations. In this example operation 0 would be applied to qubit 0, followed by operation 1 on qubit 1, and finally operation 2 on qubit 2. If the operations were aligned vertically this would represent all 3 operations being applied simultaneously instead of one after the other. This is shown in Figure 82.

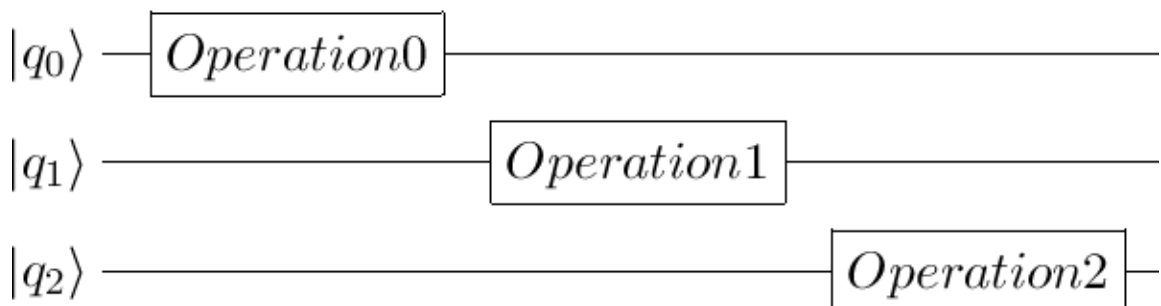

Figure 81. Quantum circuit diagram of 3 qubits and arbitrary operations.

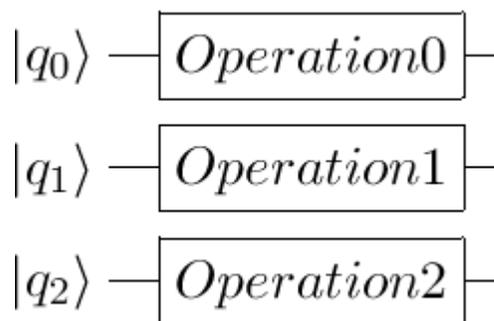

Figure 82. Applying 3 operations to 3 qubits simultaneously.

Throughout the remainder of the dissertation qubits will typically be numbered $q_0$-$q_n$ when referring to a single register. Other prefixes besides q may be used when the collection of qubits represented is broken into logical registers.

Measurement of a qubit always collapses it from any possible superposition to 0 or 1. Thus the output of any measurement operation is classical information. To denote this collapse from quantum to classical information, classical information is represented by a pair of horizontal lines after the measurement operation, as shown in Figure 83.

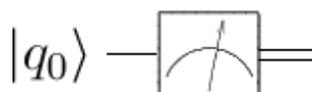

Figure 83. Measurement, resulting in classical information.

Multiple-qubit operations link various qubits together with vertical lines in the diagram. A solid circle at an intersection of horizontal and vertical lines represents a control qubit. An open circle at the intersection represents a Not operation, which is also known as the X gate. As an example, a controlled not operation with $q_0$ as control and $q_1$ as the target could be represented with a solid circle on $q_0$ and an open one on $q_1$ as Figure 84 demonstrates. Figure 85 is the equivalent: there is a control qubit ($q_0$) that then performs an X gate (Not) operation on $q_1$. In this case the X operation could be replaced by any single qubit operation. This would represent the fact that the single qubit operation is only applied when the control qubit is set.

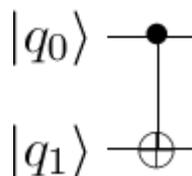
Figure 84. Quantum circuit diagram of CNot

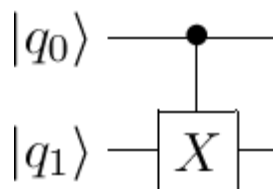
Figure 85. CNot via a control qubit and X gate operation.

Operations typically only apply to a subset of qubits in a register. If an operation does not apply to a particular qubit then the vertical line may pass through the qubits not included. If a CNot operation operates on qubits $q_0$ and $q_2$ then the diagram would look like Figure 86. Qubits included within an operation have some sort of marker at the intersection of the horizontal and vertical lines.

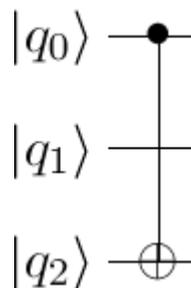

Figure 86. CNot over $q_0$ and $q_2$.

Another problem with quantum circuit diagrams is that they are meant only to express the quantum part of the computation, although Chuang allows for some integration in his quantum circuit viewer[74] (qasm2circ), which is what is used to construct the quantum circuit diagrams in this dissertation. A good way to integrate them with classical computation has not yet been encountered in the literature and is an area for further research. Quantum teleportation is a simple example that requires integration of classical and quantum computation to carry out. Recall that quantum teleportation is reconstructing the source state of a system at the target state, while destroying the state in the source as to not violate the No-cloning theorem.

Without detailing all the steps of the protocol, the destination state is reconstructed by performing various quantum operations based on the two classical bits sent. Thus there is essentially a classical *if* operation that then performs quantum operations as shown in Figure 87, which is based off the description of the protocol in [11]. In this diagram the subsets of C represent what the classical bits are: $c_{00}$ for the bits 00, $c_{01}$ for the bits 01, and so on. For the case of 00 an identity operation is applied, but no action needs to actually be performed so the identity operation in this case is used as a

---

no operation (noop). This example can be shown using quantum circuit diagrams, but the point is that there isn't an easy way to integrate this mixing of quantum and classical computation in them. Accordingly, another advantage of a framework built on a classical language is the ease that the quantum and classical computation can be integrated.

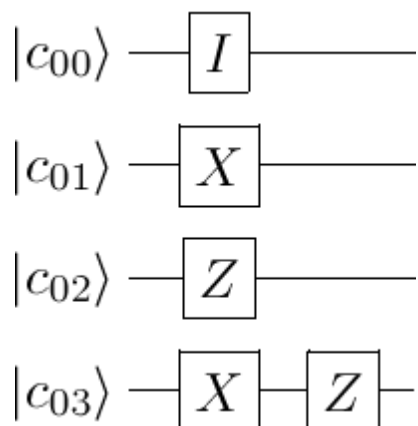

Figure 87. Quantum operations based on classical data in the quantum teleportation protocol.

### 4.3.1 Quantum Coin Toss

The quantum coin toss is a simple example when viewed as a quantum circuit, as in Figure 88 and Figure 89. This quantum coin toss is a simple example of something that does not have a classical equivalent. Recall that if it is observed after the first toss (Figure 88), the coin will be heads or tails with equal probability while if it is tossed twice before observation (Figure 89) it is always be in the same state as its starting state. The mathematics and physical details of this example have been outlined in section 2.1.2; this is only meant to show the equivalent quantum circuit.

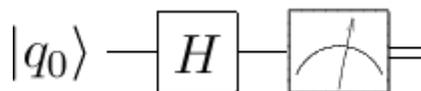

Figure 88. Quantum circuit for tossing a quantum coin once and observing.

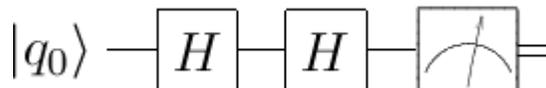

Figure 89. Quantum circuit for tossing a quantum coin twice and observing.

### 4.3.2 Entanglement

Entanglement is another example of something that is not very intuitive if one has a classical state of mind, but is a simple task when it comes to quantum computation. Entanglement is also an excellent example of several things that don't happen in classical computation. First of all, entanglement shows how measuring one qubit may affect another. Figure 90 shows the circuit needed to entangle two qubits to create the state $\frac{1}{\sqrt{2}}|00\rangle + \frac{1}{\sqrt{2}}|11\rangle$, which is also known as an EPR pair. Remember that in this case, once one qubit is measured, the other is guaranteed to be in the same state as the measured qubit since the measurement of one collapses the register to $|00\rangle$ or $|11\rangle$.

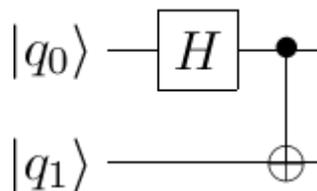

Figure 90. Quantum circuit for entanglement.

### 4.3.3   Shor's algorithm

This section describes how Shor's algorithm for factoring is carried out on a quantum computer. Much of the factoring can be carried out classically, as pointed out in 2.1.11, so here only the quantum part is examined. Many papers and texts that cover Shor's algorithm treat parts of it black boxes. For completeness all of the complex operations needed for this algorithm are built up from elementary operations of no more than a few qubits. This also shows how Cove can be considered complete by providing these operations since they could easily be pieced together to carry out yet undeveloped quantum algorithms. All of the binary numbers in this section are unsigned integers.

The first step in the quantum part of the factoring algorithm is putting all the qubits in the first register into superposition. For $n$ qubits this allows for the numbers 0 through $n - 1$ to be represented at once when treating them as an unsigned integer. To put all the qubits into superposition the Hadamard operation is applied to each as shown by the circuit diagram in Figure 91. Figure 92 shows an alternate form of the same circuit. In this second diagram $n$ qubits are represented by a line with a slash through it in the beginning as opposed to a line for each qubit as in Figure 91. A slightly alternate form is to list the circuit as "$H \otimes n$" instead of just "$H$" and is used in [42]. This just explicitly states that the matrix for the operation is constructed by tensoring the Hadamard operation $n$ times.

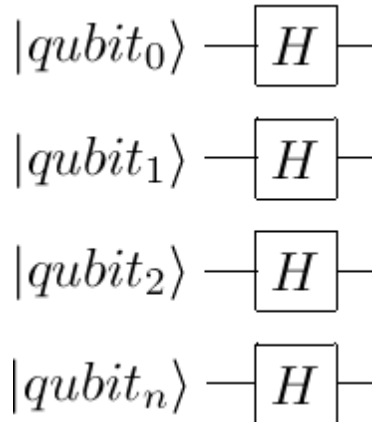

Figure 91. Putting n qubits into superposition.

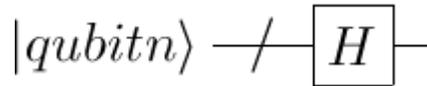

Figure 92. Abbreviated form of Figure 91.

The next step of the quantum part of factoring is the construction of the $U_f$ operation. Recall that this operation essentially takes all the possible values of $x$ represented in $|REG1\rangle$ through superposition, and places each of the results of $m^x \bmod N$ into $|REG2\rangle$. This may sound a little complicated, but we can exploit the fact that we can calculate $m^x \bmod N$ from $m^{x-1} \bmod N$ as first given in Figure 41. Further simplifying things is that $m^{2^j} \bmod N$ is a unitary operation, meaning it can be used "as is" for quantum computation. Thus we can perform $U_f$ by applying $m^{2^j} \bmod N$ for every $j$ up to the number of qubits in $|REG1\rangle$ ($2n$). If we represent $m^{2^j} \bmod N$ as the operation $Um2(j)$ then we end up with the circuit in Figure 93. The equivalent circuit is also given in [42].

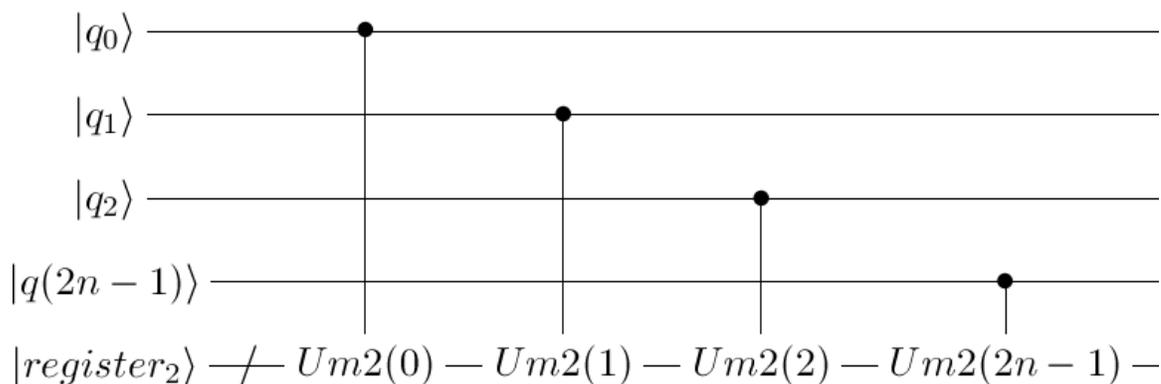

Figure 93. High level application of $U_f$ between the two registers.

Many discussions of constructing $U_f$ end at this point, not providing further details or considering it a "black box". It is surprising that this is often treated as a black box since this is the most difficult part of Shor's algorithm [131]. since  On one hand it can be viewed as simple since the output values range up to the number to be factored (N) through the function $m^x \bmod N$. However the point of including the demonstration of factoring in this example is to show how primitive quantum gates can be pieced together to create nontrivial solutions to real world problems. Hence the discussion continues with how $U_f$ can be constructed from elementary operations and is based on [23, 131].

One way to express Figure 93 is to start off by recognizing the following as given in [28]: $U_f |x, y\rangle = |x, y \oplus f(x)\rangle$, where $\oplus$ is bitwise addition mod 2. Another way to look at it is that we perform a series of control operations based on the qubits 0 through $(2n - 1)$ in $|REG1\rangle$. If the qubit is $|1\rangle$ then $|REG2\rangle$ is replaced by its modulo-N square [27].

The modular exponential function needed in $U_f$ can be built up in four steps. These four steps and the discussion creating them is largely based on the discussion in [23]. The four subparts to be created are to build up to the function are:

1. Adder
   Input: $a$ and $b$, where both are nonnegative integers.
   Output: $a + b$

2. Modular adder
   Input: $(a + b)$ and $N$
   Output: $(a + b) \bmod N$

3. Modular multiplexer
   Input: $(ab)$ and N
   Output: $(ab) \bmod N$

4. Modular exponential function
   Input: $m$, $x$, $N$
   Output: $m^x \bmod N$

The first step in building up the modular exponential function that is needed is to construct a quantum adder that operates over *n* qubits. In order to cut down on the mathematics included here, the two necessary circuits are given that are needed to build up the *n* qubit adder. For more detailed explanation, including the math, the reader is again referred to [23] which this discussion is based on. There are three perquisite gates needed to build up the n qubit adder: the sum gate, the carry gate, and the inverse carry gate (carry$^{-1}$). The circuit diagram to build these gates from elementary operations are given in Figure 94, Figure 95, and Figure 96, respectively. In these three circuit diagrams $|x\rangle$ and $|y\rangle$ are the qubits to add, $|c\rangle$ is the carry qubit, and $|a\rangle$ is a scratch qubit ("a" for ancilla). Furthermore these circuits will be denoted by "sum", "carry", and "carry$^{-1}$" in subsequent diagrams to make them more readable.

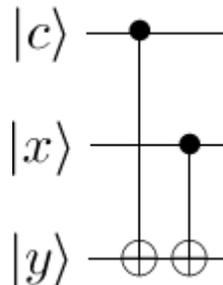

Figure 94. Circuit diagram of the sum gate.

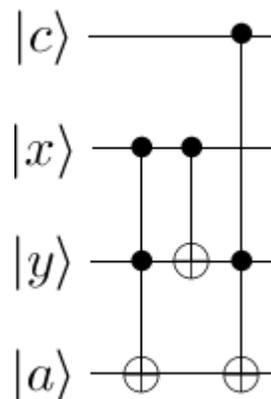

Figure 95. Circuit diagram of the carry gate.

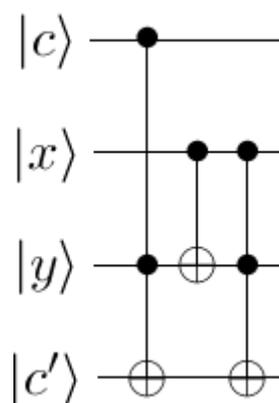

Figure 96. Circuit diagram of the inverse carry gate.

Often gates such as the preceding ones are listed as a single box with the inputs and outputs. This allows for elementary operations to be encapsulated and reduce the clutter in more complex circuit diagrams. Often when this is done the ancilla qubits are excluded from the input and output. Consequently the circuits can be more complex than they appear in this encapsulated form. Figure 97 shows the encapsulated form of the sum gate.

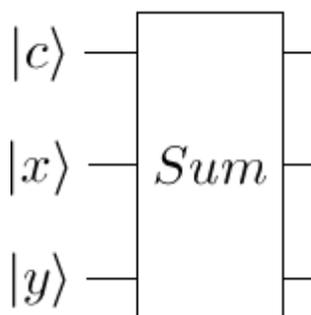

Figure 97. Encapsulated Sum gate.

The output of the gates are as follows:

- For the Sum gate, Figure 94, $|c\rangle$ and $|x\rangle$ remain unchanged after the circuit is run. The output is placed in $|y\rangle$ and can be defined as $|c \oplus x \oplus y\rangle$, where $\oplus$ defines mod 2 addition in accordance with [23].

- For the carry, Figure 95, $|c\rangle$ and $|x\rangle$ remain unchanged in the output. $|y\rangle$ will be $|x \oplus y\rangle$, and $|a\rangle$ will be $|ab \oplus ac \oplus bc\rangle$.

- For the inverse carry gate, Figure 96, $|c\rangle$ and $|x\rangle$ remain unchanged in the output. $|y\rangle$ will be $|x \oplus y\rangle$, and $|c'\rangle$ will be $|a(a \oplus b) \oplus bc \oplus c'\rangle$.

The carry (Figure 95) and the inverse carry gate (Figure 96) show the relationship between a gate and the inverse of that gate. The inverse of a gate merely runs the operations within it in reverse. Since all quantum gates must be reversible to be valid, every valid quantum gate has an inverse.

These three gates (sum, carry, carry[-1]) can then be used to piece together an add operation over $n$ qubits. With this $n$ qubit adder there are two registers, each with $n$ qubits. The add $n$ operation takes these two registers as input. As with all quantum computation, the input registers are modified to contain the result. In this case register 1 is unmodified and the result in register 2. The circuit diagram is given in Figure 98. The inverse of this add operation performs subtraction [23]. The add operation over $n$ qubits will be abbreviated "add(n)" in the subsequent diagrams, again for readability. As with the sum gate, *x* remains unchanged after the gate and *y* is replaced with the result.

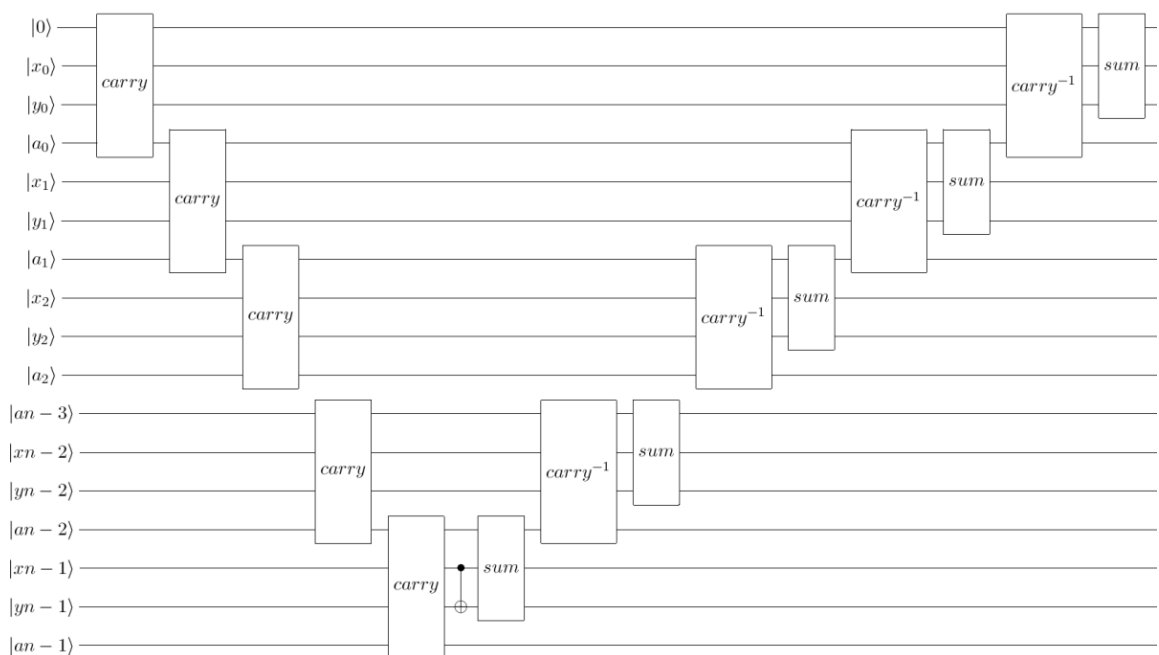

Figure 98. Circuit diagram of a *n* qubit adder.

Using the *n* qubit adder the modular adder can be constructed. The modular adder computes $x + y \bmod N$. The inputs are *x*, *y*, N, and a scratch (ancilla) qubit. The number of output qubits must match the number of input qubits, so *x*, N, and the scratch qubit are unchanged after the computation. As with adder the result, $x + y \bmod N$, is placed in the second input register (*y*), the others remain unchanged. In the diagram the "reset" means that the register is reset to 0. In this specific case it is a controlled-reset, so it is only done if the control qubit is 1. The circuit diagram for the modular adder is given in Figure 99.

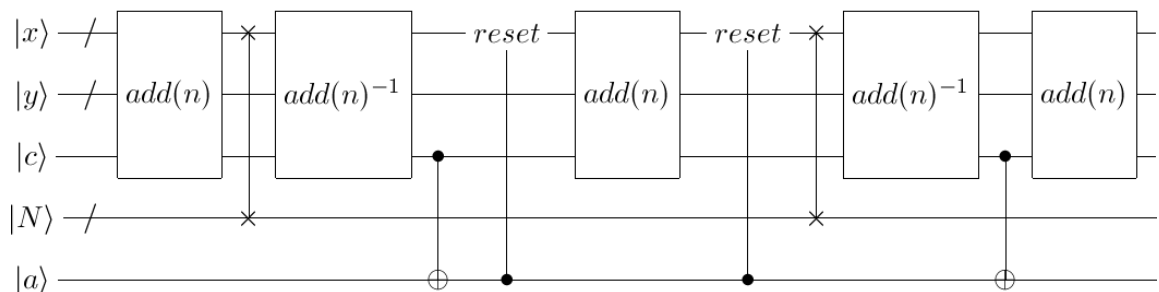

Figure 99. Circuit diagram of the modular adder.

At this point we have the pieces necessary to construct the final operation needed for U$_f$. This final piece is the controlled modular multiplexer, which is show in Figure 100. Finally this controlled modular multiplexer is used to construct the modular exponential function, U$_f$, given in Figure 101.

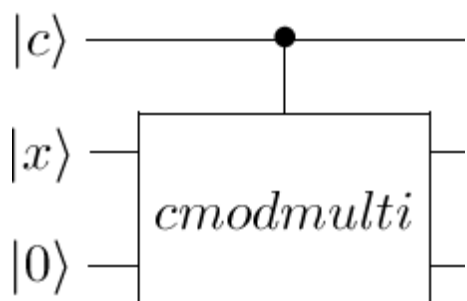

Figure 100. Circuit diagram of the controlled modular multiplexer.

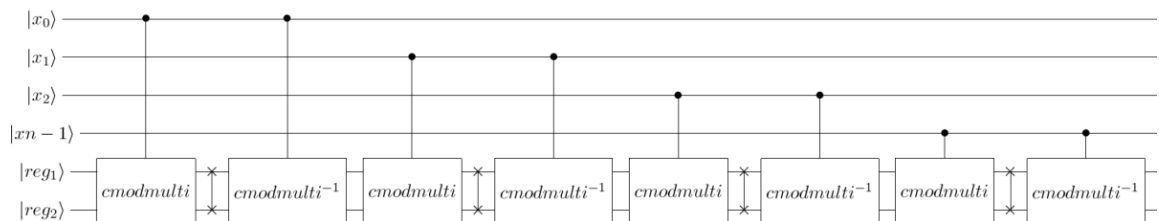

Figure 101. Circuit diagram of modular exponential function, U$_f$.

Before the circuit for the Quantum Fourier Transform can be discussed a circuit for reversing the order of $n$ qubits should be given. This circuit is used at the end of the Quantum Fourier Transform as the final step. Reversing the order of the qubits means that if the qubits are indexed 0, 1, 2, …, ($n$ -1)  then: qubits 0 and $n$ - 1 are swapped, qubits 1 and $n$ - 2 are swapped, qubits 2 and $n$ - 3 are swapped, and so on. The circuit diagram for swapping two qubits is given in Table 6, but it can also be constructed by the circuit in Figure 102 that only utilizes CNot gates. As can be seen from the case of swapping 4 qubits in Figure 103, the reversal of any number of qubits utilizing CNot gates is not complicated as the circuit is just expanded with additional swaps as necessary. For an odd number of qubits the qubit on the middle is left unaltered, as it could be considered to switch places with itself.

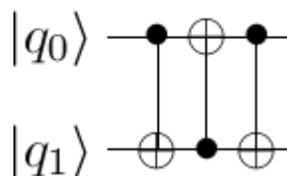

Figure 102. Reversal of 2 qubits from CNot gates.

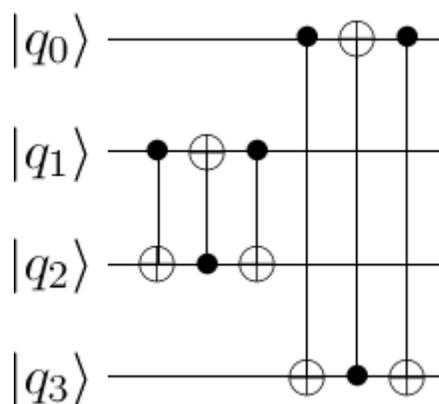

Figure 103. Reversal of 4 qubits from CNot gates.

The construction of the Quantum Fourier Transform (QFT) is the final non trivial operation that has to be implemented in order to perform factoring. Figure 104 shows the general case of the circuit diagram for the QFT and is based off the diagrams in [13, 22, 132]. The final gate is the reversing of the order of qubits. Another way to define the QFT is through the product representation, is shown in Figure 104 and is also based on [12, 13, 22, 132]. Stated another way, qubit $j_n$ after the QFT is in the state $|0\rangle + e^{2\pi i 0.j_1\cdots j_n}|1\rangle$.

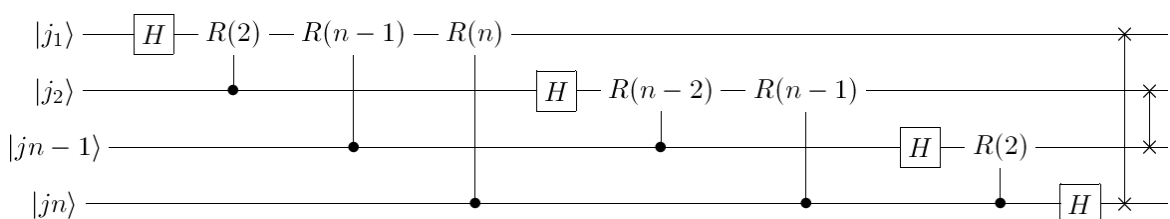

Figure 104. Circuit diagram of the Quantum Fourier Transform over $n$ qubits.

$$\left|j_1, ..., j_n\right\rangle \rightarrow \frac{|0\rangle + e^{2\pi i \cdot j_n}|1\rangle \quad |0\rangle + e^{2\pi i \cdot j_{n-1}j_n}|1\rangle \ ... \ |0\rangle + e^{2\pi i \cdot j_1 j_2 \cdots j_n}|1\rangle}{2^{n/2}}$$

Figure 105. Product representation of the Quantum Fourier Transform over $n$ qubits.

The general case of the QFT is a little abstract, so it is best stepped through for the first few values of $n$. The case of $n = 1$ is simple, as it is just the Hadamard operation. Given: $n = 2$ we have the circuit in Figure 106, $n = 3$ is the circuit in Figure 107 (also given in [13, 22]), and $n = 4$ is the circuit in Figure 108. To simplify these circuits the swapping of qubits is drawn as the wires moving from one location to another instead of the series of elementary operations required to perform this. In reality the qubits do not actually have to be reversed in the final step, a logical reordering of them suffices [12].

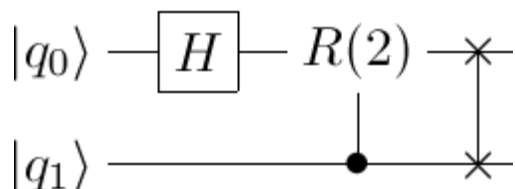

Figure 106. Circuit diagram of the Quantum Fourier Transform over 2 qubits.

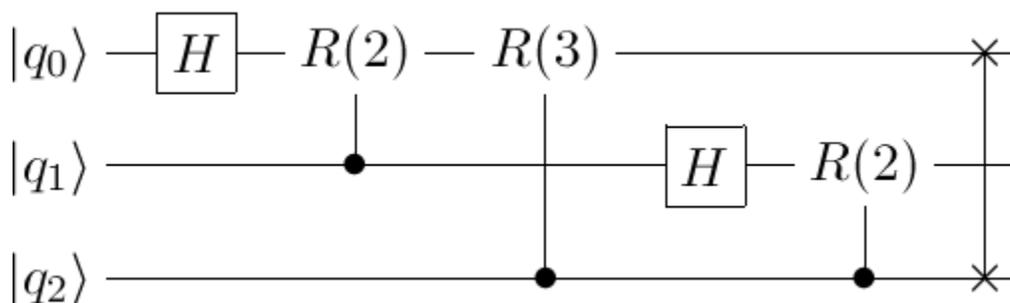

Figure 107. Circuit diagram of the Quantum Fourier Transform over 3 qubits.

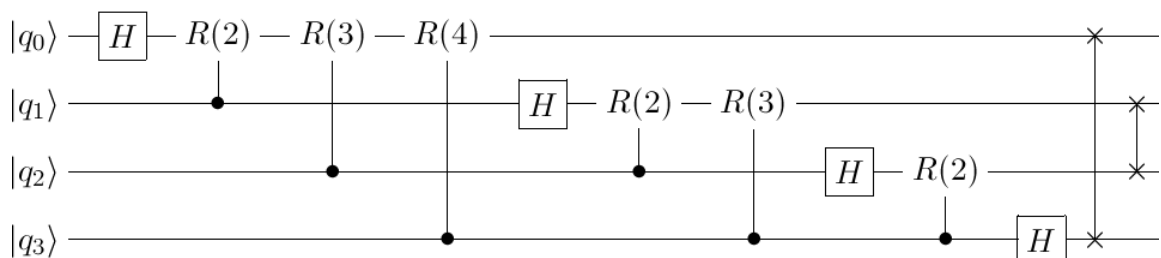

Figure 108. Circuit diagram of the Quantum Fourier Transform over 4 qubits.

Note that the R(2) and R(3) gates ($R_2$ and $R_3$) equate to the standard S and T gates, as is worked out mathematically in Figure 109. Thus all the R(2) and R(3) gates in the preceding figures may be replaced with S and T gates respectively. As Euler's formula [26] states, $e^{iy} = \cos y + i \sin y$, so for the S gate's lower right element: $e^{\pi i/2} = \cos \frac{\pi}{2} + i \sin \frac{\pi}{2} = 0 + i1 = i$. Furthermore, $S = T^2$ [13].

$$R_k = R(k) = \begin{bmatrix} 1 & 0 \\ 0 & e^{2\pi i/2^k} \end{bmatrix}$$

$$R_2 = R(2) = \begin{bmatrix} 1 & 0 \\ 0 & e^{2\pi i/2^2} \end{bmatrix} = \begin{bmatrix} 1 & 0 \\ 0 & e^{\pi i/2} \end{bmatrix} = \begin{bmatrix} 1 & 0 \\ 0 & i \end{bmatrix} = T^2 = S$$

$$R_3 = R(3) = \begin{bmatrix} 1 & 0 \\ 0 & e^{2\pi i/2^3} \end{bmatrix} = \begin{bmatrix} 1 & 0 \\ 0 & e^{\pi i/4} \end{bmatrix} = T$$

Figure 109. R(2) and R(3) equating to the S and T gates.

Another way the QFT can be represented is by the equation in Figure 110, for $n$ qubits. If the QFT is represented then the inverse QFT (QFT$^{-1}$) is the equation Figure 111. Both these figures are given in [12], and notice that the only difference between them is the negative sign in the exponent of $e$ in Figure 111. Essentially the QFT$^{-1}$ returns a binary unsigned integer encoded in a $n$ qubit state [12].

$$QFT : |x\rangle \mapsto \frac{1}{\sqrt{n}} \sum_{y=0}^{n-1} e^{2\pi i \frac{x}{n} y} |y\rangle$$

Figure 110. Quantum Fourier Transform equation.

$$QFT^{-1} : |x\rangle \mapsto \frac{1}{\sqrt{n}} \sum_{y=0}^{n-1} e^{-2\pi i \frac{x}{n} y} |y\rangle$$

Figure 111. Inverse Quantum Fourier Transform equation.

All of the necessary circuits for carrying out the quantum part of factoring have been established in this section, utilizing elementary operations. At a general level these operations are pieced together in the following order after the qubits are initialized to $|0\rangle$:

1.  The qubits in $\left| REG1 \right\rangle$ are put in superposition by applying the Hadamard operation to each.

2.  $U_f$ is performed on both registers, with the result being placed in $\left| REG2 \right\rangle$.

3.  $\left| REG2 \right\rangle$ is measured.

4.  The Inverse Quantum Fourier Transform is applied to $\left| REG1 \right\rangle$.

5.  $\left| REG1 \right\rangle$ is measured.

From this the period is obtained and the remainder of the algorithm (as detailed in 2.1.11) is carried out classically. Figure 112 details this quantum part of the factoring algorithm as a circuit diagram.

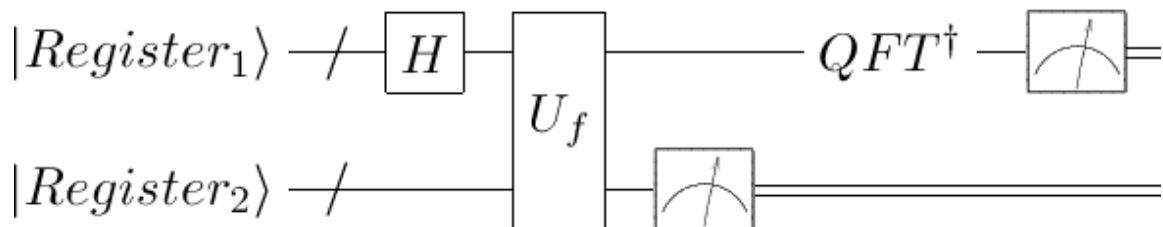

Figure 112. High level circuit for the quantum step in factoring.

Shor's algorithm is an important example of a nontrivial computation that can be carried out on a quantum computer and provides a remarkable speed up compared to a classical computer. As such this algorithm is demonstrated using Cove in section 5.4.3. The example of Shor's algorithm using Cove indirectly shows that Cove is complete enough to carry out common quantum algorithms. The necessary operations to perform this algorithm have been built from elementary operations, showing how more complex tasks can be performed from primitive operations.

This chapter has outlined the methods used to carry out the design and implementation of Cove. Additionally, circuit notation has been introduced. The circuit notation is used to detail the specific quantum operations that are needed to carry out the quantum coin toss, entanglement and factoring. These diagrams are meant to be the bridge from the mathematical working of the examples in chapter 2 to the actual source code in chapter 5. As such these circuit diagrams were consulted frequently in the design and implementation of Cove.

CHAPTER V

COVE: PROPERTIES, DESIGN AND IMPLEMENTATION

This section covers the properties, design and implementation of Cove, a framework for quantum computing that is based on existing classical languages. Specifically, the contribution is the combination of the fields of framework design and quantum computing to create a framework for quantum computation. This framework is intended to be as easy as possible for existing commercial programmers to use. The purpose of this chapter is not solely to outline the final result, but to also detail the rationale behind the design and implementation decisions. Where appropriate, some of the alternates and the flaws with alternate approaches are also detailed, and general rationale is covered throughout this chapter. This justification of design choices (at every step of the software lifecycle) in the literature is minimal and frequently nonexistent. To those studying programming techniques for quantum computers the alternate approaches and reasons why they are not viable or practical warrants discussion. Finally, the development blog and unpublished papers on implementation challenges (see Appendix A) outlines the thought process throughout the design and implementation, especially early on in the project, in even greater detail than this dissertation.

## 5.1    Assumptions, Limitations, Dependencies, and Constraints

There are several assumptions that this framework is based on. The all encompassing theme of these assumptions is optimistic: the framework will run on an ideal quantum computer[75], as the following subsections detail. As previously stated, the focus is on a practical programming approach with little focus on the physical difficulties in creating viable quantum computers. Section 5.2 details how many of these problems can be lessened from a programming perspective by relegating them to a lower level of abstraction by expanding Knill's QRAM model [3].

Some may argue that an ideal quantum computer is unattainable, but we've largely reached the stage where classical computers can be considered ideal behavior-wise from a programmer's point of view for commercial applications[76]. By ideal, it is meant that we largely do not have to be concerned with errors. As an example, 2 + 2 will always equal 4 with a high enough probably for the programmer of a commercial application to assume it will always be the case[77]. While there are different physical challenges for building quantum computers, part of the reason for assuming the quantum computer will be ideal is to avoid building in limitations based on physical implementation problems that may be solved in the future.

The assumptions, limitations, and constraints are detailed in the remainder of this section (5.1), but in summary they are:

- Cove is hardware independent, section 5.1.1.

- Users need not be concerned with error correction to maintain states of qubits, section 5.1.2

- There is no time limit states can be preserved for, section 5.1.3.

- Cove is only supports quantum operations, section 5.1.5.

---

[75] Ideal being defined as a quantum computer that functions with minimal errors, if at all, in accordance with the mathematics previously detailed.

[76] There are still challenges programming them though, distributed systems being a good example.

[77] To differentiate from something such as the Space Shuttle, where there may be multiple computers to guard against very rare cases such as this.

- No delayed execution of operations, section 5.1.6.

- Knill's QRAM model is viable, section 5.1.8.

### 5.1.1 Cove is Hardware Independent

Cove has been designed to be independent of the quantum hardware used to carry out quantum computation. It is not yet certain how quantum computers will be implemented physically, thus the assumption is that nothing specifically needs to be done within the programming environment for a particular physical implementation. This philosophy is much like that of existing high level languages where the particular processor does not matter, and in some cases such as Python not even the operating system. Obviously the physical implementation has to be dealt with at some level, but it will be in a lower level of abstraction than the framework, such as the expanded QRAM model (section 5.2).

Mathematically qubits, registers, and operations can be thought of as the matrix or Dirac notations in chapter 2 detailed.

### 5.1.2 Users not Concerned with Error Correction

Currently, physically maintaining the state of a qubit is difficult due to decoherence. Essentially the qubits must be isolated from the outside environment so that unintended measurements do not alter the state. To help alleviate this problem, quantum error correction techniques have been developed by Shor and others [28]. The classical equivalent is the addition of extra bits in classical communication to detect, and in some cases correct, errors. Qubits cannot be copied due to the no-cloning theorem, so this adds additional challenges to quantum error correction. Some quantum error correction schemes include Phase-Flip quantum error correction code, Shor's nine-qubit code, and five qubit quantum error correction code [23].

It is assumed that error correction takes place at a lower level of abstraction than the framework targets. There has been much work in quantum error correction because decoherance is a large hurdle in the physical implementation of quantum computers [13]. Nonetheless[78], not being concerned with the preservation of state is in line with many high level classical languages. As examples: a C++ programmer does not worry about their integer variables sporadically changing values due to hardware errors if the value does not span the natural word size, nor does a Python programmer have to worry about the data being read out of a file incorrectly. (While they may receive an error if the file cannot be read, the integrity of the data is essentially assured for most purposes if it is read.) In both of these cases the error correction is handled at a lower level and the programmer generally does not have to be concerned when writing high level code. As a result, the programmer of a quantum device is concerned with a single logical qubit, which in reality may consist of several physical qubits for error correction in accordance with [55].

### 5.1.3   No Time Limit for Execution

There is no time limit that quantum states can be preserved for in a generic implementation of Cove, nor is there a limit on the number of operations that can be performed. There are limits in the current physical realizations of quantum computers for both time until decoherence and number of operations that can be performed on a register [13]. Cove takes the optimistic approach that these problems will eventually be resolved. There could also be optimizations performed, such as queuing up initialization, operations, and measurements and performing them as quickly as possible– but of course queuing up operations introduces another set of problems (see 5.1.6). Solutions to the problem of a time limit would render the framework at best cumbersome and at worst obsolete if it took these limitations into account and they were later solved.

The general view of Cove is that hardware limitations, especially current ones that are not commercial implementations and still in infancy, should not impact the design. However, implementations may be created that are more restrictive as long as the interfaces are still implemented[79].

### 5.1.4   Local Simulation Not Excessively Concerned with Round Off Errors

The local simulation implementation of Cove will not be overly concerned with round off errors, which may also be considered representation errors. The simulation of a quantum system takes place on a classical system, which operates on bits. By this very nature any numeric representation with bits covers a set of discrete values– irrational numbers and numbers in a continuous range have limited accuracy when expressed on these classical systems. Much like the exponential slowdown of simulating a quantum system on a classical one, this slight loss of accuracy in certain cases will be considered a limitation. These types of errors are not limited to the application of quantum computing; they can be encountered in normal use of floating point data types. The use of more accurate floating points[80] was explored, but not utilized in Cove, it was an approach pursued by [69]. The higher the accuracy of the floating points, the slower errors accumulate on the state of simulated qubits.

---

[78] This is not to underemphasize the huge challenge of physically implementing quantum computers. Our software models should not be based on constraints present in the infancy of quantum computers.

[79] Example: An implementation may only allow for $x$ operations to be applied to a register in time slice $y$ before it collapses.

### 5.1.5    Only Quantum Operations Supplied

The framework is only concerned with carrying out quantum computation. Classical operations, including any needed for quantum algorithms, are provided by the classical language which the framework is built upon. Consequently the focus of Cove is the quantum components necessary for quantum computation. One of the downfalls of languages specific for quantum programming is that they must also replicate the huge classical power in existing classical languages.

The one exception to this is the result of measurement of a register in Cove. Any measurement of a quantum system produces a classical result, which can be considered an array of bits. In Cove this array of bits which is wrapped in a class to provide some common conversions, such as conversion to an unsigned 32 bit integer.

### 5.1.6    No Delayed Execution of Operations

Some other proposals, such as QCL [79], make use of a delayed execution stack. This stack builds up all the operations and sends them off to the quantum device. All operations are applied immediately in Cove. There are several motivations behind this. The primary motivation is that qubits may be shared between registers, and thus different computations. Thus some difficult to debug situations could arise if there was a delay in execution; imagine integers within an application having a delay before their state is changed. The second motivation is to avoid differences between implementations, since implementations should be largely interchangeable. Thus once a call to apply an operation has returned, it can be considered done.

---

[80] The standard 64 bit double data type is currently used in the local simulation prototype of Cove. More accurate floating points would be ones that consists of more than 64 bits.

### 5.1.7   *Maximum Number of Qubits in a Register for the Local Simulation*

There is a restriction on the maximum number of qubits within a register for the local simulation implementation, not the interfaces in the base library. This limitation has to do with the maximum array length within the language, which is $2^{63}$-1 elements if a signed 64 bit integer is used for addresses. The maximum number of qubits in a register is thus set to 62, as any more would not be addressable. A register of this size would be represented by a matrix with $2^{62}$ elements. Assuming the complex numbers within each element only occupy 16 bytes[81], which is an optimistic estimation, then $16(2^{62})$ = 73,786,976,294,838,206,464 bytes or nearly 74 zettabytes would be required. This is an amount of storage that will not be available on classical computers in the foreseeable future (as of 2009). So even though there is a constraint on the number of qubits in a register, the user will run into memory constraints long before that limitation[82]. To put this number in context, the total lifetime of the universe is $2^{61}$ seconds if the universe is closed [71]. Nonetheless, in order not to place arbitrary constraints on the simulation the maximum number of qubits is not set lower as in some other quantum programming techniques.

---

[81] One 8 byte double for the real part, another 8 byte double for the imaginary part.
[82] These large matrices may be sparsely populated, and therefore reduced in size. However the purpose of the local simulation implementation is allowing for working code to be written. More efficient simulations

As outlined in Chapter 2, this large amount of space is required because the simulation must be able to keep track of the probability amplitude of every possible state it may collapse to. It then becomes apparent that users of the local simulation (not actual quantum computers) should keep registers as small as possible, and break computations into multiple registers where possible. As an extreme example, 62 individual qubits can be represented by 62 matrices of 2 complex numbers each, or 124 complex numbers. Compare this to the $2^{62}$ complex numbers required if they were all within one register. This isn't to say that quantum algorithms, such as factoring, can be broken into smaller pieces– they cannot[83]. It is just to point out that for the simulation there is an exponential increase in memory required as qubits are added to a register.

As a practical example, the local simulation of Cove was run over a simple case: putting the entire register in superposition, then measuring it. The time it took to do so is graphed in Figure 113, where the time listed is the average time from 100 iterations. It isn't the specific times that matter, but the fact that the exponential slow down can be seen. There is also an exponential space requirement, so the curve will become even steeper once additional slow down is encountered due to the use of virtual memory. Given this it is easy to see how the 62 qubit limit is far beyond current means.

---

of quantum systems is another research area that has not been explored for this local simulation, but could be an area for future work.

[83] If they could be then we could crack real world crypto problems on simulated quantum computers instead of needing an actual quantum computer.

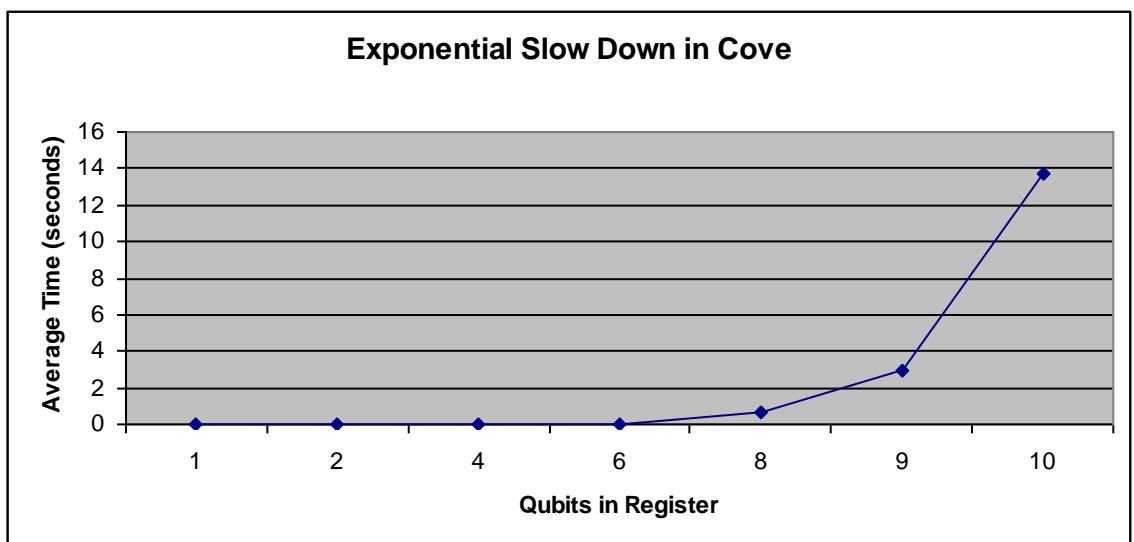

Figure 113. Exponential slow down in the local simulation of Cove in a simple case[84].

### 5.1.8 QRAM Model is Viable

Many of the existing proposals for quantum computer programming depend on Knill's QRAM model [3] being a viable interface to programming quantum computers. This dissertation makes the same assumption that this is a viable model. While this model is expanded on slightly in 5.2 the same concept of a classical computer interfacing to a quantum one (through the controller) still applies.

These assumptions include many of those already outlined in here in 5.1. In a general sense the QRAM model specifies that quantum registers can be initialized based on classical states, then manipulated by operations, and finally a classical result obtained through measurement [3]. Within [3] Knill does not make any mention of the QRAM model being tied to a specific physical implementation, thus it implicitly operates on an ideal quantum computer.

---

[84] Generated on a Intel Core Duo T2300 Processor with 1 GB of RAM.

## 5.2     Expanded QRAM Model

There is no reason why some of the limitations in section 5.1 apply only to Cove and not other quantum computer programming proposals as well. This section steps away from the constraints and limitations of the previous section and discusses an expanded QRAM model that helps to address some limitations of the existing QRAM model.

Some of the existing quantum programming techniques focus too much on solving these problems instead of relegating them to a lower level of abstraction, as has been done with many things for classical languages. The problems clearly still need to be solved at some level of abstraction, but not by the user of the framework. A good classical parallel example is logical memory management. For many cases manual memory management, as is the case with C++, is an error prone chore that distracts the programmer from writing the code that actually accomplishes the task at hand. By having memory management pushed down to a lower level of abstraction some control and performance is lost, but it comes with the benefit that more of the code is application specific instead of dealing with the more tedious tasks and infrastructure.

In Knill's quantum random access machine (QRAM) model (as first outlined in 2.1.5) there is a classical computer which controls a quantum computer. The classical computer is the master and sends elementary operations to the quantum computer[85], which is the slave. The specific interface for doing so is intentionally unspecified since it depends on the implementation[86]. The quantum computer may be local or remote. This quantum resource may also be shared, leading to competition between devices and possibly unregulated access depending on the implementation of the quantum computer. Figure 114 shows Knill's QRAM model where each client is sharing the quantum computer and must perform their own lower level tasks such as error correction since it has direct control over the physical qubits.

---

[85] This could perhaps be done by sending commands to execute specific operations, such as "execute Hadamard on qubit 3". The quantum computer would then take this command and do what is necessary to physically perform the operation.

[86] Perhaps a communication protocol is needed between classical and quantum computers. Tucci's Qubiter (outlined in Chapter 2) is a step towards this.

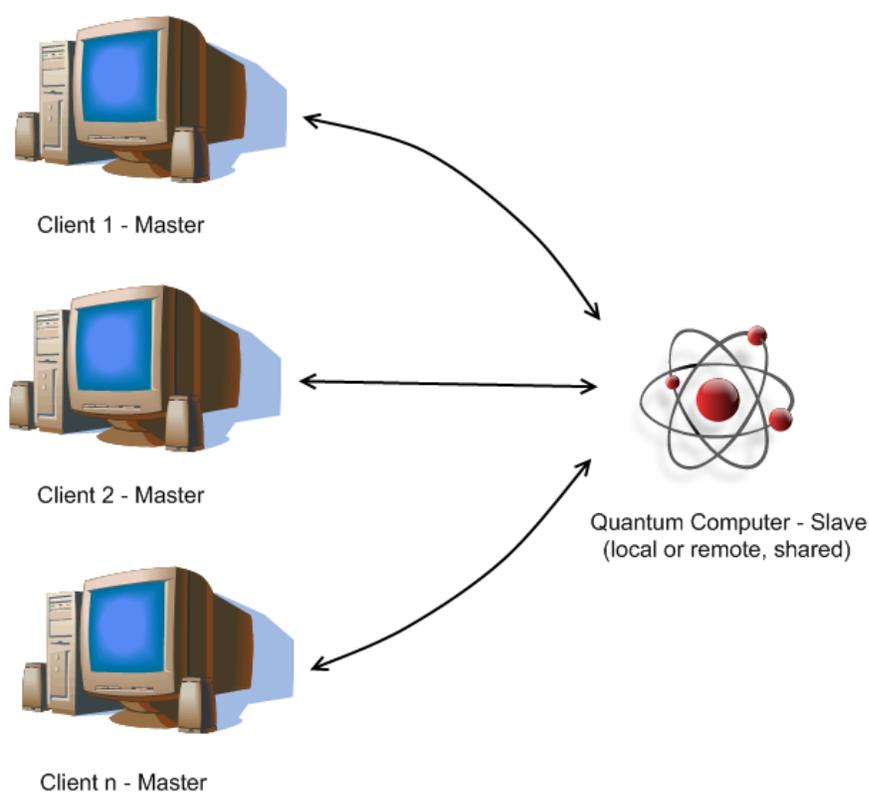

Figure 114. Knill's QRAM model.

Knill's QRAM model [3] can be expanded to better decouple tasks that users do not need to be concerned with when focusing on quantum computation. This can be done by providing an interface to the quantum computer called a "quantum controller" that all requests for the quantum resource are directed to, as shown in Figure 115. While this controller may be implicit in the QRAM model, by explicitly specifying it the appropriate tasks can be allocated to it. Figure 115 shows how this quantum controller integrates into the QRAM model. This controller receives requests from clients, which schedules all quantum requests and performs lower level tasks. Thus it is the slave to the clients, but master of the quantum computer. This quantum controller could also provide a consistent interface for different physical implementations of a quantum computer, perhaps by communicating through a yet to be specified protocol.

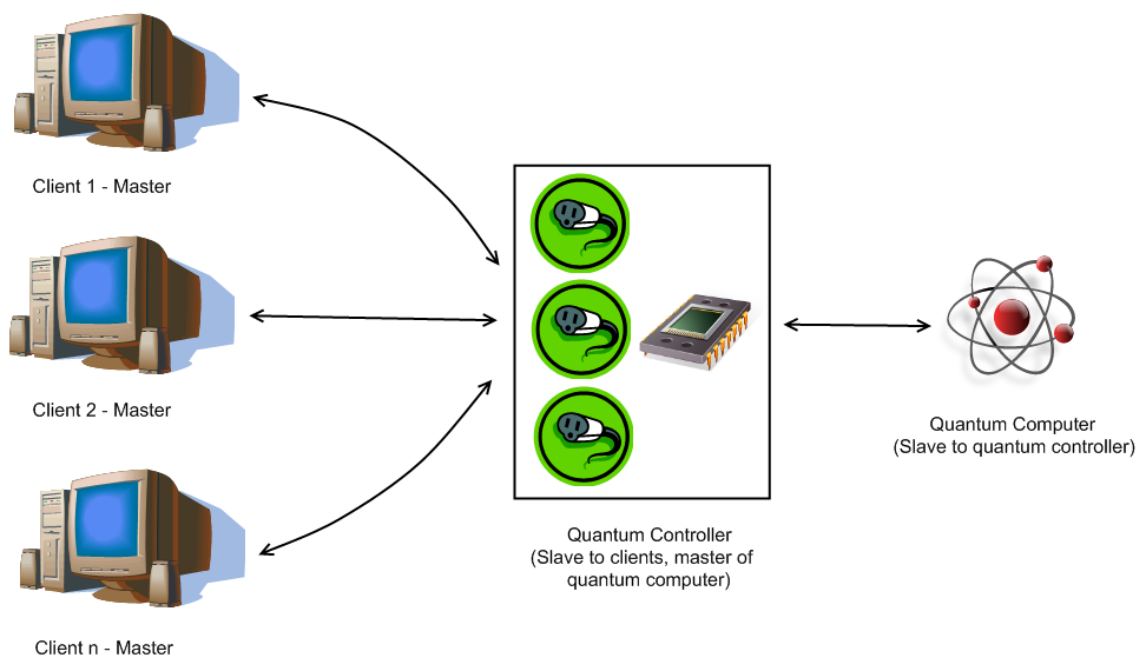

Figure 115.  The QRAM model expanded to include a quantum controller.

In the case of Cove, this means that users of the framework do not have to be concerned with dealing with these issues from within their code. In particular, the quantum controller will have the following tasks allocated to it, although these apply to the extended QRAM model and not just Cove:

- Scheduling of multiple requests for quantum computations. The quantum computer will have finite resources, so these resources have to be shared and scheduled among all users. The users may be other computers accessing the quantum computer as a remote resource, or it may be multiple programs on a single computer.

- Queuing up requests. Currently maintaining the state of quantum information over time is hard due to decoherence. The controller could queue up requests and then execute them all quickly when a measurement is needed. Take the example of initializing two qubits, performing a Hadamard on the first one, performing a CNot on them, then a measurement. If these commands each came in a minute apart[87] it would be hard to maintain the state due to decoherence with many current implementations of quantum computers. The controller would queue these up and execute them as quickly as possible and return the result once the measurement command is received.

- Performing optimizations. Instead of relying on the various clients to optimize their elementary quantum operations, the controller could do so. This means the optimization occurs in one central place and users do not have to be concerned with it. This is similar to how a modern compiler may optimize compiled code. There could be different optimizations favored as well, such as speed scheduling, although since this is an area for future exploration. Cove does not yet provide for it. The reason why this is allocated to the controller and not a quantum compiler is that the optimizations may be dependent on the physical implementation.

- Performing error correction. The controller is in charge of error correction, such as allocating extra physical qubits as needed while only presenting one logical qubit. The benefit is that users can work with an "ideal" quantum device by not having to write error correction into their programs– it is handled at a lower level as far as they are concerned.

- Hardware independence. As Bettelli has pointed out [29], quantum programming methods should be hardware independent. The quantum controller presents a consistent interface to any physical realization of a quantum computer. The specifics of this interface are beyond the scope of this dissertation.

- Dealing with any nonfatal communication or execution issues. A nonfatal execution issue might be the system unintentionally collapsing due to decoherence, and having to start over from the beginning. This starting over from the beginning would require the operations to be queued, although this would present problems if a partial measurement were already delivered in which case it might be a fatal error. This approach is much like how a packet in an FTP transfer can be resent without the knowledge[88] or intervention of the high level programmer. The resending of the packet is considered nonfatal, while the loss of the physical connection (say the network interface goes out) would be fatal.

For the case of a single user system, this quantum controller and computer could be built into the system. A classical example of the same concept might be using web services locally. Typically there is a good deal of infrastructure in place for handling remote requests, much like there would be for the quantum controller in the expanded QRAM model. That infrastructure could be in place on the local machine and used there instead of creating a new infrastructure just to run it locally. Hence the only difference between running locally or remotely is that running locally merely loops back to the machine on which the request was issued.

---

[87] A low rate might be caused by outside factors such as waiting for user input.

This expanded QRAM model does however prevent the use of quantum states prepared by other devices as originally envisioned by Knill [3]. For the purpose of most computations this limitation of is outweighed by the benefits of a clear decoupling between physical device performing computation and the program utilizing that resource. Multiple references to qubits are also dealt with by the controller, and the communication scheme between the clients and controller will take this into account since multiple registers from a single client may contain references to the same qubits as Bettelli describes [29]. A practical case would be creating a second logical register of one qubit from an existing two qubit register. There are then two logical registers, but one of the qubits is shared between them.

One way this could be done by assigning a unique identifier to each qubit, so the controller knows which qubits are being shared. So register 1 may consist of qubits $x$ and $y$, while register 2 consists of qubits $y$ and $z$. So any manipulation on qubit $y$ would in fact impact both registers. This is just one potential option and the details and alternates will not be explored further in this dissertation, but it is a good area for future work.

---

[88] Although decreased performance may be evidence of packet loss. The key point is that if performance isn't a concern then the high level programmer doesn't really have to do anything about it.

Thus an extra layer of abstraction, the quantum controller, is placed into Knill's QRAM model to deal with these various issues for all clients. By pushing these issues from the particular programming method (not even necessarily Cove) into the quantum controller they are reliably handled in one location instead of relying on the various clients to implement things such as error correction in the proper manner. Furthermore this means that users of Cove, or any other client, can deal with a quantum computer that behaves in a consistent and predictable[89] manner and are freed from the more tedious infrastructure tasks. This follows the behavior of existing classical computers where users of high level languages are not concerned with things such as CPU scheduling, memory errors, and nonfatal communication problems.

### 5.3 Common Tasks and Philosophy

This section discusses how some important common tasks are generally accomplished in Cove: applying operations and ordering of operations and registers. In addition to how these are accomplished, some of the philosophy behind the design is also discussed.

#### 5.3.1 Applying Operations in Cove

The prototype simulation supplied uses `GeneralSimulatedOperation` derived objects for operations instead of the more general `IQubitOperation` or `IQuantumOperation` interface from the base library. The reason for this is that the classical implementation wouldn't necessarily know how to apply operations from other implementations since they may be specified differently in different implementations.

---

[89] This is not meant to imply that the quantum computer is predictable!

The simulation uses matrices of complex numbers to maintain the state of operations and qubits. This means that the operations derived from `GeneralSimulatedOperation` can have their matrix extracted by the register and applied as was shown in 2.1.1.

This means that the `ApplyOperation()` method implemented in the prototype simulation may throw an exception at run time if it is passed an object not derived from `GeneralSimulatedOperation`. This behavior runs slightly counter to strongly typed languages such as C# which prefer to catch these types of errors at compile time instead of run time. This is not necessarily a bad thing since this is how dynamically typed languages such as Python operate. If a method accepting only `GeneralSimulatedOperation` derived classes were provided then it would catch invalid objects at compile time. However, doing so would mean that the classical simulation would no longer be interchangeable with an actual one because they would no longer share the same interfaces. Thus the decision has been made to defer some errors to run time in order to maintain interchangeability, but more important consistency, between different implementations.

The quantum register has a method to apply arbitrary operations, `ApplyOperation()`. As an example one would apply a Hadamard operation with the snippet `TestRegister.ApplyOperation(`Operations`.Hadamard)`. A more concise notation is also provided, as in `TestQubit.OperationHadamard()`. Some reasons on why the two different methods are supported:

- The `ApplyOperation()` method allows the user to pass in any object that implements `IQuantumOperation`. This means that the register works as is with user defined operations– no subclassing is needed to work with user defined operations. One can imagine users stringing together operations for various purposes, meaning that an algorithm could be performed just by passing in that user defined operation.

- If the operations were on only on a register itself as in `TestRegister.OperationHadamard()`, then the user would have to derive their own register class and implement the new operation– a potentially daunting refactoring task if the user has already written code using the supplied quantum register class.

- The extensible method reduces the coupling between the operations and qubits, per commonly accepted object oriented design principles where low coupling is preferred [106]. Although it should be noted that the two are still coupled through the concise method.

- By specifying "Operation" in the syntax, all available operations show up in intellisense tools and alphabetical lists together– showing the programmer exactly what is available for quantum operations on a register. If the operations were just methods on a register the user would have to be intimately familiar or consult the documentation to know which methods were applying quantum operations. This detracts from the goal of usability for commercial programmers.

Thus users are provided with an extensible method to call arbitrary operations, and a concise way to call the common ones.

Some other approaches pass the registers to operations, instead of operations to registers as in Cove. As is elaborated on in 6.2, this violates some object oriented principles and makes creating new operations more difficult.

### 5.3.2   Ordering of Registers and Operations in Cove

When manipulating registers and operations, it is often convenient to alter the order of qubits (registers) or the targets (operations). A good example of this is the CNot operation, which has a control qubit and a target qubit. Given two qubits, $q_0$ and $q_1$, the user may elect to use either as the control qubit, the two possibilities are shown in Figure 116. This can be done in a multiple of ways in Cove.

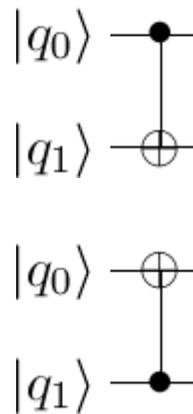

Figure 116. The two ways CNot can be applied to two qubits.

The following explains the various ways the control and target can be changed, as shown in Figure 117. A register of two qubits can be created (line 1), and the initial order will be $q_0$ as the first qubit and $q_1$ is the second. The register interface contains methods to apply all elementary operations to the register, so the case of the control qubit being the first is the default as shown in line 3. The method to apply the CNot operation is also overloaded to explicitly specify the control and target qubits (line 5). (All numeric parameters in this example are integers.) An alternate is to logically reverse the qubits in the register and then apply the default CNot (line 7). Instances of operations can also be created (line 9), and then manipulated before applying them to the register (line 10). Line 13 shows that the contructor is overloaded to explicitly set the control at target when instantiated. The control and target can also be changed after construction, as shown in line 16.

```
1.      IQuantumRegister TestRegister = new QuantumRegister(2);

2.      //default, control qubit is q0
3.      TestRegister.OperationCNot();

4.      //the control qubit is q1
5.      TestRegister.OperationCNot(1, 0);

6.      //change the order, then apply the operation. control qubit is q1
7.      (TestRegister.SliceReverse()).OperationCNot();

8.      //create a default instance of CNot, so control is q0
9.      IOperationCNot CNot = new OperationCNot();
10.     TestRegister.ApplyOperation(CNot);

11.     //create an instance of CNot, but change the targets when
12.     //created, so control is q1
13.     CNot = new OperationCNot(1, 0);
14.     TestRegister.ApplyOperation(CNot);

15.     //change the control back to q0 after it is created.
16.     CNot.SetIndexes(0, 1);
17.     TestRegister.ApplyOperation(CNot);
```

Figure 117. Various ways to switch the control and target of CNot.

Clearly there are many ways the control and target can be set, but why so many? The ways are broken into three categories. The first way is to call the method on the register (line 5). These allow for elementary operations to be applied without the need to create instances of operations. This reduces the amount of code and keeps things clear. The second way deals with logically changing the ordering of qubits, in this case reversing (line 7). If one thinks of the register as an array of qubits then this isn't too surprising: the order of elements in an array is frequently altered in classical computation, so why not also in quantum computation? The final method is to create instances of operations, manipulate them, and then apply them to the register (lines 13 and 14). This one requires the most code, but it has an important use– it can be used to build up quantum algorithms. Thus methods can be created whose output is an ordered list of operations that can be applied to any register. This is the technique used to build up complex algorithms such as factoring and is elaborated on in 5.4.3.

**5.4      Examples of Quantum Computation in Cove**

Three examples have been a theme throughout this dissertation: the quantum coin toss, entanglement, and Shor's algorithm. In chapter 2 these examples were worked through to understand what is physically happening and why these are not possible classically, while 4.3 detailed the quantum circuits for these examples. These quantum circuit diagrams explicitly detail what elementary operations must be applied to registers to carry out the examples. As stated in the proof criteria, these examples are used to show the completeness and correctness of Cove as opposed to tediously working through examples for each functionality property. The coin toss and entanglement examples are meant to be simple cases, while Shor's algorithm demonstrates how Cove could be used to carry out a practical task– in this case factoring.

All of these examples make use of the prototype implementation of the Cove local simulation, which is a simulation of a quantum computer that runs on the local computer. Nonetheless the code is implementation independent, the local simulation implementation is merely used to execute it on to show that the interfaces can be utilized effectively. The preferred method of writing code in Cove is to "program to the interfaces" in order to make the code that follows as implementation independent as possible. That is, implementation objects are created but assigned to variables of the interface type, as shown in Figure 118.

```
1.    IQuantumRegister MyRegister = new QuantumRegister(2);
```

Figure 118. Assigning an implementation object to an interface type.

All of these examples are written using this "programming to interfaces" approach.

### 5.4.1 *Quantum Coin Toss*

Figure 119 shows the code snippet for carrying out the tossing of any number of coins within Cove, as specified by the local variable `NumberOfCoins`. Unlike Figure 123, this is broken into multiple lines for readability. The `Result` is initialized to null and is assigned in line 9.

```
1.    int NumberOfCoins = 8;
2.    IQuantumRegister Coins = new QuantumRegister(NumberOfCoins);
3.    ClassicalResult Result = null;

4.    //toss the coins once
5.    Coins.OperationHadamardAll();

6.    //toss the coins twice
7.    Coins.OperationHadamardAll();

8.    //observe and display results
9.    Result = Coins.Measure();
10.   Console.WriteLine(Result.ToString());
```

Figure 119. Tossing quantum coins in Cove.

As this is the first detailed example of quantum computation using Cove, it will be dissected line by line. The numbers in this list correspond to the line numbers in Figure 119.

1. Creates a local variable for the number of coins in the register that will be tossed. For simulations an exponential amount of memory is required, per the limits of quantum a simulation (section 2.1.9), so larger values should be avoided. The maximum number of qubits allowed in a register in the supplied prototype simulation is 62 due to addressing limitations. However, as far as the interface is concerned, there is no limitation.

2. An instance of a simulated quantum register is created that contains 8 qubits, and is initialized to the default state of all qubits being 0: $\left|00000000\right\rangle$. This object is then assigned to the local variable `Coins`, which implements `IQuantumRegister`. Since the type is an interface from the Cove base class, only operations defined in the base interfaces (`Cove.Base`) can be accessed without an explicit cast back to the simulated type, `QuantumRegister`. If only the methods in the interface are used from here forward on `Coins` then any implementation could be substituted at this line[90], including one that runs on a quantum computer.

3. A variable for the result, `Result`, of this computation is created and initialized to null. This `ClassicalResult` type is essentially a wrapper for an array of classical bits, which are the output of measurement of any register. This wrapper allows these bits to be converted and manipulated in a variety of ways. This class can also be extended by users to add additional functionality if desired.

4. A comment for the next line of code.

5. Coins (qubits) are tossed by performing a Hadamard operation on them. All of the coins in the register are tossed at the same time by calling the `OperationHadmardAll()` method on the register. This is one of several ways that the Hadamard operation could be performed on all qubits. One example that is longer but more flexible, would be to call `ApplyOperationAll(new OperationHadamard())` on the register, which applies a single qubit operation to all qubits– that is Hadamard operation in this case. If a user defined operation were created it could also be applied to all qubits using this alternate method. With these "All" methods an operator is constructed that applies the single qubit operation to all applicable qubits simultaneously.

An alternate to using the "All" operations is to use the iterator to iterate over the register and apply the single qubit operation to each qubit in a `foreach` loop, as shown in Figure 120. The difference between this and the "All" operation is that this applies the operation to the qubits one at a time. Figure 121 gives the circuit diagram for applying Hadamard to four qubits at once, while Figure 122 shows how the operation is applied one at a time using the iterator, which also takes longer.

---

[90] Alternately one could use a `using` statement and just change that to make it even easier to swap implementations.

At first glance one might think that a method `OperationHadmard()` would apply the operation to all the qubits. This method exists, but it only applies the Hadamard operation to a single qubit; so the register must be of 1 qubit or the target qubit must be specified. This convention makes the single qubit operations, such as Hadamard, more consistent with multiple qubit operations such as CNot. There is not corresponding "All" call for multiple qubit operations since it would be ambiguous: what would a CNot (2 qubit operation) over 3 qubits mean? So multiple qubit operations are similar to `OperationHadmard()`, the user must apply it to a two qubit register or explicitly specify the target and control qubits as parameters (the method is overloaded).

6. A comment for the next line of code.

7. The coins are tossed a second time by performing the Hadamard operation on all of them again.

8. A comment for the next line of code.

9. The register is measured, which collapses it to a state that can represent by classical information. The return from this `Measure()` method is a new instance of a `ClassicalResult` object, which is assigned to the local variable created earlier. Since the register has collapsed to a state that is not in superposition, the `Measure()` method can be called any number of times to generate new `ClassicalResult` objects. (Although there is little point in actually doing so.)

10. The classical results are converted to a string and displayed to the console. The default string output is the string of bits, so in this example the output will be eight zeros: "00000000".

```
1.    IQuantumRegister TestRegister = new QuantumRegister(4);
2.    foreach(IQuantumRegister CurrentQubit in TestRegister)
3.        CurrentQubit.OperationHadamard();
```
Figure 120. Iterating over a register to apply a single qubit operation.

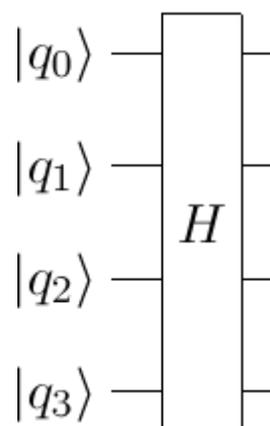

Figure 121. Applying the Hadamard operation to four qubits at once.

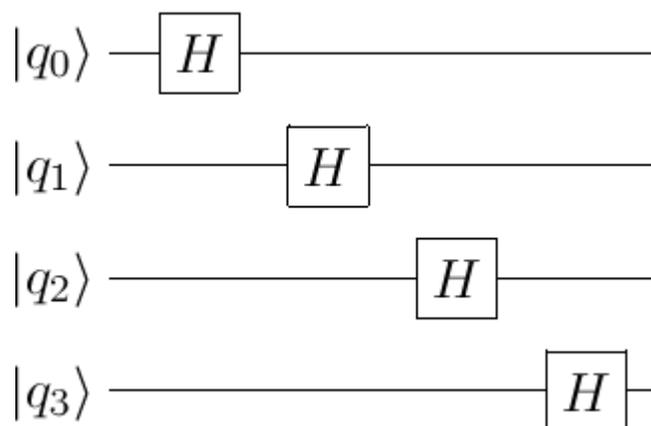

Figure 122. Apply the Hadamard operation to four qubits one at a time.

This quantum coin toss example demonstrates how Cove can create a number of qubits and put them into and take them out of superposition. All of the operations within Cove can also be chained together, meaning it is possible to carry out the above example in one line of code. Chaining essentially means that a reference a modified object is returned after a method call. Hence `Register.X().Y().Z()` means that method `X` is applied, then `Y`, and finally `Z`. This means that the result from each method is the register after the operation is applied (Figure 123). The output to the console could also be included.

```
1.    ClassicalResult Result = ((IQuantumRegister)
2.        new QuantumRegister(NumberOfCoins))
3.        .OperationHadamardAll()
4.        .OperationHadamardAll()
5.        .Measure();
```
Figure 123. Carrying out quantum coin tossing in one line of code.

### 5.4.2 Entanglement

A second simple example of quantum computation that has been a theme is entanglement. Entanglement illustrates several things that are not possible in classical computation. One thing that is not possible classically is that an operation on a qubit, such as measurement, can influence the state of other qubits. Another example is that the CNot operation can do more than just toggle the second qubit: it puts it into superposition in this example:

```
1.    IQuantumRegister cTestRegister = new QuantumRegister(2);
2.    ClassicalResult cResult = null;

3.    //Entangle via Hadamard followed by CNot
4.    (cTestRegister.SliceTo(0)).OperationHadamard();
5.    cTestRegister.OperationCNot();

6.    //measure and display the result
7.    cResult = cTestRegister.Measure();
8.    Console.WriteLine("Result: " + cResult.ToString());
```
Figure 124. Entangling two qubits in Cove.

As with the prior example of tossing quantum coins, this example is best worked through line by line in detail. The numbers in this list correspond to the line numbers in Figure 124.

1. An instance of the simulated register with two qubits is created.

2. A local variable is created to capture the result of measurement.

3. A comment for the following two lines of code.

4. The term "slice" comes from Python and means to get a subset of a list. So the various slice operations are `SliceTo`, `SliceFrom`, `Slice`, `SliceReverse`, `SliceSubset`, and `SliceReorder`. As their names imply `SliceTo` and `SliceFrom` get a slice of the register from the beginning to the specified index and from the specified index to the end, respectively. `Slice` gets a slice of a register by specifying the start and ending indexes. `SliceReverse` simply returns the register with the qubits reversed. `SliceSubset` returns an arbitrary subset by specifying exactly which indexes should be included in the slice. `SliceReorder` reorders the register in an arbitrary order.

   An alternate way to perform this same line of code would be to call `cTestRegister.OperationHadamard(0)`. The `OperationHadamard()` method on the register is overloaded where you can also specify an index, effectively doing the slice and application in one call. It is also overloaded to take an array of indexes to apply the single qubit operation to multiple qubits.

5. The CNot operation is applied to the register. Since the control qubit is in superposition, this is what entangles the two qubits. The entanglement example in section 2.1.4 walks through the details of what is happening. After this operation is applied the register is in the state $\frac{1}{\sqrt{2}}|00\rangle + \frac{1}{\sqrt{2}}|11\rangle$.

   Similar to the application of the Hadamard, this is also overloaded to take parameters. Specifically it is overloaded where the control and target indexes can be specified. So the same call utilizing the overloaded method would be `cTestRegister.OperationCNot(0, 1)`, since the first integer is the index of the control qubit in the register and the second is the index of the target qubit in the register.

6. The register is measured, which collapses it to $|00\rangle$ or $|11\rangle$. Even if one of the qubits were measured, it would still collapse to this state due to entanglement–the state of the two qubits are not independent. This measuring of one of the qubits, in this case the first, would be: `(cTestRegister.SliceTo(0)).Measure()`. Just as operations can be applied to slices, measurement can also be performed on slices.

   Like the application of operations in the previous steps, the measurement method is also overloaded where the indexes to be measured can be specified. It is overloaded to take a single integer to measure one qubit in the register, or an enumerable list of integers. So the equivalent of measuring one qubit would be `cTestRegister.Measure(0)` and measuring both qubits (0 and 1)

explicitly would be

```
cRegister.Measure(new int[] {0, 1}).
```

7. The result of measurement is displayed to the console as a bit array, either "00" or "11".

The slicing of registers and applications of operations to them raises the question: what if the sizes don't match? In Figure 124 the Hadamard operation, which is a single qubit operation, is applied to a register that consists of a single qubit. What if there is more than one qubit? What if the CNot operation is applied to three qubits, or one? The behavior is to apply the operations to the lowest indexed qubits. So a CNot over three qubits would just perform the CNot over the first two qubits, the third would be left untouched. Of course, the user can also explicitly specify the targets of the register. If the register is too small for the operation than a `SizeMismatchException` is thrown.

Users can also create instances of operations and specify the explicit targets in those instances, then apply them to registers. So it is possible for a user to create a CNot instance whose targets are indexes 2 and 3. If this is applied to a 2 qubit register, a `SizeMismatchException` will be thrown because indexes 2 and 3 do not exist.

For the case of applying a CNot operation to two qubits of a three qubit register, the user has several choices[91]. The first is that they can take a slice of the desired two qubits and then apply the CNot operation, shown in Figure 125 line 3. The second option is that the user can create an instance of the CNot operation, then change it to specify the indexes it applies on, then apply it. By using this method an instance of the CNot operation is being created, and that instance is changed to set the control and target index. This is shown in Figure 125 lines 7 − 9. Finally, the user can use the overloaded method that takes the indexes it applies too, as given by line 12 in Figure 125. In these alternate cases the `SizeMismatchException` exception is thrown if the indexes specified are out of bounds for the register. Figure 125 illustrates all these possible ways of apply a CNot operation to a 3 qubit register, in the same manner, with the control qubit being at index 0 of the register and the target register being at index 2. Of course the simplest case is line 14, just applying it to the first two qubits implicitly.

```
1.    IQuantumRegister TestRegister = new QuantumRegister(3);

2.    //apply by taking a slice of the last two qubits.
3.    (TestRegister.SliceFrom(1)).OperationCNot();

4.    //apply by creating an instance of CNot and expanding it
5.    //to three qubits before applying it. The control will
6.    //be index 0 and the target will be index 2.
7.    IOperationCNot ExpandedCNotOperation = new OperationCNot();
8.    ExpandedCNotOperation.SetIndexes(0, 2);
9.    TestRegister.ApplyOperation(ExpandedCNotOperation);

10.   //finally the overloaded method to apply CNot to a register
11.   //can be used to explicitly specify the control and target.
12.   TestRegister.OperationCNot(0, 2);

13.   //simplest case, apply it to indexes 0 and 1
14.   TestRegister.OperationCNot();
```

Figure 125. Examples of apply CNot to a three qubit register.

### 5.4.3   *Shor's Algorithm*

The construction of the complex operations necessary to perform Shor's algorithm for factoring have been given in 4.3.3. This section details how the quantum part of the factoring is carried out and built using Cove. However, for brevity's sake the entire quantum portion is not detailed. Once the general philosophy is outlined it can easily be seen how the more complex operations are constructed from the simpler ones. The full source of the current state of the algorithm is available on the website (see appendix A).

---

[91] Under the covers in the local simulation prototype the CNot operation is created, reordered and expanded to match the register size, and then applied.

When examining circuit diagrams it is clear that quantum algorithms are nothing more than the application of elementary operations to specific qubits in a particular order. Cove takes the approach that these common algorithms are encapsulated into methods (in the `IQuantumAlgorithms` interface), which return those operations. Those arrays of operations can then be applied to any register. The inputs to these methods are the particular targets of the operations.

This is best illustrated by example. Figure 126 shows how the sum operation is constructed from two CNot operations. Thus there are three inputs when getting the operations to perform sum: carry, *x*, and *y*. Once sum is applied the result is in the *y* qubit. Carry (Figure 127) is a slightly more complex example, as it needs a fourth ancilla (scratch) qubit.

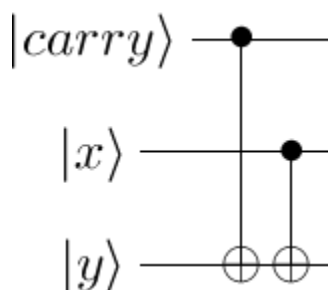

Figure 126. Sum, from elementary operations.

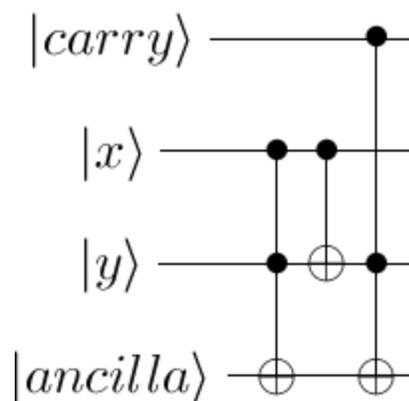

Figure 127. Carry, from elementary operations.

Notice that the targets of these operations could really be any qubits in a register. There is no specific requirement that for a sum operation that the first qubit must be carry for example[92]. The $33^{rd}$ qubit in the register could very well be the carry, and $x$ the $5^{th}$ qubit and $y$ the $19^{th}$. Since all operations in Cove target specific qubits of a register, the output from the methods to generate these are instances of operations that target those specific qubits. The method signatures for carry and sum are given in Figure 128 and Figure 129, respectively, along with the XML documentation that is part of all the source code.

---

[92] Complicating things further, the "first" qubit could very well be any other qubit due to logical reordering of the register.

```
1.    /// <summary>
2.    /// Return the operations that perform Sum.
3.    /// </summary>
4.    /// <param name="CarryIndex">The index of the carry qubit.
5.    /// This remains unchanged
6.    /// after the operations are applied.</param>
7.    /// <param name="XIndex">The index of the X qubit.
8.    /// This remains unchanged
9.    /// after the operations are applied.</param>
10.   /// <param name="YIndex">The index of the Y qubit.
11.   /// CarryIndex + XIndex + YIndex
12.   /// (mod 2 addition)</param>
13.   /// <returns>The quantum operations that perform
14.   /// sum over the specified qubits.</returns>
15.   /// <exception cref="DuplicateIndexesException">Thrown
16.   /// if any of the indexes passed in are the same. All
17.   /// indexes must be unique.</exception>
18.   List<IQuantumOperation> Sum(int CarryIndex, int XIndex,
19.   int YIndex);
```

Figure 128. Interface signature of sum.

```
1.    /// <summary>
2.    /// Return the operations to perform the carry gate.
3.    /// </summary>
4.    /// <param name="CarryIndex">The index of the carry qubit.
5.    /// Remains unchanged after
6.    /// the operations are applied.</param>
7.    /// <param name="XIndex">The index of the X qubit.
8.    /// Remains unchanged after
9.    /// the operations are applied.</param>
10.   /// <param name="YIndex">The index of the Y qubit. On
11.   /// output this will be a + b (mod 2
12.   /// addition)</param>
13.   /// <param name="AncilliaIndex">The index of the ancillia
14.   /// (scratch) qubit. On output this will be
15.   /// (CarryIndex)(XIndex) + (XIndex)(CarryIndex)
16.   /// + (YIndex)(CarryIndex)
17.   /// (mod 2 addition)</param>
18.   /// <returns>The operations to
19.   /// perform carry.</returns>
20.   /// <exception cref="DuplicateIndexesException">
21.   /// Thrown if any of the indexes are
22.   /// duplicates.</exception>
23.   List<IQuantumOperation> Carry(int CarryIndex, int XIndex,
24.   int YIndex, int AncilliaIndex);
```

Figure 129. Interface signature of Carry.

While specific to the local simulation prototype implementation, it is useful to see how these two algorithms are implemented. The implementation illustrates how a user could encapsulate any set of operations in a method, and thus make the operation set applicable to any register. The XML comments are the same as the preceding figures, and have been omitted from these, although they appear in the source. Figure 130 and Figure 132 give the implementations. The inverse operations, denoted sum[-1] and carry[-1], are constructed by reversing the output from these methods[93] (Figure 131).

```
1.    public List<IQuantumOperation> Sum(int CarryIndex, int XIndex,
2.    int YIndex)
3.    {
4.        List<IQuantumOperation> listRetVal
                = new List<IQuantumOperation>();

5.        //make sure duplicates are not specified
6.        if ((CarryIndex == XIndex)
7.        || (CarryIndex == YIndex) || (XIndex == YIndex))
            throw new DuplicateIndexesException(
                "Unique indexes must be specified.");

8.        //build up the operations to apply.
9.        listRetVal.Add(new OperationCNot(CarryIndex, YIndex));
10.       listRetVal.Add(new OperationCNot(XIndex, YIndex));

11.       return listRetVal;
12.   }
```

Figure 130. Implementation of Sum.

```
1.    /// <summary>
2.    /// Return the operations to perform the inverse carry.
3.    /// </summary>
4.    /// <param name="CarryIndex">The index of the carry qubit.
5.    /// Remains unchanged once the operations are applied.</param>
6.    /// <param name="XIndex">The index of the X qubit.
7.    /// Remains unchanged once the operations are applied.</param>
8.    /// <param name="YIndex">The index of the Y qubit.
9.    /// After the operations are applied this will be
10.   /// X + Y (mod 2 addition)</param>
11.   /// <param name="CarryPrimeIndex">The index of the carry
12.   /// prime qubit. Will be  x(x + y) + yc + c' on output.</param>
13.   /// <returns>The operations to apply the carry inverse.</returns>
14.   /// <exception cref="DuplicateIndexesException">Thrown if
15.   /// any of the indexes specified are duplicates.</exception>
16.   public List<IQuantumOperation> CarryInverse(int CarryIndex,
17.   int XIndex, int YIndex, int CarryPrimeIndex)
```

---

[93] As the inverse of an operation is just the same steps performed in reverse.

```
18.   {
19.       List<IQuantumOperation> listRetVal
20.           = Carry(CarryIndex, XIndex, YIndex, CarryPrimeIndex);

21.       //The carry inverse is just the carry in reverse.
22.       listRetVal.Reverse();

23.       return listRetVal;
24.   }
```

Figure 131. Implementation of carry$^{-1}$.

```
1.    public List<IQuantumOperation> Carry(int CarryIndex, int XIndex,
2.    int YIndex, int AncilliaIndex)
3.    {
4.        List<IQuantumOperation> listRetVal
5.            = newList<IQuantumOperation>();

6.        //make sure all the indexes are unique
7.        if ((CarryIndex == XIndex) || (CarryIndex == YIndex)
8.        || (CarryIndex == AncilliaIndex) || (XIndex == YIndex)
9.        || (XIndex == AncilliaIndex) || (YIndex == AncilliaIndex))
10.           throw new DuplicateIndexesException(
11.               "All indexes passed must be unique.");

12.       //build the operations to return
13.       listRetVal.Add(new OperationToffoli(XIndex,
14.           YIndex, AncilliaIndex));
15.       listRetVal.Add(new OperationCNot(XIndex, YIndex));
16.       listRetVal.Add(new OperationToffoli(CarryIndex,
17.           YIndex, AncilliaIndex));

18.       return listRetVal;
19.   }
```

Figure 132. Implementation of Carry (method description omitted).

These sum, carry, and carry inverse operations can then be pieced together to create an *n* qubit adder. The circuit diagram is given in Figure 133 while the local simulation implementation is given in Figure 134.

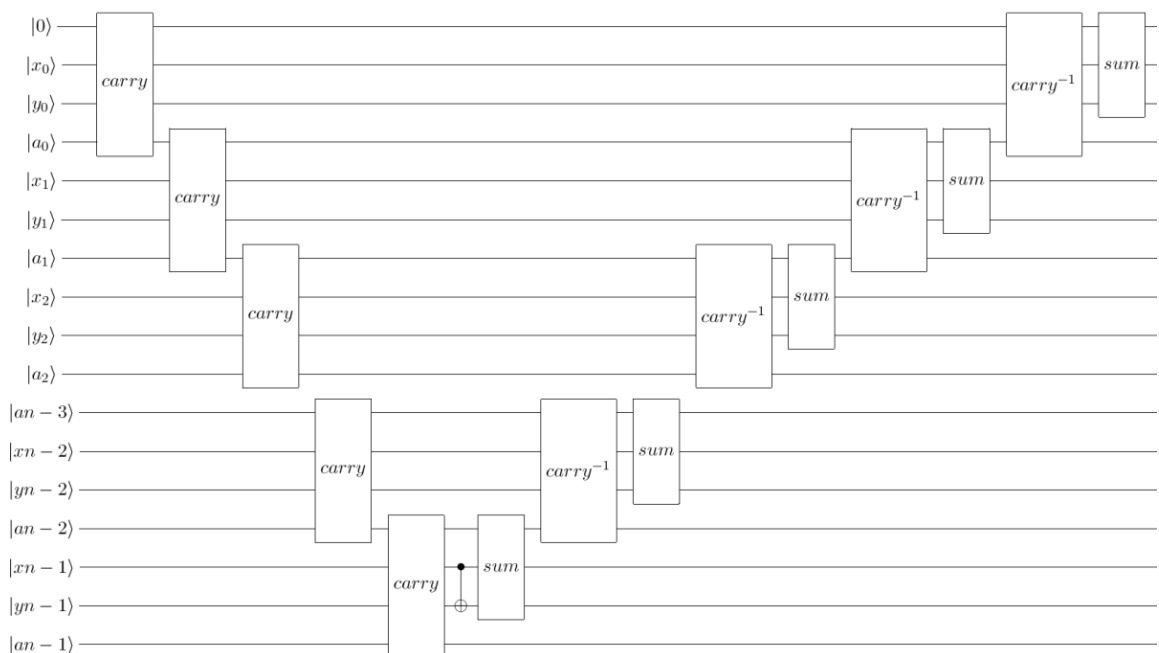

Figure 133. The n qubit adder circuit diagram.

```
1.    /// <summary>
2.    /// Return the operations needed to perform Add
3.    /// over two registers of equal size.
4.    /// </summary>
5.    /// <param name="XIndexes">The indexes of the X register.
6.    /// These remain unchanged
7.    /// once the operations are applied.</param>
8.    /// <param name="YIndexes">The indexes of the Y register.
9.    /// These contain the result after the operations are applied,
10.   /// along with the last ancillia qubit</param>
11.   /// <param name="AncilliaIndexes">The indexes of the ancillia
12.   /// qubits, which should be initialized to |0>. There should
13.   /// be one more ancillia qubit than there are
14.   /// X or Y qubits. The result will be in the YIndexes and
15.   /// the last ancillia index.</param>
16.   /// <param name="ResultIndexes">The YIndexes and last
17.   /// ancillia index contain the result,
18.   /// but this parameter will be populated with them explicitly
19.   /// for ease of use.</param>
20.   /// <returns>The operations to apply add n.</returns>
21.   /// <exception cref="ArgumentNullException">Thrown if any of
22.   /// the indexes passed
23.   /// in are null.</exception>
24.   /// <exception cref="DuplicateIndexesException">Thrown if any
25.   /// of the indexes specified
26.   /// are duplicates. All indexes must be unique.</exception>
27.   /// <exception cref="SizeMismatchException">Thrown if
28.   /// XIndexes and YIndexes are not of
29.   /// equal length, or if AncilliaIndexes is not one larger
```

```csharp
30.    /// than XIndexes and
31.    /// YIndexes.</exception>
32.    public List<IQuantumOperation> AddN(int[] XIndexes,
33.    int[] YIndexes, int[] AncilliaIndexes, out int[] ResultIndexes)
34.    {
35.        List<IQuantumOperation> listRetVal
36.            = new List<IQuantumOperation>();
37.        Dictionary<int, bool> dictUsedIndexes
38.            = new Dictionary<int, bool>();
39.        List<int> listResultIndexes = new List<int>();

40.        //first verify nulls and sizes
41.        if ((XIndexes == null) || (YIndexes == null)
42.        || (AncilliaIndexes == null))
43.            throw new ArgumentNullException(
44.                "Cannot pass null indexes");
45.        if (XIndexes.Length != YIndexes.Length)
46.            throw new SizeMismatchException(
47.                "XIndexes and YIndexes must contain
48.                an equal number of elements.");
49.        if ((XIndexes.Length + 1) != AncilliaIndexes.Length)
50.            throw new SizeMismatchException("The AncilliaIndexes
51.                must have one more index than the XIndexes
52.                and YIndexes.");

53.        //next make sure there are no duplicate indexes
54.        //by going through them  and keeping track of what has
55.        //been used in a dictionary. (Also populate the
56.        //result indexes)
57.        foreach (int iCurIndex in XIndexes)
58.        {
59.            if (dictUsedIndexes.ContainsKey(iCurIndex) == true)
60.                throw new
                    DuplicateIndexesException(
                    string.Format("The index {0} is specified
                    more than once. All indexes must be unique.",
                     iCurIndex));
                else
                    dictUsedIndexes[iCurIndex] = true;

61.        }
62.        foreach (int iCurIndex in YIndexes)
63.        {
64.            if (dictUsedIndexes.ContainsKey(iCurIndex) == true)
65.                throw new
66.                    DuplicateIndexesException(string.Format(
67.                    "The index {0} is specified more than once. All
68.                    indexes must be unique.", iCurIndex));
69.            else
70.                dictUsedIndexes[iCurIndex] = true;
71.            //keep track of Y indexes, as they are part
72.            //of the result.
73.            listResultIndexes.Add(iCurIndex);
74.        }
75.        foreach (int iCurIndex in AncilliaIndexes)
```

```
76.          {
77.              if (dictUsedIndexes.ContainsKey(iCurIndex) == true)
78.                  throw new
79.                      DuplicateIndexesException(string.Format(
80.                      "The index {0} is specified more than once.
81.                      All indexes must be unique.", iCurIndex));
82.                  else
83.                      dictUsedIndexes[iCurIndex] = true;
84.          }                //end of check for duplicate indexes

85.          //populate the out indexes
86.          listResultIndexes.Add(AncilliaIndexes[
87.              AncilliaIndexes.Length - 1]);
88.          ResultIndexes = listResultIndexes.ToArray();

89.          //construct the necessary gates to the half way point.
90.          for (int iCurIndex = 0; iCurIndex < XIndexes.Length;
91.          iCurIndex++)
92.          {
93.              listRetVal.AddRange(this.Carry(
94.                  AncilliaIndexes[iCurIndex],
95.                  XIndexes[iCurIndex],
96.                  YIndexes[iCurIndex],
97.                  AncilliaIndexes[iCurIndex + 1]));
98.          }                //end for iCurIndex, to the half way point

99.          //CNot and Sum after the half way point
100.         listRetVal.Add(new OperationCNot(
101.             XIndexes[XIndexes.Length - 1],
102.             YIndexes[YIndexes.Length - 1]));
103.         listRetVal.AddRange(this.Sum(
104.         AncilliaIndexes[AncilliaIndexes.Length - 2],
105.             XIndexes[XIndexes.Length - 1],
106.             YIndexes[YIndexes.Length - 1]));

107.    //now carry inverses and sums to complete the gate
108.    //NOTE: the look starts at XIndexes.Length - 2, as
109.    //there is one less carry inverse
110.    //gate than there is carry. If looking at the circuit
111.    //diagram the bottom
112.    //(highest indexes) are Carry CNot Sum, then Carry
113.    //inverse starts at the next highest (lowest indexes)
114.    for (int iCurIndex = (XIndexes.Length - 2);
115.    iCurIndex >= 0; iCurIndex--)
116.    {
117.        listRetVal.AddRange(this.CarryInverse(
118.            AncilliaIndexes[iCurIndex],
119.            XIndexes[iCurIndex],
120.            YIndexes[iCurIndex],
121.            AncilliaIndexes[iCurIndex + 1]));
122.        listRetVal.AddRange(this.Sum(
123.            AncilliaIndexes[iCurIndex],
124.            XIndexes[iCurIndex],
125.            YIndexes[iCurIndex]));
126.     }
```

```
127.        return listRetVal;
128.   }
```

Figure 134. Implementation of the *n* qubit adder.

The remainder of the factoring algorithm is omitted and available on the website (Appendix A). From the examples given it can be seen that factoring is nothing more than building up more complex operations from simpler ones. The end result is merely a series of elementary operations that are applied to the register. Figure 135 shows the quantum circuit diagram for the quantum part of factoring, while Figure 136 shows how that circuit diagram is implemented.

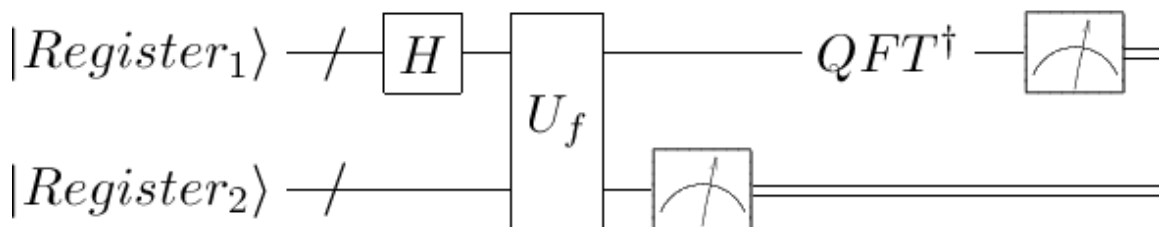

Figure 135. Circuit diagram of the quantum step in factoring.

```csharp
/// <summary>
/// This implementation of Shor's algorithm is based off
/// of pages 229 - 237 of S. Loepp and W. K. Wootters,
/// Protecting Information: From Classical Error Correction
/// to Quantum Cryptography, 1 ed. New York, NY: Cambridge
/// University Press, 2006.
/// </summary>
/// <param name="N">Number to factor.</param>
public ClassicalResult ShorsAlgorithm(int N)
{
    int MSize = Utilities.BitsToExpress(N);
    IQuantumRegister CompleteRegister
        = new QuantumRegister(3 * MSize);
    IQuantumRegister Register1
        = CompleteRegister.SliceTo(2 * MSize);
    IQuantumRegister Register2
        = CompleteRegister.SliceFrom((2 * MSize) + 1);
    List<IQuantumOperation> Operations = null;
    IQuantumAlgorithms Algorithms = new QuantumAlgorithms();

    //Step 1: Set registers to intitial state |00....0>
    CompleteRegister.SetAllQubitsTo(false);

    //Step 2: apply the Hadamard op to each qubit in
    //the X Register
    Register1.OperationHadamardAll();

    //Step 3: apply Uf
    Operations = Algorithms.FactoringUf(
        CompleteRegister.GetIndexes(0, 2 * MSize),
        CompleteRegister.GetIndexes((2 * MSize) + 1,
        (3 * MSize)),
        N);
    CompleteRegister.ApplyOperations(Operations);

    //Step 4: measure register 2
    Register2.Measure();

    //Step 5: Inverse QFT on register 1
    Operations = Algorithms.QuantumFourierTransform(
        Register1.GetIndexes());
    Register1.ApplyOperations(Operations);

    //Step 6: Measure register 1
    return Register1.Measure();
}
```

Figure 136. Interface signature of the quantum step in factoring.

## 5.5 Implementation of Cove in C#

Cove is a framework for quantum computing that extends classical languages to allow them to carry out quantum computation. By extending existing classical languages the focus of the framework is quantum computation itself; the designer does not have to spend resources trying to mimic already existing classical functionality that classical languages already provide. Part of the challenge of implementing the framework is preserving its generic nature and not imposing unnecessary constraints due to the language and implementation. The proof criteria (3.2) detail the requirements for what Cove must accomplish. The concept of programming to interfaces laid out first in section 5.4 have not been utilized in quantum computer programming previously. This concept helps to make implementations interchangeable, which may be important if there are different physical implementations of quantum computers.

In accordance with the idea of good readability, the code contains plenty of comments. Furthermore, the documentation itself (see APPENDIX A) is generated from many of these comments. By creating the documentation from the comments within the code it is much easier to keep the two in sync. This also provides complete documentation within the code itself.

### 5.5.1 Selection of C#

Cove is written in C#. The language C# was selected for several reasons. Currently C# is a popular language for developing commercial applications in its own right, but it is also closely related to C++ and Java. This means that the syntax is close to those languages, making the examples understandable to a large audience of programmers[94]. This framework is also built on version 3.5 of the .NET framework, which is the latest available at the time of this writing (2008-2009). The first widely distributed version of C# was in July 2000, and the latest standard is 2006 [133].

Aside from popularity, another reason C# was selected is that with .NET any assembly (essentially a .NET library) can be used by any compatible language. As an example, even though Cove is implemented in C# it can also be used from Visual Basic or IRON Python. Since the language and framework have been recently updated they provide some features not available in older languages. Finally, the author has years of professional experience with C derivatives. This has allowed for resources to be spent tackling the problem of quantum computation and not the nuances of a particular language.

The framework also provides a prototype implementation, which is a simulation of a local quantum computer. By providing an implementation instead of just the interfaces, users can write code that can be executed– although the exponential slow down still applies. It should be emphasized that this implementation is just a prototype in that there are certain pieces that have not yet been completed. Those methods that are not yet implemented are marked as throwing `NotImplementedException` in the documentation. This is a more beneficial method than a static list provided in this dissertation as a user can see what is not implemented as the framework evolves after this writing.

Nonetheless all of the examples within this dissertation can be executed successfully[95]. Since the factoring example is far from trivial, this shows that the

---

framework is complete enough to carry out some practical examples. One example of part of the implementation that is incomplete at the time of writing is the verification that operations are unitary before application, although this is clearly marked in the code as a missing component. The focus has been in implementing those parts that carry out computation in a generic manner instead of trying to optimize the simulation. However this unitary check is a high priority for future work, as it is needed to ensure that user defined operations are valid.

*5.5.2   C# Specifics*

Namespaces are also utilized in the C# implementation, helping to encapsulate various components along with standard object oriented coding practices [106]. When practical quantum computers are realized this helps minimize the effort needed to switch implementations, as Figure 137 illustrates. Ideally, the user merely has to change the "using" statement from the simulation to another implementation. Since all implementations support the interfaces, the code should be interchangeable. This is one of the primary reasons why the programming-to-interfaces concept is advocated throughout this dissertation.

---

95 Although the complete factoring example is still in progress, the components necessary to construct it are implemented as shown in 5.4.3.

```
1.    using Cove.Base;
2.    //this next line would be changed to the desired implementation
3.    using Cove.LocalSimulation;
4.
5.    //...
6.
7.    //Entangle via Hadamard followed by CNot
8.    (cTestRegister.SliceTo(0)).OperationHadamard();
9.    cTestRegister.OperationCNot();
```

Figure 137. Example of how namespaces can minimize effort to switch implementations.

The random number generation for the simulation warrants mention as it plays a key role. Almost always in computers the random number generators supplied are not truly random but pseudo-random, that is they give the appearance of being random but are not truly random. These random number generators typically take a seed value. Given the same seed value, two random number generators will give the same sequence of random numbers. This could clearly be a problem in the simulation. To help alleviate this problem there is a static singleton instance of the random number generator that is shared across all simulated qubits.

Static classes are those that do not need to be instantiated to use. C# does not allow for static classes to implement interfaces or derive from classes other than `System.Object`. This means that one cannot build an inheritance hierarchy for static classes. Thus regular classes have to be written in order to implement types specified in the base library. One idea was to create a static class that generated instances of the operations, so that a user could call `Operations.Hadamard()` to get an instance of the Hadamard operation. The intention was to have this static class implement an interface available in the base library so that it would be available in every implementation. This inheritance for a static class cannot be done, as has been described. This static class to generate the common operations has been supplied in the local simulation, but is not specified in the base class. Thus any operation can easily be applied to a register, and the argument could also be a user defined operation. Figure 138 shows how this static class can be used to apply an operation in line, which is really just a more condensed form of line 2. Methods to apply the common operations (listed in 3.2.4) are also provided, as in line 3.

```
1. ExampleRegister.ApplyOperation(Operations.Not);
2. ExampleRegister.ApplyOperation(new OperationNot());
3. ExampleRegister.OperationNot();
```

Figure 138. Various ways operations can be applied.

While this static `Operations` class is currently supported, the preferred method is still programming to interfaces, using lines 2 and 3 in Figure 138 (although this method has not been depreciated). As the framework evolves this static class needs to be specified somehow in the base class, or potentially removed– but this is another area for future work.

### 5.5.3 Custom Operations

To apply custom operations the user has two choices: create new operation classes, or extend the register class. When creating new operations they could be applied to registers using the `ApplyOperation()` method as shown in lines 1 and 2 of Figure 138. The alternate would be to derive a new quantum register class and add a method to create and apply that new operation, as in `OperationFoo()`. It is anticipated that creating custom operations would be the preferred method as the register class does not become bloated as new operations are created. This also has the added benefit of encapsulating the new operations. Of course there is no reason why a user couldn't create a new operation and then wrap the `ApplyOperation()` method to do both as shown in Figure 139. In this example the operations and derived register are inherited from the local simulation, as they would likely contain code that is specific to that implementation.

```
1.      public class OperationFoo :
        Cove.LocalSimulation.AbstractSimulatedQubitOperation
2.      {
3.          //class implementation here
4.      }

5.      public class UserDerivedRegister :
        Cove.LocalSimulation.QuantumRegister
6.      {
7.          public IQuantumRegister OperationFoo()
8.          {
9.              return this.ApplyOperation(new OperationFoo());
10.         }
11.         //rest of implementation here
12.     }
```

Figure 139. Wrapping a custom operation in a derived register

### 5.5.4    Primary Components of Cove

There are two primary components of Cove: the interfaces, and the implementations. Users should write their code against the interfaces in order to make it as implementation independent as possible. For this dissertation the interfaces have been designed and provided in a base library, referred to by the namespace `Cove.Base`. There is also a prototype implementation supplied. This is a simulation that runs on the local machine as opposed to remote machines. This implementation is referred to by the namespace `Cove.LocalSimulation`. The local simulation needs some components that are not provided by the language or the .NET framework, but are not necessarily specific to quantum computing. This library is referred to by the namespace

`Cove.ClassicalUtilities` and contains classes such as an arbitrary size matrix of complex numbers. In order to help ensure correctness, each of these three libraries also has a corresponding unit test library, which is referred to by the same name with ".`Test`" appended to the namespace and assembly. The final library is `Cove.Examples`, which contains much of the example code used within this dissertation and provides a console program that can be used for debugging. The minimum set of libraries needed to carry out the example code is: `Cove.Base`, `Cove.LocalSimulation`, and `Cove.ClassicalUtilities`. Figure 140 details these various components (libraries) and their dependencies. An arrow to a component means that is a dependency of the source of the arrow.

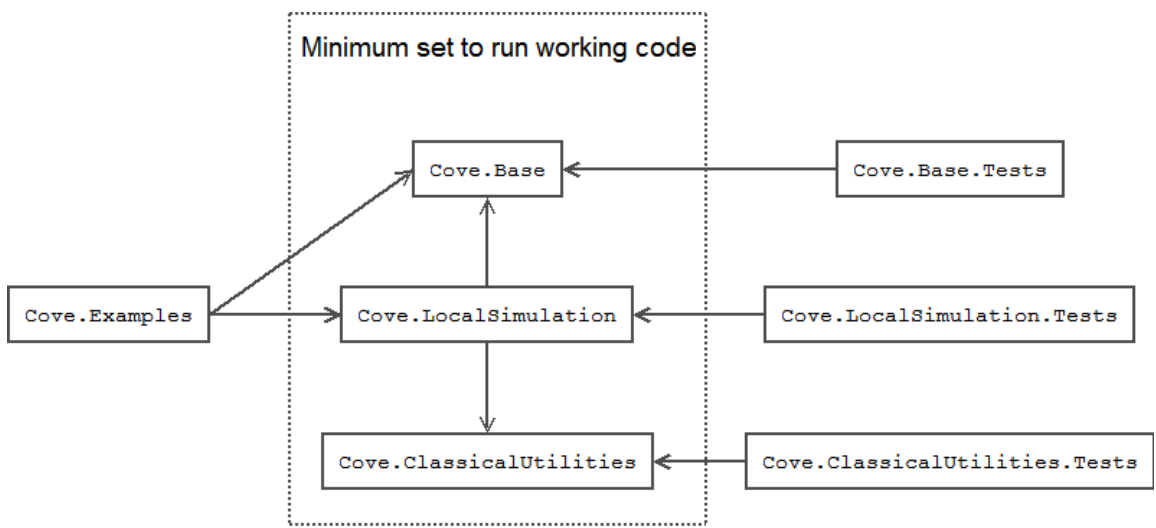

Figure 140. Cove components and dependencies.

UML diagrams are a commonly accepted method of detailing classes and their relations within the software engineering community. The following figures are the UML diagrams of the three core libraries. Admittedly these diagrams are hard to view within the space constraints of the pages of this dissertation. However, the entire image is embedded, so one viewing an electronic copy of this dissertation may zoom in or copy them into the image view of their choice. An alternate way to view these UMLs would be to download the images of these class diagrams from the website or within the code, how to access both is provided in the appendix.

Figure 141. Cove.Base UML diagram.

Figure 142. Cove.LocalSimulation UML.

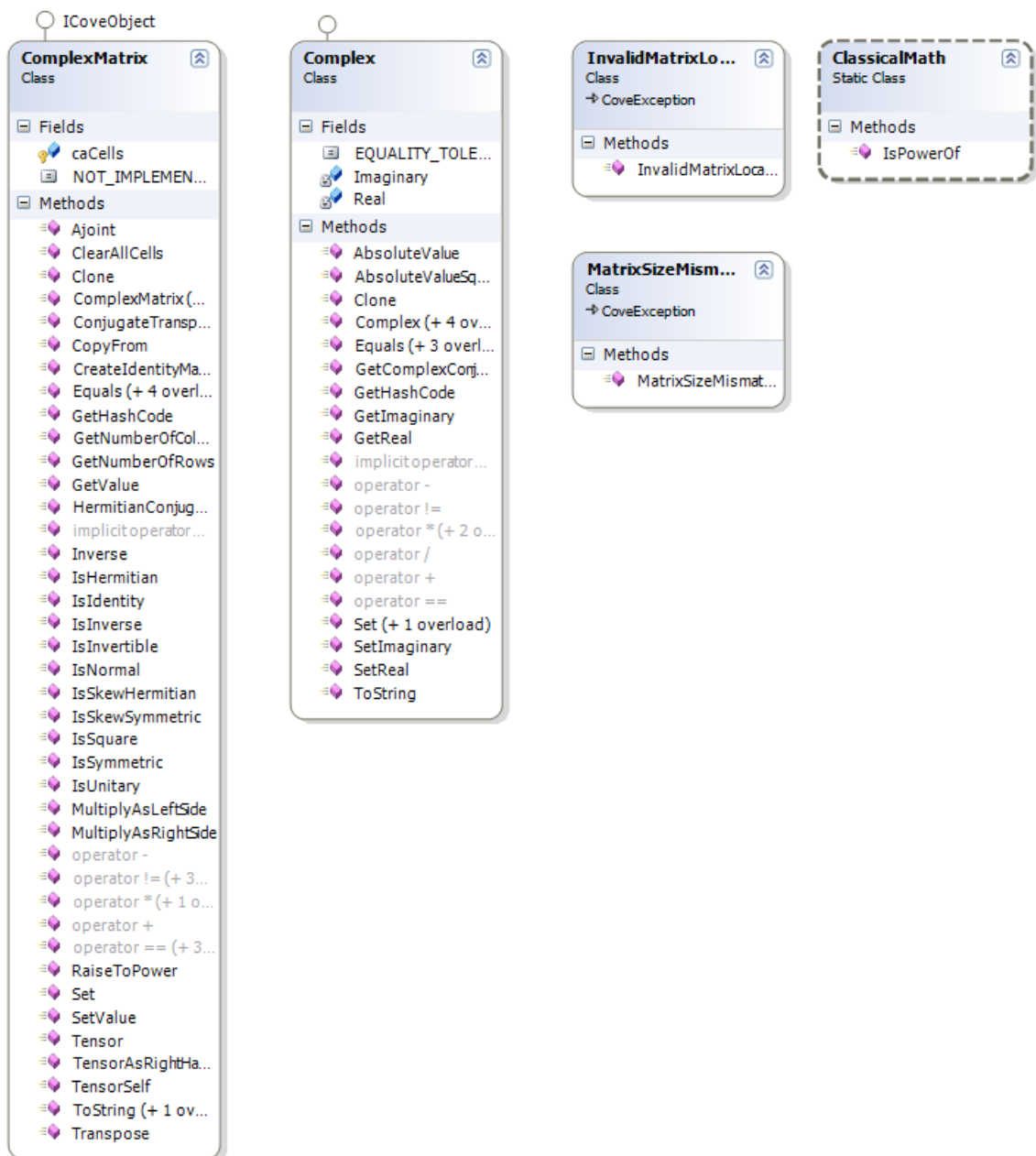

Figure 143. Cove.ClassicalUtilites UML diagram[96].

---

[96] A generic version of the matrix class was attempted at first, with the intention of utilizing instances of it with complex numbers. This became too unwieldy, so it was abandoned and replaced with the simpler, but less elegant, `ComplexMatrix` class.

*5.5.5    Simulation Implementation Details*

There are several implementation details of the simulation worthy of discussion as they may help others implementing simulations and provide insight into what is happening at lower levels of abstraction. This section is not meant to provide complete details on the simulation, just highlights. Detailed discussion of the major implementation challenges are covered in unpublished papers on the website as to not bog down the dissertation with those details.

Within the simulation quantum registers and operations are represented by matrices of complex numbers. At the time the project was started (late 2007) there was not an adequate complex number class or complex matrix class available, so they were written from scratch. As of late 2008 there were plans to include more useful numeric data types in Microsoft's F#, but they were not mature enough to utilize in Cove. The decision was made to only utilize .NET features to avoid users having to hassle with adding addition libraries– perhaps even of specific versions. For this reason the use of parallel for loops were also not utilized.

The underlying data type utilized in the `Complex` and `ComplexMatrix` classes is the double data type, which is an eight byte floating point that conforms to the IEEE 754 standard [134]. When writing tests for these two classes some accuracy problems were encountered, thus a tolerance was added when testing for equality. For the complex matrix class (a matrix of complex number) a number of operations necessary for quantum computation such as tensoring and conjugate transpose were also implemented.

Quantum registers represent logical collections of qubits. As has been mentioned previously it is possible for these registers to share qubits. Obviously if registers share qubits manipulations to one register may impact others. This presents an implementation challenge: how does manipulation of one register impact another? The solution to this is that each register contains its own list of exposed qubits, and any registers that share qubits share a reference to a single complex matrix that represents all of the qubits that might be shared between them. The list of exposed qubits is essentially a mapping from the qubits presented to the user to the potentially shared qubits.

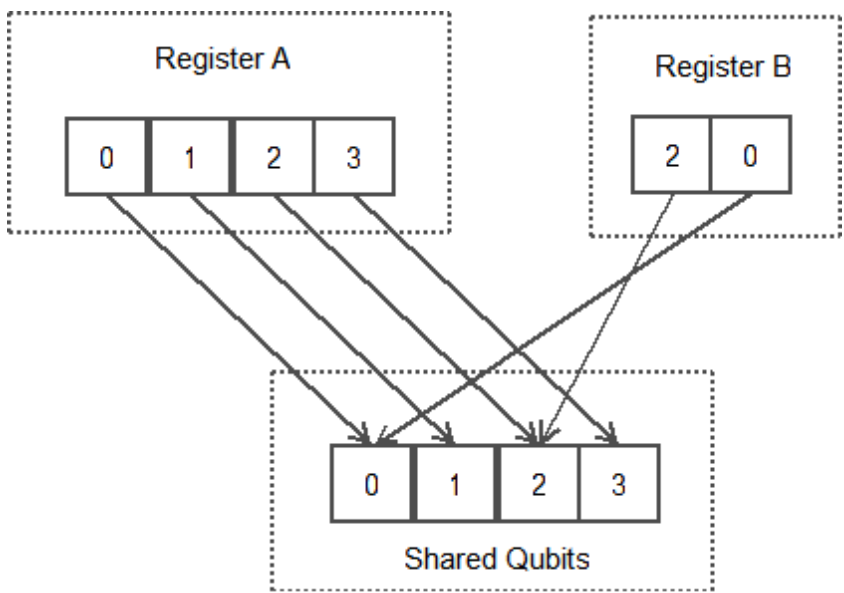

Figure 144. Exposed qubits to shared qubits.

Figure 144 illustrates an example of this mapping from exposed qubits to shared ones. For this figure Register A and B are simply different views of the shared qubits. In this example the shared qubits are always in the same order, the lowest index is the first qubit and the highest index is the last. Each register then contains a list of integers. The indexes in this list are the exposed qubits, while their values point to the indexes in the shared qubits. In this example register A is in the exact same order as the actual qubits.

So the zero-th exposed qubit maps to the zero-th shared qubit. Register B is not a direct mapping. In register B there are two qubits exposed. To the user these exposed qubits have index 0 and 1. The values in the list mean that the first (element 0) exposed qubit is actually the third (element 2) of the shared qubits. Likewise the second exposed qubit in register B (element 1) is actually the first (element 0) of the shared qubits. Accordingly all of the slicing operations available in the local simulation for Cove are really just creating new lists of exposed qubits and manipulating them. As an example if a `SliceReverse()` operation were called on register A then a new register would be returned with the qubits in the opposite order. This would be done by copying the list of exposed registers from register A then reversing it, as shown by extending Figure 144 to include a third register, register C, in Figure 145. The original register A and B from Figure 144 still exist with the same mappings, but are not shown.

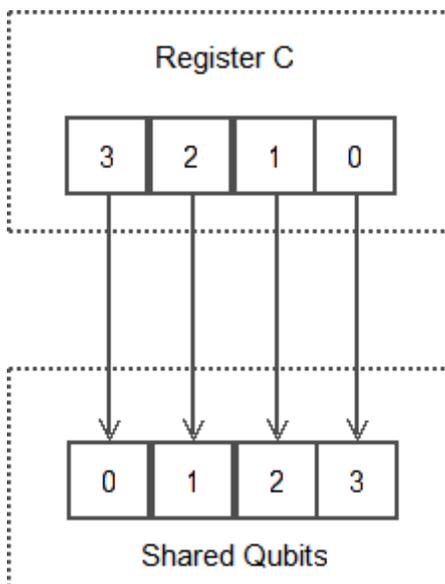

Figure 145. Creating a third register by reversing register A from Figure 144.

Since all of the registers actually share the same qubits, this solves the problem of manipulations to one register propagating to others that share the qubit. For instance, if the first exposed qubit in register A was put into superposition via a Hadamard operation then the second exposed qubit in register B would be in superposition, as would the fourth exposed register in register C.

### 5.5.6   *Enforcing Limitations of Quantum Computing in the Simulation*

The various limitations of quantum computing within the simulation are satisfied in the following ways:

- No copying qubits: All copies of quantum register classes are by reference and not value. No copy operation is supplied, so it is not possible to copy a qubit and violate the no-cloning theorem.

- Measurements cannot be undone: The measurement operation collapses the qubit to an absolute state, just like would happen in an actual register. No previous state information is preserved, making it impossible to undo this collapse. Additionally the only operation to get the value of the register is the `Measurement()` operation, meaning that there is no public backdoor to examine the qubit through the interfaces[97].

- Reversible operations: All operations are checked to ensure they are reversible before they are applied. A `NotUnitaryOperationException` is thrown when any nonreversible operation is applied to a register. In this case the state of the register is unaltered.

- Quantum resource must be local: In the current implementation the quantum resource is local. Trying to change the location to anything other than `localhost` results in a `ArgumentException` being thrown.

---

[97] The exception is that the local simulation does have  a implementation specific method to get the state. This is supplied largely for testing and learning purposes. If one adheres to the programming to interfaces method, this method will not be exposed. Furthermore the debugger will show the state as a private member of the register. While the debugger could be used to view the state, the method allows for tests to be written that don't inadvertently alter the state.

## 5.6    Cove Usability Properties

The usability properties of Cove outline goals that have been targeted for ease of use and are outlined in this section. Together with the proof criteria (section 3.2), which are the function properties, these largely represent the requirements Cove has been designed to satisfy.

- Consistent naming: To reduce the learning curve names of things such as classes, methods, and parameters should be consistent.

- Consistent ordering: Ordering of parameters should be consistent between similar calls and in overloaded methods the common parameters should be the first ones.

- Complete names: Abbreviations and the like can be ambiguous or confusing, the framework should always err on the side of detailed and complete names.

- Common prefixes: Similar methods should share common prefixes so they show up next to each other in alphabetical listings such as intellisense tools.

- Static methods: Methods that do not utilize the state of an instance should be static.

- Methods versus operators: Methods should modify the state of the object they are called on while operators should return new objects. An example would be add. `Foo.Add(Bar)` manipulates `Foo`, while `(Foo + Bar)` returns a new object and leaves `Foo` and `Bar` unaltered.

- Chaining operations: Where possible users should be able to "chain" together method calls for ease of use. So instead of calling `X()` on `Foo`, then `Y()`, and finally `Z()` all on separate lines, a user can call `Foo.X().Y().Z()`.

- No violations of quantum computing limitations[98].

- Programming to interfaces.

---

[98] Not violating limitations of quantum computing is also listed as a usability property because it is possible for simulations to violate them such as peeking at the state. While this is useful for debugging and testing, it shouldn't be supported by the language or framework since it cannot be done on an actual quantum computer.

- Base interface: All classes in the framework should inherit from a common interface to support encapsulation of common behavior and maintenance of the framework.

- Names not tied to physical behaviors: As an example "T gate" should be used as opposed to specifying the rotation of $\frac{\pi}{8}$.

- Users can create arbitrary operations: To be considered a framework as opposed to a library, the framework must be extendable– operations are one such area.

- Users can work with subsets of quantum registers: Oftentimes a user may only need to work with a subset of a register, the framework should support various ways of obtaining those subsets to work with.

- Changing ordering of common operations: As an example users should be able to specify arbitrary qubits for the control and target of a CNot operation.

- Exceptions instead of error codes.

- Operators applied to qubits, not vice versa: Registers should take operations as parameters to methods and modify their own state accordingly. Having operators modify the states of registers is a violation of core object oriented principles.

- Flexible initialization and measurement to classical types: Quantum computation typically starts with a classical state (usually $|0\rangle$) and results in classical data. The framework should make this easy to do in a variety of ways.

- Documentation: The importance of detailed and accurate documentation cannot be overemphasized.

This chapter has outlined the design and implementation of Cove in addition to usability properties that have been targeted. Cove has been designed so that working code can be written using the prototype simulation, but that implementation could be swapped out with another that perhaps runs on an actual quantum computer as easily as possible.

Cove itself may not end up being the actual way that quantum computers are programmed, but it is hoped that the focus on usability in commercial environments does.

CHAPTER VI

ANALYSIS OF RESULTS

This chapter is largely a reflection and analysis of Cove. As such this chapter consists of several sections: how common flaws are avoided (6.1), comparison to a few other proposals (6.2), and finally lessons learned (6.3).

**6.1    How Common Flaws are Avoided**

As mentioned in previous sections, existing proposals suffer from one or more common flaws that make them difficult to use in a commercial software development environment. The following is a list of those fatal flaws, and how Cove avoids them:

- Foreign techniques – Cove is designed for use with popular object oriented languages, which are familiar to most commercial programmers, unlike functional languages. No mathematical notation is utilized that cannot be expressed on a keyboard. Furthermore no proofs are required.

- Not scalable – While not a solution to all scalability problems, object oriented approaches help alleviate some of the problems such as bloat and complexity that arise in large scale systems developed with procedural techniques. By utilizing the object oriented approach scalability is emphasized since a user can focus only on the specific parts that will be utilized or extended.

- Proprietary language – Cove is built as a framework upon existing classical languages. In the current incarnation it is implemented in C#, which means the libraries can be used by any other .NET language such as Visual Basic.

- Difficult to integrate with existing software– By their very nature frameworks are built on top of existing languages. By utilizing objects in those languages, Cove can integrate with existing software. Furthermore Cove has been designed with various implementations in mind, such as swapping out a local simulation with one that runs on an actual quantum computer. This is possible because the interfaces and other common components are encapsulated in a common library which implementations are built against.

- Usability/Unconventional framework design– Common framework conventions have been followed wherever possible. Unlike many other quantum programming proposals, there is emphasis on usability. An example of this is the operations that can be applied to qubits. Methods are supplied that easily apply common operations, as in `OperationHadamard()`. For extendibility an `ApplyOperation()` method is also provided that works with any operation, including user defined ones. Furthermore this method is overloaded to also take an array of operations so that a list of operations can be applied in one call. The static class Operations is also provided so that users can reference operations by names instead of their mathematical representations, which are more verbose and hard to express in code.

- Runs only on a quantum computer– The fact that Cove has been implemented as a simulation on a classical computer via a framework built on classical languages clearly shows that it is not limited to quantum computers.

## 6.2    Comparison of Cove to Other Proposals

The following is a brief comparison of how various quantum programming tasks are carried out in Cove compared to a few other quantum programming proposals. The first proposal Cove is compared to is Omer's QCL. QCL is selected because it is one of the most complete and recognized quantum programming method. QCL and Bettelli's work are also the closest existing proposals to Cove. Figure 146 shows the allocation of two qubits, followed by a rotation on the first and Hadamard on the second.

```
qcl> qureg a[1]; qureg b[1]; // allocate 2 qubits
qcl> Rot(-pi/3,a);           // rotate 1st qubit
qcl> Mix(b);                 // Hadamard Transformation
```
Figure 146. Simple example in Omer's QCL, from [81].

The notation of QCL is easy enough to read compared to more mathematical methods. There are still a few areas for improvements however:

- A procedural approach is utilized in this instead of an object oriented one. While this itself is not a problem, it does mean that user created methods could incorrectly modify the state of a qubit. With an object oriented approach only valid methods are exposed. With a procedural approach it is easier to write errors that won't be detected until run time. An example of this is the application of an operation to a register: an object oriented approach can validate the operation is unitary before application. This is checking and exposure of valid methods one of the strong arguments for utilizing the object oriented approach, as Cove does. In short it helps to reduce programmer errors.

- The method names are abbreviated, decreasing readability. "Rotation" is clearer than the abbreviation "Rot". Furthermore, it isn't obvious what kind of rotation this is: about the x, y, or z axis, or maybe something else? The name "Mix" for performing the Hadamard transformation is also unclear and ambiguous, especially when there are a large number of functions available. Cove avoids the use of abbreviations as in the `ApplyOperation()` method on the qubit. The names utilized in Cove could be considered to be ambiguous, but the use of an object oriented approach lessens this since the methods apply to specific classes. While the method `OperationNot()` in could be considered ambiguous by itself, the fact that it exists as a method on the qubit breaks this ambiguity.

- The –pi/3 is could be considered a "magic number" and would better be replaced by a constant. The purpose of this rotation is unclear unless one is very familiar with the quantum computing[99]. This however merely concerns the example and not the language. In Cove the common rotations are given names such as `OperationYGate`, which has a clearer meaning: rotation ($\pi$ radians) about the Y axis.

The next example to briefly compare Cove to is Sanders and Zuliani's qGCL.

Figure 147 shows a single quantum coin toss in qGCL.

$$\textbf{var } \chi : q(\mathbb{B}), \ i : \mathbb{B} \bullet$$
$$In\,(\chi) \ \fatsemi$$
$$Fin[\Delta]\,(i)$$

Figure 147. A single quantum coin toss in Sanders and Zuliani's qGCL, from [74].

There are several usability flaws with this example in qGCL:

- Without being very familiar with qGCL or a good deal of comments, it is very hard to tell what this code snippet carries out– and it is only 3 lines of code.

- The mathematical notation is difficult to input on a traditional computer without a special interface. In a case such as this a programmer may spend time trying to express what they want to carry out in the notation at the expense of focusing on the steps to solve the problem. An easy to use method of programming should do as little to slowdown the user as possible.

For comparison, here is an initialization of a qubit and a single coin toss in Cove:

```
1. Qubit Test = new Qubit("Heads", "Tails");
2. Test.OperationHadamard();       //first toss
3. Test.Measure()                  //collapse and discard result.
```

Figure 148. Single coin toss in Cove.

These examples show that while existing techniques can functionally carry out quantum programming, they suffer from problems in usability. As the examples help to illustrate, this is the focus of Cove: a *usable* quantum programming framework.

## 6.3    Lessons Learned

There were many lessons learned throughout this project; this section just outlines some of the major ones that may be of use to others. For a more informal and complete discussion of issues encountered during the development of Cove, see the development blog on the website (see Appendix A for how to access the blog).

It was anticipated that the design of the interfaces would be one of the most difficult parts of the project, and it turned out to be such. That is because the interfaces are what will determine if the framework satisfies the list of functional properties, and how usable it is. While the initial design of the interfaces wasn't too difficult, the writing of the unit tests helped tremendously in evolving them into a hopefully more complete and usable set of interfaces. Writing the unit tests essentially forced a good deal of code to be written. While writing the test code it often became apparent that things would be easier if there was additional or altered functionality. In those cases work on the test code was paused while the interfaces (and implementation) were updated.

---

[99] Or mathematics for that matter. Math plays a much smaller role in computer science today than it did in the earlier days when computer science was typically part of math departments at universities.

The implementation of the simulation prototype was much more time consuming than anticipated. This in part was due to several challenges that aren't well covered in the literature. Two prime examples are the expansion of operators to apply to arbitrary sized registers and the reordering of operation targets. These are elaborated on in detailed in a rough draft form in a paper available on the website.

Carrying out factoring was also more difficult than anticipated. While the Quantum Fourier Transform is covered in great detail throughout the literature, $U_f$ was not. In most of the literature $U_f$ is treated as a black box and how it is built up from elementary operations is not covered. As the discussions about building it up in 2.1.11, 4.3.3, and 5.4.3 this is far from trivial. In fact most of the implementation of $U_f$ in Cove was based on Nakahara [23] and Vedral [131], although there were still challenges with some pieces that weren't explicitly outlined such as the "reversible reset". At the completion of this dissertation (fall 2009) all of the pieces necessary to carry out factoring are believed to be present, but the completion of $U_f$ is still incomplete. Obviously completion of the factoring example is a clear and immediate area for future research.

Initially Cove was started as an implementation of a single qubit as part of the proposal phase. Obviously a single qubit is of limited use, but the intent was to provide a start to the project and show that it was viable. This single qubit approach did not scale up, and was replaced with the existing quantum register. However, some of the ideas such as applying operations several ways did live on to become part of the register.

CHAPTER VII

CONCLUSION

## 7.1 Implications of the Research

There are several implications of this research. Perhaps most noteworthy is the focus on making Cove usable. Admittedly the judgment that something is usable or not is largely subjective– which makes it hard to conclude that Cove is usable with any certainty at this point. Nonetheless, it hoped that the focus on usability will raise awareness of the issue as far as quantum computer programming is concerned. As the review in chapter 2 illustrates, many existing proposals have severe usability hurdles. While classical computation has fundamentally been the same for decades, we have also spent that time refining our software methods to become more productive. By focusing on usability now, we have the chance to hit the ground running with quantum computers as opposed to learning along the way as we have done with classical computers.

A second implication of Cove is that it is a framework, and as such it builds on an existing classical language. The closest thing to Cove in this respect is Bettelli's C++ extensions. As Bettelli [2] and this dissertation advocate, the desired approach is an extension of a classical language. Most other proposals are languages developed specifically for quantum computation. Languages specifically for quantum computing lack the facilities for classical computation. Even with quantum computers, classical computing will be the majority of computation performed.

This is evident even in factoring (Shor's algorithm): a quantum computer is only exploited at one step, the rest is carried out classically.

Cove is unique in another aspect in that it draws a clear distinction between the interfaces, *what* must be provided for quantum computation, and the implementation, *how* that is provided. This separation allows for implementations to be swapped, with no code changes other than the using statements if the interfaces are strictly adhered to. This may be especially important as quantum computers are developed: this approach allows code written and tested with a simulation to be swapped out and run on an actual quantum computer.

As a framework, Cove has also taken into account that users may want to extend the functionality in ways not anticipated at design time. This allows users to more easily modify the framework as they see fit. This extendibility is more difficult if the proposal is something such as a proprietary language.

## 7.2    Areas for Future Work

One obvious area for future work is to complete the local simulation: making it more robust, adding more documentation, completing the factoring example, and implementing the rest of the methods specified by the interfaces[100]. Increased accuracy via more precise floating point numbers is also a possibility. Another area would be to have the simulation support remote users, perhaps through the use of web services to promote platform independence. However this also means that many of the multi-client issues that must be dealt with by the expanded QRAM model (section 5.2) must be dealt

---

[100] There are methods still not implemented in the local simulation that are specified by the interfaces. These methods are anticipated to be less commonly used and aren't required to implement the three examples used through this dissertation: tossing quantum coins, entanglement, and factoring.

with by the simulation. Of course, improvements based on user feedback are an obvious area as well.

As is outlined earlier, there is a hard limit of 62 qubits in the current local simulation constructed as part of this work. While any simulation suffers the exponential slow down in worst case, the simulation could be expanded over a grid of computers. Figure 149 shows the high level view of how a grid implementation would work. Doing a grid simulation would allow for simulations of larger number of qubits. Of course in order for this to be possible, the simulation must be partitioned in a way it can be spread across the grid. This in itself may be some work, and all the issues that apply to any grid still apply.

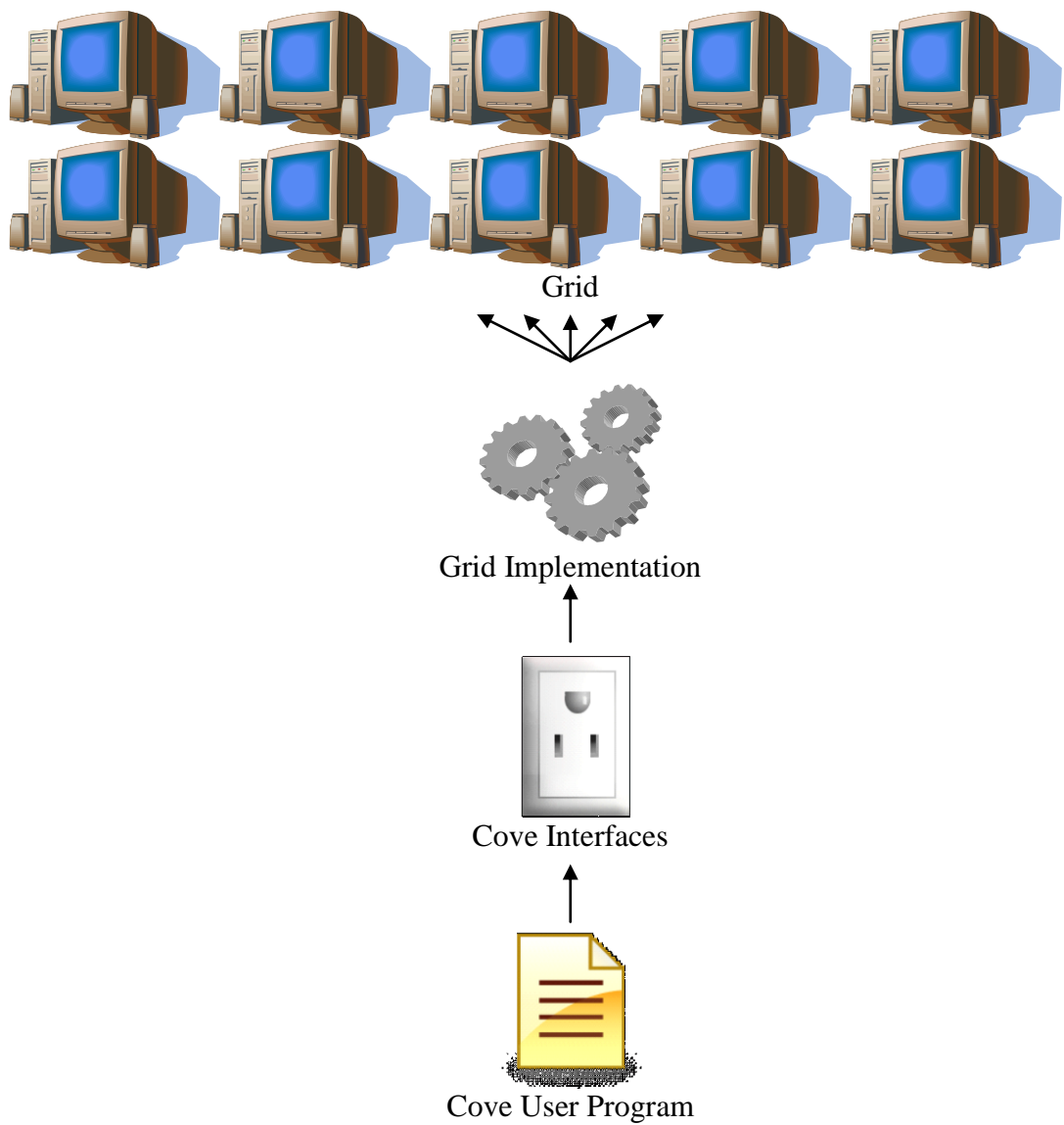

Figure 149. High level view of a grid implementation of Cove.

Similar to the grid idea, the simulation could also be scaled up by using the disk for memory instead of using RAM. Of course disk IO is much slower than RAM, but this would get around the memory limits.

There has also been work done in more efficient simulations of quantum computers. This work could also be leveraged to create a more efficient simulation implementation for Cove. While the worst case is still the same, exponential slow down, this may allow for users to more easily work with more simulated qubits.

The expanded QRAM model (section 5.2mulation implementation for Cove. While the worst case is still the same, exponential slow down, this may allow for users to more easily work with more simulated qubits.

The expanded QRAM model (section 5.2) and the QRAM model itself are also areas for futures work. While not a specific focus of this dissertation and literature review within it, there are many details to be worked out on how quantum primitives will interface with the physical quantum computer. There are many issues to work out such as optimization of requests[101], what to do when classical computers fail in the middle of computations, quantum memory management, and so on.

As mentioned in Chapter 4, integration of quantum circuit diagrams within a classical computing framework is also an area for work. This would be useful since the quantum part of a computation is merely one piece of the overall task. What would the circuit diagram be for all Shor's algorithm, not just the quantum part[102] [103]?

Another area for future work would be better graphical notation of quantum computation itself. The Bloch sphere is the standard choice for a single qubit, but how would multiple qubits be visualized? Being able to visualize the computation may help students grasp what quantum computation is more quickly. Related to graphical notation of the computation itself, there could also be more work done towards diagramming large quantum computer systems.

---

[101] Perhaps by queuing up operations until a measurement one is received. At that point all the operations could be quickly applied and measured. This would reduce problems with decoherence due to long delays between operations being sent.
[102] Surprisingly, even the entire diagram of the quantum part is often lacking in the literature. As an example the "$U_f$" part of the diagram is often listed as a black box, even though it is the most complex piece.
[103] As stated in 2.2.2 graphical programming techniques are not practical, but the circuit diagrams (or subsets) are useful in understanding the algorithms.

## 7.3    Summary

Quantum computers mimic nature more accurately than classical computers and represent a fundamental change in the way computing is done– quantum computers do not fall under the classical computing machines Alan Turing envisioned in the 1930s that have been the model for computation ever since. As such, they are able to perform certain types of computations that are not efficient on classical computers. Some examples of problems efficiently solvable by quantum computers include unsorted search, factoring, and perhaps not surprisingly, simulation of quantum systems.

While much work remains in order for quantum computers to become viable commercially, it is largely believed that this will occur in the early decades of the twenty first century [7]. Nonetheless, quantum computers and algorithms are of little use if they cannot be utilized effectively in software. This dissertation proposes a practical programming framework for quantum computing called Cove. The practical part is key– the goal is for the framework to be as simple to use as possible for current commercial programmers. The existing proposals for quantum computer programming suffer from flaws which make them impractical for existing commercial programmers, who write a majority of classical software in use and would perhaps do the same for quantum computers. Furthermore, Cove is a framework and as such leverages all the classical computing capabilities of the language it is built on. This frees Cove to focus only on quantum computation without trying to provide classical functionality as well, which languages specific for quantum computing must do.

It is the goal that the framework proposed here does not suffer from the usability flaws that have been outlined in the introduction. By doing so we have a chance to experiment with and write software for quantum computers before they appear commercially. This allows us to identify and resolve software problems as they pertain to quantum computers ahead of their introduction instead of learning as we go, as has largely been the case with classical computers.

## APPENDIX A

## ELECTRONIC RESOURCES

Electronic resources pertaining to this project can be obtained from [https://cove.purkeypile.com/trac/](https://cove.purkeypile.com/trac/). The home page (accessible via the "wiki" tab) also contains some links to specific items:

- Latest drafts of dissertation, proposals and presentations.
  - This also includes sections that have been cut from this dissertation, including simulation challenges and a more in depth discussion of the usability properties.
- All source code, including examples.
- Online documentation of Cove ([https://cove.purkeypile.com/Help](https://cove.purkeypile.com/Help)).
- A blog on the development of Cove.

All of the documents and source code are also available for download from the Subversion repository, which is an easy way to receive any updates. The URLs are case sensitive and are https, not http. The latest can always be obtained via Subversion:

- Documents: [https://cove.purkeypile.com/PurkSVN/Trunk/ResearchDocs](https://cove.purkeypile.com/PurkSVN/Trunk/ResearchDocs)
- C# source: [https://cove.purkeypile.com/PurkSVN/Trunk/Cove-CSharp](https://cove.purkeypile.com/PurkSVN/Trunk/Cove-CSharp)

Tortoise SVN is an excellent Subversion client for Windows that will allow you to download the files. You can obtain Tortoise from [http://tortoisesvn.tigris.org/](http://tortoisesvn.tigris.org/). The "Browse Source" from the website also lets you view the source code in a browser. Furthermore periodic builds are made of the source code and available to for download as a single compressed file for those who wish not to deal with pulling or building from the Subversion repository.

Following much of this work a blog on quantum computer programming and related topics has been started and can be accessed at http://mpurkeypile.blogspot.com/.

The author can also be followed on Twitter at http://twitter.com/mpurkeypile or contacted via email at mpurkeypile@acm.org.